\documentclass[prb,
reprint,
superscriptaddress,
amsmath,amssymb,
aps,twocols, tabular
]{revtex4-2}

\usepackage{graphicx}
\usepackage{dcolumn}
\usepackage{bm}
\usepackage{hyperref}
\usepackage[mathlines]{lineno}
\usepackage{comment}
\usepackage{graphicx,epsfig} 
\usepackage{amsmath}
\usepackage{amsfonts}
\usepackage{bm}
\usepackage{amssymb}
\usepackage{amsmath,lipsum}
\usepackage{orcidlink}
\usepackage[normalem]{ulem}

\newcommand{\mcurv}{D}
\newcommand{\aspv}{u}
\newcommand{\cratio}{K_0}

\begin{document}

\title{Charge-tunable acoustic surface plasmons in topological nodal-line semimetals}

\author{Francesco Buccheri}
\email{f.buccheri@ifw-dresden.de}
\affiliation{Leibniz Institute for Solid State and Materials Research,\\ IFW Dresden, Helmholtzstrasse 20, 01069 Dresden, Germany }
\author{Francesco M. D.  Pellegrino}
\affiliation{Dipartimento di Fisica e Astronomia “Ettore Majorana”, Università di Catania, Via S. Sofia 64, I-95123 Catania, Italy}
\affiliation{INFN, Sez. Catania, I-95123 Catania, Italy}
\author{Jeroen van den Brink}
\affiliation{Leibniz Institute for Solid State and Materials Research,\\ IFW Dresden, Helmholtzstrasse 20, 01069 Dresden, Germany }
\affiliation{W\"urzburg-Dresden Cluster of Excellence ctd.qmat}

\date{\today}

\begin{abstract}
    Topological nodal-line semimetals host  drumhead surface states, carrying a large surface density of states. We show that these states support an acoustic surface plasmon, whose velocity is strongly enhanced by the Coulomb interaction.
    At sufficiently small wavevectors, the mode lies above the surface particle-hole continuum and below both mixed surface-bulk absorption thresholds, leaving bulk particle-hole creation as the leading damping channel.
    Near drumhead depletion, the dispersion is highly sensitive to the surface charge, which suggests applications of plasmon spectroscopy of nodal-line semimetals to sensing technologies.
\end{abstract}

\maketitle


Nodal-line semimetals (NLS) are characterized by a one-dimensional manifold in which valence and conduction bands come in contact \cite{Burkov2011,Fang2015,Fang2016,Chiu2014}. The resulting toroidal Fermi surface generates characteristic transport signatures and anisotropic optical responses \cite{Syzranov2017,Rui2018,Biderang2018,Yang2022,Mukherjee2017,Pronin2021,Anand2026,Rhim2016}.
Notably, hyperbolic bulk plasmons have been predicted \cite{Singh2023} and observed \cite{Shao2022,Santamariagarcia2026,Chang2025} in NLS.
While surface plasmons have been studied in metals \cite{Nagao2001,Diaconescu2007,Silkin2008}, topological insulators \cite{Stauber2017,DiPietro2020} and Weyl semimetals \cite{Song2017,Andolina2018,Weylcone,Pellegrino2025}, the theory of light-matter interactions in NLS has been so far primarily concerned with bulk properties \cite{Losic2022}. Still, NLS host topological interface states, observable in ARPES and QPI experiments \cite{Biderang2018,Muechler2020}, which can enter optical IR responses \cite{Buccheri2024}. In this work, we explore the half-space geometry and study theoretically a collective surface charge oscillation carried by these states.

Crossings, including nodal lines, are actually a recurring band feature across many material families.
In centrosymmetric space groups, every nonzero spinless symmetry indicator corresponds to nodal lines at generic momenta \cite{Song2018}. Indeed, high-throughput computations identify thousands of materials in which crossings take place on high-symmetry lines in the Brillouin zone, and few hundreds more exhibit nodal lines at generic momenta \cite{Zhang2019c,Vergniory2022}.
The underlying reason is that, neglecting spin-orbit coupling (SOC), combined inversion and time-reversal symmetry permits a real Hamiltonian: for two bands, a crossing then requires two conditions in three momentum dimensions, allowing stable nodal lines once a suitable band inversion occurs \cite{Fang2015,Kim2015,Tiwari2020}. 
When SOC is included, the most common scenario is the opening of a gap \cite{Vergniory2022}. Mirror or nonsymmorphic protection nevertheless can allow nodal lines to survive SOC \cite{Chiu2014}, and the physics associated with a nodal line still applies whenever the gap is small compared with the relevant energy scales.
In fact, theoretical predictions and experimental probes have uncovered a growing number of NLS representatives \cite{Huang2016,Weng2016,Yang2018s,Chang2025}, including Ca$_3$P$_2$ \cite{Chan2016a}, the ZrSiX family \cite{Hosen2017}, CaAgAs \cite{Takane2018},
and various others \cite{Belopolski2019,Liu2018TiB2,Yi2019TiB2,Lou2018,Hosen2020,Mori2022,Song2023,Yang2024,Akiba2024}.
In addition to those, synthetic analogues include ultracold fermions \cite{Song2019}, phononic crystals \cite{Deng2019} and photonic lattice platforms \cite{Park2022}. 

Acoustic surface plasmons (ASP) are low-energy collective oscillations of surface electrons screened by bulk carriers, with a frequency linear in the in-plane wave number at long wavelengths \cite{Silkin2004,Pitarke2004}. Predicted for metallic surfaces, they were  observed in Be(0001) \cite{Diaconescu2007}, Cu(111) \cite{Pohl2010}, and very recently in the NLS ZrSiS and ZrSiSe \cite{Xue2026}.
In contrast to conventional surface plasmon-polaritons, which are hybrid electromagnetic and charge-density waves, whose long-wavelength dispersion approaches the light line, ASPs remain essentially electrostatic, with propagation velocities governed by electronic rather than optical scales \cite{Pitarke2004,Pitarke2007}. Their microscopic origin is also different, relying on the presence of a partially filled surface band, rather than on a sign change of the permittivity across an interface. Compared to bulk plasmons, surface confinement changes the plasmon dispersion from gapped to gapless, while bulk screening turns the energy dispersion from $\omega\propto\sqrt{q}$, characteristic of 2d materials \cite{Rahimpoor2024}, to a linear dispersion $\omega\propto q$ \cite{Pitarke2004}.

Here, we establish ASPs as coherent surface modes of NLS, originating from a partially filled topological drumhead state.
We find a strongly nonretarded linear dispersion relation in the long-wavelength limit, with a phase velocity \eqref{eq:u0} determined by the surface Fermi velocity and dressed by the screened interaction. 
We predict that Landau damping from the surface and from surface-bulk terms is suppressed at small $q$ by kinematic constraints associated with the band structure of the NLS, with the leading source of dissipation arising from the bulk particle-hole continuum, resulting in a linewidth $\sim\mathcal{O}\left(q^{2}\right)$.
Near the drumhead band edge, the rapid variation of the surface Fermi velocity produces a strong plasmon tunability, with the ASP dispersion reacting strongly to variations of the chemical potential and of the surface charge. 
Our results connect tunable surface collective dynamics to the distinct localization and screening scales of nodal-line semimetals. Moreover, they pave the way to applications of topological NLS to charge sensing in the THz regime, based on the detection of the surface plasmon resonance. 
Although the surface states have a different, nontopological origin in this material, our findings are consistent with the recent observation of a surface plasmon mode in ZrSiSe \cite{Xue2026}.
	
\emph{Model Hamiltonian and surface states.} To retain qualitative features without sacrificing generality, we employ a minimal low-energy Hamiltonian, which describes a pair of spinful valence and conduction bands and an energy-degenerate circular nodal line \cite{Chan2016a}.
The half-space problem has been studied in \cite{Rudi2024}, who showed that the dispersion of the drumhead state depends not only on the bulk particle-hole symmetry breaking, but on details of the surface as well, e.g., the surface composition. The Hamiltonian is diagonal in spin and acts in orbital space as \cite{Chiu2016,Schnyder2008,Buccheri2024}
\begin{equation} \label{eq:Hk}
    H_{\boldsymbol{k}}=- i\hbar v\tau^{y}\partial_{z}+m_{\boldsymbol{k}}\tau^{z}+g\,m_{\boldsymbol{k}}\tau^{0},
\end{equation}
with $m_{\boldsymbol{k}}=\mcurv(k^{2}-k^{2}_{\mathrm{NL}})$, $\bm{k}=(k_x,k_y)^{\rm T}$, and $k=\lvert\boldsymbol{k}\rvert$. Here $v>0$ is a velocity along $z$, $D>0$ parameterizes the in-plane dispersion, $k_{\mathrm{NL}}>0$ is the nodal-ring radius; $g$ a dimensionless particle-hole-asymmetry parameter, for a type-I semimetal $\lvert g\rvert<1$. The model is invariant under out-of-plane mirror transformation, represented by the unitary matrix $\tau^z$.

We study a setting in which a dielectric extends for $z<0$, while the NLS occupies the half-space $z>0$. 
In these conditions, the Zak phase quantized by the mirror symmetry protects a two-dimensional band of exponentially localized states \cite{Burkov2011,Chan2016a,Biderang2018,Buccheri2024}. The energy of these "drumhead" bands is
\begin{equation}\label{eq:ES}
    E_{S}(\boldsymbol{k})=
    C(k^{2}-k^{2}_{\mathrm{NL}}),
\end{equation}
in terms of the surface band curvature \mbox{$C=\mcurv(g+\cos \vartheta )$}. This is parameterized by a phenomenological angle $ \vartheta $, encoding the details of the surface via a general boundary condition \cite{Rudi2024}.
From \eqref{eq:ES}, we read the surface DOS \mbox{$\nu_S=1/(2\pi | C|)$} and the surface Fermi velocity \mbox{$v_S=2|C|k_F/\hbar$}, where $k_F$ is the radius of the drumhead Fermi contour.
The position of the chemical potential is conveniently expressed as a ratio to the bandwidth \mbox{$\mathcal{Y} =|\mu/Ck_{\mathrm{NL}}^{2}|$}. A necessary condition for the observation of the ASP is that the Fermi energy lies within the drumhead band, namely $\mathcal{Y} \in ]0,1[$.
The drumhead state wavefunction is exponentially localized around $z=0$, the penetration length being a function of the in-plane momentum $\boldsymbol{k}$ and diverging at the nodal line. At the Fermi energy, the probability density profile is
\begin{equation}
    w_{F}(z)=2\kappa_{F}e^{-2\kappa_{F}z} \label{eq:wF(z)},
\end{equation}
in which $\kappa_{F}$ is the inverse penetration length at the Fermi energy.
We consider a diagonal, anisotropic permittivity tensor in the material for $z>0$, with background permittivity $\boldsymbol{\epsilon}_{b}=\mbox{diag}(\epsilon_{\parallel},\epsilon_{\parallel},\epsilon_{\perp})$. All components are positive. The relative permittivity in the dielectric, extending for $z<0$, is denoted as $\epsilon_{2}$. 
The relevant frequencies in this work lie well below the hyperbolic regime studied in \cite{Singh2023}.

\emph{The acoustic surface plasmon.} 
In order to identify the plasmon resonance, we consider the dielectric operator $\mathcal{E}$. In terms of the bare Coulomb interaction potential $v_q$ and of the proper polarization $\Pi^R$, it is given by \mbox{$\mathcal{E}=\mathbb{I}-v_q\Pi^R$} within the random phase approximation (RPA) \cite{Giuliani}. 
A plasmon is the self-sustained solution of the homogeneous integral equation $\mathcal{E}U=0$ for the electrostatic potential profile $U$ \cite{Newns1970,Pitarke2007}. In particular, a nontrivial solution exists when the dielectric operator has a zero eigenvalue, and can be identified as a surface plasmon when its induced charge density and electric field decay away from the interface.
The relevant bands in our analysis are the drumhead surface band (S), see \eqref{eq:ES} and the bulk (B) valence and conduction bands, described by \eqref{eq:Hk}; the boundary eigenproblem is reviewed in App. \ref{app:Eigenstates}. The creation of particle-hole excitations in the combined B and S channels, on the one hand, produces damping of the plasmon mode, on the other hand, screens the Coulomb interaction felt by the surface electrons.
To isolate the collective surface response while retaining bulk and mixed screening, we factorize the dielectric operator as \cite{Aryasetiawan2004}
\begin{equation}
    \mathcal{E}=
    \left(\mathbb{I}-v_{q}\Pi_{r}\right)\left(\mathbb{I}-W_{r}\Pi^{SS}\right).\label{eq:factorization}
\end{equation}
Here the regular-sector polarization $\Pi_{\mathrm{r}}$ contains the BB, SB and BS contributions. The interaction \mbox{$W_{\mathrm{r}}=\left(\mathbb{I}-v_{q}\Pi_{\mathrm{r}}\right)^{-1}v_{q}$} therefore includes screening by these sectors, while the second factor, containing the surface polarization $\Pi^{SS}$, describes the surface intraband response. Details about the calculation of the polarization sectors are in App. \ref{app:Polarization} and App. \ref{app:StaticRegular}.
Projection onto the drumhead band produces a scalar collective mode condition
which can be studied analytically in the nonretarded limit. 
The drumhead band allows an ASP branch to appear, which was studied in metals in \cite{Pitarke2004,Silkin2004}.
This mode is in addition to the surface-plasmon-polariton branch, whose dispersion is $\omega\approx c q$ at small wavevectors. 

The relevant length scales are the bulk screening length $q_{\rm TF}^{-1}$, expressed via the Thomas-Fermi wavevector $q_{\rm TF}\propto\sqrt{|\mu|}$, and the penetration length of the drumhead state at the Fermi energy $\kappa_F^{-1}\propto|\mu|^{-1}$. Their ratio is conveniently expressed via the dimensionless variable \mbox{$\rho={\kappa_{F}}/{q_{\mathrm{TF}}}$}. The limit $\rho\ll1$ is associated with a large penetration of the drumhead states in the bulk and strong screening, while the opposite limit $\rho\gg1$ with a small penetration and weak bulk screening.
The details of the material enter through the dimensionless combinations
\begin{equation}
    \begin{aligned}
        \rho_{0}
        &=\mcurv k_{\mathrm{NL}}\sin \vartheta 
        \sqrt{\frac{2\pi\varepsilon_0\epsilon_\perp(1-g^2)^{3/2}}
            {e^2\hbar v|g+\cos \vartheta |}},\\
         \cratio &=\frac{\nu_S\kappa_F}{2\nu_B}=\frac{\sin \vartheta (1-g^2)^{3/2}}{(g+\cos \vartheta )^2},
    \end{aligned}	
    \label{eq:scaling-material-constants}
\end{equation}
where $\nu_B$ denotes the bulk density of states per spin at the Fermi energy. 
Band structure and dielectric screening jointly determine $\rho_0$, which is the maximum value of $\rho=\rho_0\sqrt{\mathcal{Y}}$ achievable within the drumhead band. For $\rho_0\gg1$ a weak screening window exists, while in the opposite limit $\rho_0\ll1$ bulk screening is strong on the penetration scale throughout the whole band.
The compressibility ratio $ \cratio $ compares the surface compressibility with the bulk compressibility per unit area, integrated over a surface penetration depth.
For drumhead states, it is independent of the chemical potential.

The screening from surface and bulk enters the ASP dispersion through the dimensionless effective interaction \cite{Pitarke2007}
\begin{equation}
     \zeta =\nu_S \mathcal{V}_0.
    \label{eq:beta}
\end{equation}
Here, $\mathcal{V}_0$ is the static screened interaction projected onto the drumhead charge profile, see App. \ref{app:finite-green}. 
The parameter $ \zeta $ is the ratio between the quantum capacitance of the surface \cite{Luryi1988,Ando1982,Mukherjee2011} and the effective charging capacitance associated with the screening from the regular sectors \cite{Christen1996}.
When screening from mixed bulk-surface excitations is neglected, the coupling \eqref{eq:beta} reduces to the simple form
\begin{equation}\label{eq:bulk-only-coupling}
     \zeta _B= \cratio \frac{1+4\rho}{(1+2\rho)^2},
\end{equation}
with $ \cratio $ in \eqref{eq:scaling-material-constants}.
The mixed-term corrections are always negative, i.e., $\zeta\leq\zeta_B$.
For $\rho\ll1$, the drumhead charge extends over many bulk screening lengths and $ \zeta _B\to \cratio $.
For $\rho\gg1$, on the contrary, the surface charge is confined within a small region compared to the bulk screening length and acts as a charge sheet: 
$\zeta_B\to\zeta_{\mathrm{sh}}\equiv\cratio/\rho=\nu_S e^2/(\varepsilon_0\epsilon_\perp q_{\mathrm{TF}})$, the coupling of a surface charge sheet screened by the bulk \cite{Pitarke2004}.
The inequality $5/9\leq\zeta_B/\min(\cratio,\zeta_{\mathrm{sh}})\leq1$ holds for any $\rho$.

The complex solution of the scalar collective mode condition, obtained from \eqref{eq:factorization}, can now be written in the form
\begin{equation}\label{eq:oSP}
    \omega_{\mathrm{SP}}(q)=\aspv q+c_2 q^2-i\Gamma_2q^2+o(q^2),
\end{equation}
in which we read the ASP dispersion from the real part, and the linewidth from the imaginary part.
The phase velocity at $q=0$ is
\begin{equation}
\aspv =v_S\frac{1+ \zeta }{\sqrt{1+2 \zeta }}
\label{eq:u0}
\end{equation}
and is renormalized upwards by the Coulomb interaction, via the parameter $ \zeta $ introduced above.
The dispersion bends downwards at larger momenta, as $c_2<0$.
Additional details and analytic expressions of the coefficients are in App. \ref{app:finite-green} and App. \ref{app:Landau-damping}.

\emph{Damping and resolution of the ASP mode.}
Interactions cause the ASP mode to emerge from the surface particle-hole continuum, but a bulk branch of massless excitations is nevertheless still present. A natural question is whether the mode can still be resolved under such conditions. To answer, consider the screened interaction 
\begin{align}\label{eq:WRgsurf}
    &W^{R}\left(q;z<0,z'<0;\omega\right)\\
    &\qquad=\frac{e^{2}}{2\varepsilon_{0}\epsilon_{2}q}\left[e^{-q\left|z-z'\right|}-g_{\rm surf}\left(q,\omega\right)e^{q\left(z+z'\right)}\right],
    \nonumber
\end{align}
for a source and a probe in the dielectric. The first term is the direct Coulomb interaction in the exterior medium of relative permittivity $\epsilon_{2}$. Surface electrons respond to this screened field, producing an additional charge fluctuation and a reflected field in the dielectric:
here $g_{\mathrm{surf}}(q,\omega)$ describes the complete response of the surface, including the effect of the bulk within the adopted approximations.
\begin{figure}[h]
    \centering
    \includegraphics[totalheight=0.21\textheight]{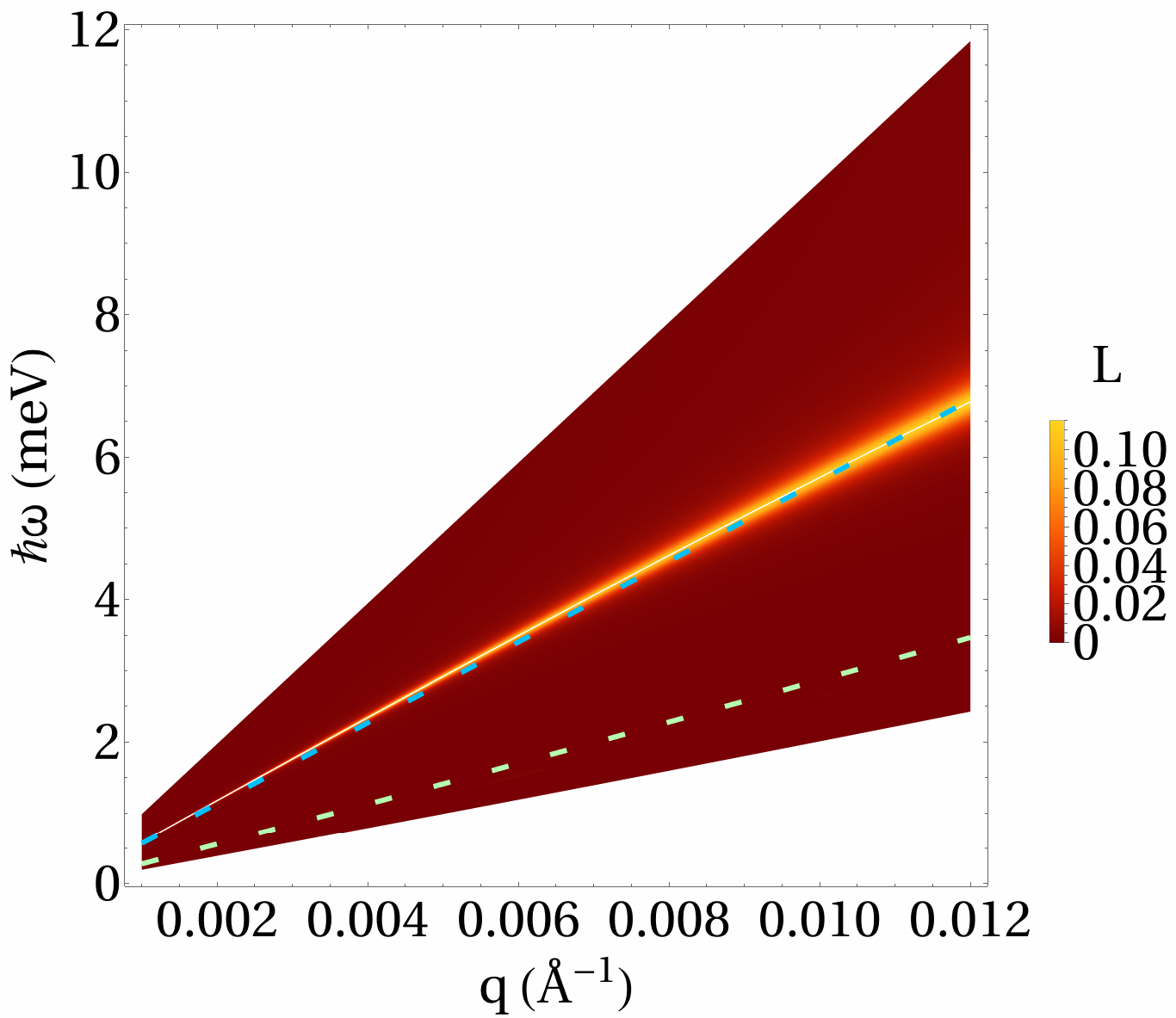}
    \caption{
    Loss function $L(q,\omega)$.  Its maximum tracks the dispersion of the ASP in the energy-wavevector plane. The blue dashed line is the analytical small-$q$ dispersion \eqref{eq:oSP}. The resonance broadening originates, in the displayed regime, from bulk particle-hole creation. 
    The mode remains resolved from the upper edge of the drumhead particle-hole continuum, marked with a green dashed line.
}
\label{fig:Loss}
\end{figure}
Its poles in the complex plane provide the dispersion and the lifetime of the
surface plasmons. 
The loss function
\begin{equation}
    L(q,\omega)=\Im\left[g_{\rm surf}(q,\omega)\right]\label{eq:Loss}
\end{equation}
is typically the experimentally measured signal, e.g., in reflection EELS \cite{Xue2026}.
The maximum of $L$ traces the ASP dispersion, while its linewidth measures its damping \cite{Ibach}. 
The full $g_{\rm surf}$ is determined by the interaction of the surface electrons with the dielectric environment and with the surface and bulk bands.

We describe the response of bulk electrons screening the Coulomb interaction within the static Thomas-Fermi approximation and retain the leading low-frequency absorption
responsible for the ASP resonance broadening coming from both bulk and surface terms.
We solve the integro-differential equation for the electrostatic potential by projecting it onto a functional subspace \cite{BrennerScott2008}. This is spanned by two trial exponentials \cite{Ando1982}, with inverse decay lengths $q_{\rm TF}$ and $2\kappa_F$, and is exact when the mixed surface-bulk terms are neglected.
To validate our approach, we tested the analytic slope against a converged numerical implementation \cite{Boyd2001,Atkinson1997} of the same model
, consistently finding relative errors below 5\%.

The gapless bulk intraband sector gives the leading contribution to the linewidth 
\mbox{$\Gamma_B(q)=-\Im\left[\omega_{\mathrm{SP}}(q)\right]=\Gamma_2 q^{2}$}, while the surface intraband and the mixed surface-bulk sectors only contribute away from $q=0$, due to kinematic constraints. We obtain, up to corrections that vanish for $q\to0$, a quality factor 
\begin{equation}
    Q_{SP}(q)=\frac{\Re\left[\omega_{\mathrm{SP}}(q)\right]}{2\Gamma_B(q)}
    =\frac{(1+2 \zeta )^{3/2}}{2\nu_Sv_S \zeta \mathcal C_r}\frac{1}{q},
    \label{eq:QSP}
\end{equation}
with the absorption coefficient $\mathcal{C}_{r}$ quantifying how efficiently an oscillating drumhead charge density can lose energy by creating particle-hole pairs.  
In this long wavelength limit, our zero-temperature RPA calculation predicts that the acoustic mode becomes asymptotically sharp. 

The ratio of the ASP resonance shift from the particle-hole continuum to the linewidth also diverges as $1/q$, so we conclude that the ASP should be observable as a well-defined resonance for $q\to0$. The conclusion holds for generic experimentally relevant parameters, within the stated approximations and regimes. As an example, we plot the loss function in Fig. \ref{fig:Loss}, in which the parameters have been modeled on Ca$_3$P$_2$ \cite{Chan2016a}. The maximum of this function tracks an approximately linear resonance, whose slope is predicted by the dispersion relation \eqref{eq:oSP}. The bright region never overlaps with the threshold of the surface particle-hole continuum, and the linewidth shrinks for $q\to0$. Additional details about the resonance broadening can be found in App. \ref{app:Landau-damping}.

A relevant aspect concerns the dependence of the ASP velocity on variations of the Fermi energy and surface charge, at fixed material parameters.
The slope $\aspv $ depends on the Fermi energy through both the surface Fermi velocity and the screened coupling, as shown in \eqref{eq:u0}. Near the drumhead band edge, the surface Fermi contour contracts to a point, while the coupling approaches a finite value $ \zeta _*$. 
At fixed material parameters near depletion, the sensitivity to the charge density $\sigma$ carried by drumhead electrons is determined by the derivative
\begin{equation}
    \left|\frac{d\aspv }{d\sigma}\right|=\frac{k_{\mathrm{NL}}|C|}{\hbar\sigma_{\mathrm{D}}}\frac{1+ \zeta _*}{\sqrt{1+2 \zeta _*}}\frac{1}{\sqrt{1-\mathcal Y}}\left[1+O(1-\mathcal Y)\right],
    \label{eq:band-edge-speed-speed}
\end{equation}
in which $\sigma_{\mathrm{D}}={ek_{\mathrm{NL}}^2}/{2\pi}$ is the charge density of a full drumhead band.
The growing magnitude of the derivative when $\mathcal{Y}\to1$
reflects the small carrier population remaining near depletion,
as a given charge transfer changes an increasingly large fraction
of the electron or hole population in that pocket.
For charge variations small compared with the charge of the residual pocket,
$|\delta\sigma|\ll\sigma_{\mathrm{D}}(1-\mathcal Y)$,
the scaling $u\propto\sqrt{1-\mathcal Y}$ implies a relative frequency shift
\mbox{$|\delta\Re\omega_{\mathrm{SP}}|/\Re\omega_{\mathrm{SP}}
\simeq |\delta\sigma|/[2\sigma_{\mathrm{D}}(1-\mathcal Y)]$}.
Using \eqref{eq:oSP} and \eqref{eq:QSP}, we obtain the resonance shift relative to the half-width,
\mbox{$|\delta\Re\omega_{\mathrm{SP}}(q)|/\Gamma_B(q)
\simeq Q_{\mathrm{SP}}(q)\,|\delta\sigma|/
[\sigma_{\mathrm{D}}(1-\mathcal Y)]$}.
The shift is enhanced both by the quality factor, which grows
as $1/q$, and by the proximity to drumhead depletion.
By comparison, the deeply occupied metallic surface band supporting the ASP in Be(0001) undergoes a much smaller fractional occupation change under a small charge transfer \cite{Pitarke2004}.
The shallow drumhead bandwidth makes the depletion regime accessible. 
Additional properties of the resonance in limiting regimes can be found in App. \ref{app:scaling}.

\emph{Materials.}
To quantify how the described effects combine in actual materials, we show in Table \ref{tab:u0v-ratios} the values of the relevant material parameters, of $ \zeta $ and of the ensuing ratio of the phase velocity to the surface Fermi velocity $\aspv /v_S$ for representative materials. The estimates are obtained from available literature as described in App. \ref{app:Materials}.
\begin{table}[h!]
	\centering\small
	\resizebox{0.95\columnwidth}{!}{%
            \begin{tabular}{@{}l c c c c c c@{}}
			Material & $|C|/D$ & $ \cratio $ & $\rho$
			& $ \zeta _B$ & $ \zeta $ & $\aspv /v_S$ \\
			\hline
			Ca$_3$P$_2$ (doped) & $0.219$ & $17.1$ & $0.432$
			& $13.4$ & $6.84$ & $2.05$ \\
			BaSn$_2$ (SOC-free)& $0.980$ & $0.585$ & $0.338$
			& $0.490$ & $0.445$ & $1.05$ \\
			CaAgAs (SOC-free)& $0.157$ & $26.1$ & $0.917$
			& $15.2$ & $8.42$ & $2.23$ \\
			\hline
			ZrSiSe (FSS) & $-$ & $-$ & $\infty$
			& $2.73$ & $2.73$ & $1.47$ \\
			Be(0001) & $-$ & $-$ & $-$ & $-$ & $-$ & $1.03$ \\
		\end{tabular}	}
	\caption{Four NLS representatives: drumhead curvature in \eqref{eq:ES} in units of the bulk mass curvature,
            compressibility ratio $ \cratio $ in \eqref{eq:scaling-material-constants}, 
		screening-to-penetration length ratio $\rho$ at the chosen chemical potential,
            projected interaction parameter $ \zeta $ \eqref{eq:beta}, its upper bound $\zeta _B$ \eqref{eq:bulk-only-coupling}, 
            and ASP velocity ratio from \eqref{eq:u0}.
		ZrSiSe hosts floating surface states (FSS) instead of drumhead states; its estimate uses the surface charge sheet, $\kappa_F\to\infty$ and $\zeta=\zeta_B\to\zeta_{\mathrm{sh}}$.
		Be(0001) appears for comparison with a conventional metal; estimate from \cite{Pitarke2004}.}
	\label{tab:u0v-ratios}
\end{table}
The table shows that larger values of $ \zeta _B$ accompany larger velocity enhancement. Equation \eqref{eq:bulk-only-coupling} identifies then two favorable characteristics: a large $ \cratio $ and a large $\zeta_{\mathrm{sh}}$. The former parameter is only determined by the band dispersions; its $(D/C)^2$ dependence in \eqref{eq:scaling-material-constants} favors drumhead bands with small curvature compared with the bulk mass curvature. 
The latter is set by electrostatics: 
moving the Fermi level toward the nodal line lowers the bulk DOS and screening, raising $\zeta_{\mathrm{sh}}$. 
%
To observe the ASP resonance in NLS, these data suggest searching for materials whose drumhead band is flat compared with the bulk mass curvature, a criterion with no counterpart in metals 
\cite{Pitarke2004,Yan2012,Xue2026}.
For ZrSiSe, whose surface band is a floating band of nontopological origin, the charge-sheet limit of our model predicts $\aspv/v_S = 1.47$, within the measured bracket $1.3-1.6$~\cite{Xue2026}.
        
\emph{Discussion.}
We have presented a theory of acoustic surface plasmons in nodal-line semimetals, using a minimal model which describes two spin-degenerate bands meeting along a circular nodal line.
Using available literature, we fitted the parameters of the minimal model to representative NLS materials: the calculated ASP velocities show that the drumhead bands can produce a large effect on the dispersion for experimentally relevant values of the parameters.
%
The mechanism differs qualitatively from metal surfaces:
in metals, a large ASP velocity enhancement requires weak
bulk screening. In NLS,
as the chemical potential approaches the nodal line ($\rho \to 0$),
the drumhead penetration depth grows as the bulk DOS decreases: the two trends combine
and the magnitude of the effect is instead set by the band structure alone.
In particular, the upper bound on the effective interaction $\zeta_B$, approaches the finite value $K_0$, which scales as $(D/C)^2$. 
Flatness of the drumhead band relative to the bulk can therefore produce a large enhancement even under strong bulk screening.

Our analysis points to the bulk particle-hole creation as the main source of Landau damping. The contribution of the surface and mixed surface-bulk terms is instead suppressed at small wavevectors by dynamical thresholds, which are determined by the characteristic surface and bulk bands of NLS. 
The net effect is a large quality factor around the long-wavelength limit, even if inclusion of scattering with impurities and phonons, not considered in our estimate, can be expected to ultimately limit its value.

Near the drumhead band edge, the steep dependence of the ASP velocity on chemical potential enhances the frequency shift produced by a small change in surface charge.
The drumhead bandwidth sets the energy scale for this tunability and differentiates the ASP in NLS from the known examples of ASP resonances in metals, while the quadratic bulk linewidth favors spectral resolution at long wavelengths. Together, these properties suggest ASP spectroscopy as a sensitive probe of surface charge in nodal-line semimetals.

\textit{Acknowledgments}. We thank A. De Martino for stimulating discussions. We acknowledge financial support by the Deutsche Forschungsgemeinschaft (DFG) through the Collaborative Research Center SFB 1143, and the Würzburg-Dresden Cluster of Excellence on Complexity, Topology and Dynamics in Quantum Matter—ctd.qmat, EXC 2147, Project ID390858490.

	\appendix

\section{Eigenstates}\label{app:Eigenstates}
We provide here details about the spectrum and eigenstates of the model Hamiltonian used in this work. The most general momentum-independent boundary condition that ensures the vanishing of the charge current matrix elements at the interface is \cite{Witten2016}
\begin{equation}\label{eq:BC}
	\left(\cos \vartheta \,\tau^{z}+\sin \vartheta \,\tau^{x}\right)\psi(0)=\psi(0).
\end{equation}
The normalized eigenvectors of the boundary matrix in \eqref{eq:BC} are 
\begin{equation}
	\xi_{+}=\begin{pmatrix}\cos\frac{ \vartheta }{2}\\
		\sin\frac{ \vartheta }{2}
	\end{pmatrix}
	\;,\qquad\xi_{-}=\begin{pmatrix}\sin\frac{ \vartheta }{2}\\
		-\cos\frac{ \vartheta }{2}
	\end{pmatrix}.
	\label{eq:boundary-spinors}
\end{equation}
The boundary condition \eqref{eq:BC} can then equivalently be written as $\xi^{\dagger}_{-}\psi(0)=0$. 

At fixed $\boldsymbol{k}$, use the notation $m=m_{\boldsymbol{k}}=\mcurv(k^{2}-k^{2}_{\mathrm{NL}})$ and denote the energy measured from the identity contribution as $\varepsilon=E-gm$.
Note that the term $g\,m_{\boldsymbol{k}}\tau_{0}$ in the Hamiltonian is proportional to the identity matrix, so it changes the energy but not the spinor or the penetration depth.
With "zero-current" boundary conditions, the system exhibits a surface drumhead band \cite{Burkov2011,Rudi2024}
\begin{equation}\label{eq:Drumhead}
	\Phi_{S,\boldsymbol{k}}(z)=\sqrt{2\kappa_{\boldsymbol{k}}}\,e^{-\kappa_{\boldsymbol{k}}z}\xi_{+}\;.
\end{equation}
The orbital spinor is given in \eqref{eq:boundary-spinors}, the energy in the main text,
and the inverse penetration depth is 
\begin{equation}\label{eq:kappa}
	\kappa_{\boldsymbol{k}}=-\frac{m_{\boldsymbol{k}}\sin \vartheta }{\hbar v}\;,
\end{equation}
representing the inverse of the penetration depth. 
From the drumhead state dispersion relation from the main text
\begin{equation}
	E_{S}(\boldsymbol{k})=
	C(k^{2}-k^{2}_{\mathrm{NL}})
\end{equation}
and \eqref{eq:kappa}, the penetration parameter at the surface Fermi contour takes the form
\begin{equation}
	\kappa_F=\frac{|\mu|\sin \vartheta }{\hbar v|g+\cos \vartheta |}.
	\label{eq:kappaF}
\end{equation}

In addition to surface-localized states, the Hamiltonian supports propagating bulk states.
Their on-shell condition is
\begin{equation}
	\varepsilon^2=m^2+\hbar^2v^2p^2,
	\label{eq:bulkonshell}
\end{equation}
with positive normal wave number
\begin{equation}
	p(E,m)=\frac{\sqrt{(E-gm)^2-m^2}}{\hbar v}.
	\label{eq:p}
\end{equation}
Propagating states exist for $E<gm-|m|$ or $E>gm+|m|$.
We denote their support by
\begin{equation}
	\mathcal{A}(E,m)
	=\Theta\!\left((E-gm)^2-m^2\right),
	\label{eq:AbulkStates}
\end{equation}
and their band index by $\eta=\mathrm{sgn}(\varepsilon)$.
The continuum thresholds are understood as limits from $p>0$.

A real bulk eigenstate satisfying the boundary condition
\eqref{eq:BC} is
\begin{equation}
	\begin{aligned}
		\Phi_{E,m}(z)
		={}&\sqrt{\frac{2\eta p}{\varepsilon-m\cos \vartheta }}
		\Bigg[
		\begin{pmatrix}
			\cos\frac{ \vartheta }{2}\\
			\sin\frac{ \vartheta }{2}
		\end{pmatrix}\cos(pz)\\
		&+\frac{1}{\hbar vp}
		\begin{pmatrix}
			(\varepsilon+m)\sin\frac{ \vartheta }{2}\\
			(m-\varepsilon)\cos\frac{ \vartheta }{2}
		\end{pmatrix}\sin(pz)
		\Bigg].
	\end{aligned}
	\label{eq:psibulk}
\end{equation}
Here $\Phi_{E,\boldsymbol{k}}(z)=\Phi_{E,m_{\boldsymbol{k}}}(z)$.
The prefactor is real because $\varepsilon-m\cos \vartheta $
has the sign of $\varepsilon$ on the propagating continuum.
Direct substitution gives
$(H_{\boldsymbol{k}}-gm)\Phi_{E,m}=\varepsilon\Phi_{E,m}$,
while $\Phi_{E,m}(0)$ is proportional to $\xi_+$ and therefore
satisfies \eqref{eq:BC}.
When passing from wavenumber to energy as label for bulk states,
the needed Jacobian is
\begin{equation}
	J(E,m)=\left|\frac{\partial p}{\partial E}\right|
	=\frac{|\varepsilon|}{\hbar^2v^2p}.
	\label{eq:Jacobian}
\end{equation}
On either continuum branch, the coefficient of \mbox{$\delta(p-p')$} in the overlap of \eqref{eq:psibulk} is $2\pi J(E,m)$.
Since \mbox{$\delta(p-p')=\delta(E-E')/J(E,m)$} on that branch,
the states obey
\begin{equation}
	\int_0^\infty dz\,
	\Phi^\dagger_{E,m}(z)\Phi_{E',m}(z)
	=2\pi\delta(E-E').
	\label{eq:normalizationbulk}
\end{equation}
Thus the continuum measure is $dE/(2\pi)$.
For $m<0$, orthogonality to the drumhead state at the same
in-plane momentum follows
\begin{equation}
	\begin{aligned}
		&\int_0^\infty dz\,
		\Phi^\dagger_{S,\boldsymbol{k}}(z)\Phi_{E,m}(z)\\
		&=\sqrt{\frac{4\kappa_{\boldsymbol{k}}\eta p}
			{\varepsilon-m\cos \vartheta }}\,
		\frac{\kappa_{\boldsymbol{k}}+m\sin \vartheta /(\hbar v)}
		{\kappa_{\boldsymbol{k}}^2+p^2}
		=0,
	\end{aligned}
\end{equation}
where the last equality uses \eqref{eq:kappa}.

\section{The polarization function}\label{app:Polarization}

In this section, we compute the contributions to the proper polarization from the various surface and bulk sectors. The formalism is common to all interface problems, but we make explicit use of the eigenstates computed in App. \ref{app:Eigenstates}.

\subsection{Density-density kernels}
We begin by reviewing some definitions to be used here below.
An electron at depth $z$ with orbital index $a$, spin $s$, and parallel momentum $\boldsymbol{k}$ is annihilated by the field operator $\hat{\psi}_{a s,\boldsymbol{k}}(z)$. The $\boldsymbol{q}$ component of the density reads
\begin{equation}
	\hat{n}_{\boldsymbol{q}}(z)=\sum_{s=\uparrow,\downarrow}\sum^{2}_{a=1}\int\frac{d^{2}k}{(2\pi)^{2}}\,\hat{\psi}^{\dagger}_{a s,\boldsymbol{k}+\boldsymbol{q}}(z)\hat{\psi}_{a s,\boldsymbol{k}}(z).
\end{equation}
The retarded density response is defined by \cite{Giuliani,Jishi}
\begin{equation}
	\chi^{R}_{nn}(\boldsymbol{q};z,z';t)=-\frac{i}{\hbar}\Theta(t)\left\langle \left[\hat{n}_{\boldsymbol{q}}(z,t),\hat{n}_{-\boldsymbol{q}}(z',0)\right]\right\rangle .
\end{equation}
The noninteracting Matsubara Green function in the electronic problem is 
\begin{equation}
	G^{0}_{ab}(\boldsymbol{k};z,z';i\nu_{n})=\sum_{\lambda}\frac{\varphi_{\lambda\boldsymbol{k},a}(z)\varphi^{*}_{\lambda\boldsymbol{k},b}(z')}{i\nu_{n}-\xi_{\lambda\boldsymbol{k}}}.\label{eq:G0}
\end{equation}
Here $\nu_{n}=(2n+1)\pi k_{\mathrm{B}}T$ is a fermionic Matsubara energy and the indices $a,b=1,2$ label the orbital components. 
As the Hamiltonian is spin-independent, the full noninteracting Green function is diagonal in spin and consists of two identical copies of this $2\times2$ internal-space Green function. In this problem, the Green function \eqref{eq:G0} is constructed from eigenstates satisfying the boundary condition \eqref{eq:BC}. The label $\lambda$ includes both bulk-like states $B$ and drumhead surface states $S$.

In RPA, the proper polarization is the bubble formed from two noninteracting Green functions 
\begin{align}
	&\Pi_{0}(\boldsymbol{q};z,z';i\Omega_{m})=-2k_{B}T\sum_{n}\int\frac{d^{2}\boldsymbol{k}}{(2\pi)^{2}} \\
	&\qquad\
	\times\mbox{tr}_{2}\left[\boldsymbol{G}_{0}(\boldsymbol{k};z,z';i\nu_{n})\boldsymbol{G}_{0}(\boldsymbol{k}+\boldsymbol{q};z',z;i\nu_{n}+i\Omega_{m})\right]
	\nonumber
\end{align}
where $\Omega_{m}$ is a bosonic Matsubara energy, $m$ an integer, $\mbox{tr}_{2}$ the trace over the two internal components and the factor $2$ in front accounts for spin degeneracy. The subscript $0$ reminds that the RPA proper polarization constructed from the noninteracting $G_{0}$, see \eqref{eq:G0}. Evaluating the Matsubara sum \cite{Mahan} and performing the analytic continuation $i\Omega_{m}\to\hbar\omega+i0^{+}$, one obtains 
\begin{align}
	\Pi^{R}_{0}(\boldsymbol{q};z,z';\omega)&=2\sum_{\lambda,\lambda'}\int\frac{d^{2}k}{(2\pi)^{2}}\frac{f(\xi_{\lambda\boldsymbol{k}})-f(\xi_{\lambda',\boldsymbol{k}+\boldsymbol{q}})}{\hbar\omega+\xi_{\lambda\boldsymbol{k}}-\xi_{\lambda',\boldsymbol{k}+\boldsymbol{q}}+i0^{+}} \nonumber\\
	&\qquad\quad
	\times F^{*}_{\lambda'\lambda}(\boldsymbol{k},\boldsymbol{q};z)F_{\lambda'\lambda}(\boldsymbol{k},\boldsymbol{q};z')
	\label{eq:PiRbubble}
\end{align}
with real angular frequency $\omega$. Here \mbox{ $F_{\lambda'\lambda}(\boldsymbol{k},\boldsymbol{q};z)=\varphi^{\dagger}_{\lambda',\boldsymbol{k}+\boldsymbol{q}}(z)\varphi_{\lambda,\boldsymbol{k}}(z)$} is the local density matrix element between states $\lambda'$ and $\lambda$.
As both bulk-like and surface states are present, the RPA polarization splits into four transition sectors, see main text: the four terms follow from using \eqref{eq:G0} in \eqref{eq:PiRbubble}, by restricting $\lambda'$ and $\lambda$ to the indicated sectors.

As seen above, the polarization contains a sum of transitions from an initial state $i$ to final state $f$, with kernel
\begin{equation}\label{eq:K}
	\mathcal{K}_{fi}(z,z')=\rho^{*}_{fi}(z)\rho_{fi}(z').
\end{equation}
Let us now write explicitly the four types of contributions.

\subsubsection{Surface-surface}
For a surface initial state at $\boldsymbol{k}$ and a surface final state at $\boldsymbol{k}'$, the transition density is 
\begin{equation}
	\rho_{S'S}(z)=\Phi^{\dagger}_{S,\boldsymbol{k}'}(z)\Phi_{S,\boldsymbol{k}}(z)=2\sqrt{\kappa\kappa'}\,e^{-(\kappa+\kappa')z},\label{eq:rhoS'S}
\end{equation}
with $\kappa=\kappa_{\boldsymbol{k}}$, so one writes \eqref{eq:K} as
\begin{equation}
	\mathcal{K}_{SS}(z,z')=4\kappa\kappa'e^{-(\kappa+\kappa')(z+z')},\label{eq:KSS}
\end{equation}
having used $\xi^{\dagger}_{+}\xi_{+}=1$.

\subsubsection{Mixed part}
Consider a bulk state specified by $E,m$ and a surface state
whose inverse penetration depth is $\kappa_s$.
For a bulk final state and a surface initial state, direct
multiplication of the wavefunctions gives
\begin{equation}
	\begin{aligned}
		\rho_{BS}(z)
		={}&\sqrt{\frac{4\kappa_s\eta p}
			{\varepsilon-m\cos \vartheta }}\,e^{-\kappa_s z}\\
		&\times\left[
		\cos(pz)+\frac{m\sin \vartheta }{\hbar vp}\sin(pz)
		\right].
	\end{aligned}
	\label{eq:rhoBS}
\end{equation}
Both wavefunctions are real in this convention, so the reverse
transition density is
\begin{equation}
	\rho_{SB}(z)=\rho_{BS}(z).
	\label{eq:rhoSB}
\end{equation}
The two transitions therefore give the same kernel:
\begin{equation}
	\begin{aligned}
		\mathcal{K}_M(E,m;\kappa_s;z,z')
		={}&\frac{4\kappa_s\eta p}{\varepsilon-m\cos \vartheta }
		e^{-\kappa_s(z+z')}\\
		&\times\left[
		\cos(pz)+\frac{m\sin \vartheta }{\hbar vp}\sin(pz)
		\right]\\
		&\times\left[
		\cos(pz')+\frac{m\sin \vartheta }{\hbar vp}\sin(pz')
		\right].
	\end{aligned}
	\label{eq:KM}
\end{equation}

\subsubsection{Bulk-bulk part}
For an initial bulk state labeled by $E,m$ and a final bulk
state labeled by $E',m'$, define
$\varepsilon'=E'-gm'$, $p'=p(E',m')$, and
$\eta'=\mathrm{sgn}(\varepsilon')$.
Using the standing-wave states \eqref{eq:psibulk}, their
transition density is
\begin{equation}
	\begin{aligned}
		&\rho_{B'B}(z)
		=\Phi^\dagger_{E',m'}(z)\Phi_{E,m}(z)\\
		&\quad=\sqrt{\frac{2\eta'p'}{\varepsilon'-m'\cos \vartheta }}
		\sqrt{\frac{2\eta p}{\varepsilon-m\cos \vartheta }}\Bigg[
		\cos(p'z)\cos(pz)\\
		&\qquad+\frac{m\sin \vartheta }{\hbar vp}
		\cos(p'z)\sin(pz)+\frac{m'\sin \vartheta }{\hbar vp'}
		\sin(p'z)\cos(pz)\\
		&\qquad+\frac{\varepsilon'\varepsilon+m'm
			-(\varepsilon'm+m'\varepsilon)\cos \vartheta }
		{\hbar^2v^2p'p}\sin(p'z)\sin(pz)
		\Bigg].
	\end{aligned}
\end{equation}
Since this transition density is real, the kernel
\eqref{eq:K} becomes
\begin{equation}
	\mathcal{K}_{BB}(z,z')
	=\rho_{B'B}(z)\rho_{B'B}(z'),
\end{equation}
see also \cite{Newns1970}. 
In App. \ref{subsec:Polarization-in-polar} we will write the same expressions in cylindrical coordinates and then in App. \ref{subsec:Absorptive-parts} we will consider their imaginary parts, while in App. \ref{subsec:Long-wavelength-surface} we will consider the small-momentum expansion, relevant to our problem.

\subsection{Polar coordinates}\label{subsec:Polarization-in-polar}
The continuum model is rotationally invariant in the surface plane,
which makes it convenient to use cylindrical coordinates \cite{Rhim2016}.
We take the reference axis as $\boldsymbol{q}=q\hat{\boldsymbol{x}}$, with 
$\hat{\boldsymbol{x}}$ the unit vector along the $x$ direction, and parameterize the initial in-plane momentum as $\boldsymbol{k}=k(\cos\phi,\sin\phi)$,
where $\phi\in[0,2\pi)$ is the polar angle. The magnitude of the final in-plane momentum is then
\begin{equation}
	k'=|\boldsymbol{k}+\boldsymbol{q}|=\sqrt{k^{2}+q^{2}+2kq\cos\phi}.\label{eq:k'}
\end{equation}
Furthermore, using the definition of $m$ in App. \ref{app:Eigenstates}, we can write
\begin{equation}
	m'=\mcurv(k'^{2}-k_{\mathrm{NL}}^2)=m+\mcurv(q^{2}+2kq\cos\phi).
	\label{eq:mprime}
\end{equation}

\subsubsection{Surface - surface contribution}
Both the initial and final surface states must lie inside the projected nodal ring, which limits the integration over the radial momentum. We obtain
\begin{align}
	&\Pi^{SS}(q;z,z';\omega) \label{eq:PiSS0}\\
	&=
	2\intop^{k_{\mathrm{NL}}}_{0}\frac{k\,dk}{(2\pi)^{2}}
	\intop^{2\pi}_{0}d\phi
	\frac{\Theta(k_{\mathrm{NL}}-k')\left[f_{S}(k)-f_{S}(k')\right]}{\hbar\omega+E_{S}(k)-E_{S}(k')+i0^{+}}
	\mathcal{K}_{SS}(z,z')\nonumber
\end{align}
with $k'$ in \eqref{eq:k'} and the factor $2$ in front accounting for spin degeneracy. The surface-state energy difference in the denominator is 
\begin{equation}
	E_{S}(k')-E_{S}(k)
	=E_{\mathrm{S}}(|\boldsymbol{k}+\boldsymbol{q}|)-E_{\mathrm{S}}(k)=C\left(q^{2}+2kq\cos\phi\right).\label{eq:ES(kpq)-ES(k)}
\end{equation}

\subsubsection{Surface - bulk contribution}
With the initial surface momentum obeying $k<k_{\mathrm{NL}}$, the final continuum state is labeled by its energy $E'$, because the momentum is fixed by $\boldsymbol{q}$. The surface-bulk contribution is 
\begin{align}
	&\Pi^{BS}(q;z,z';\omega)=2\intop^{k_{\mathrm{NL}}}_{0}\frac{k\,dk}{(2\pi)^{2}}\intop^{2\pi}_{0}d\phi\intop^{\infty}_{-\infty}\frac{dE'}{2\pi}\mathcal{A}(E',m')
	\nonumber\\
	&\qquad\qquad\times\frac{f_{S}(k)-f_{B}(E')}{\hbar\omega+E_{S}(k)-E'+i0^{+}}\mathcal{K}_{M}(E',m';\kappa_{\boldsymbol{k}};z,z')\label{eq:PiBS}
\end{align}
where $\mathcal{A}$ in \eqref{eq:AbulkStates} selects the integration domain of bulk states.
It explicitly identifies the intervals 
\begin{equation}
	\intop^{\infty}_{-\infty}dE'\mathcal{A}(E',m')=\intop^{gm'-\lvert m'\rvert}_{-\infty}dE'+\intop^{\infty}_{gm'+\lvert m'\rvert}dE',
\end{equation}
see App. \ref{app:Eigenstates}. The factor $2$ in front accounts for spin degeneracy.

\subsubsection{Bulk - surface contribution}
In the reverse process, the final surface state must satisfy $k'<k_{\mathrm{NL}}$. The corresponding contribution is 
\begin{align}
	&\Pi^{SB}(q;z,z';\omega)=2\intop^{\infty}_{0}\frac{k\,dk}{(2\pi)^{2}}
	\intop^{2\pi}_{0}d\phi\intop^{\infty}_{-\infty}\frac{dE}{2\pi} \Theta(k_{\mathrm{NL}}-k') \nonumber\\
	&\quad\times
	\frac{\mathcal{A}(E,m) \left[f_{B}(E)-f_{S}(k')\right]}{\hbar\omega+E-E_{S}(k')+i0^{+}}
	\mathcal{K}_{M}(E,m;\kappa_{\boldsymbol{k}'};z,z')
	\label{eq:PiSB}
\end{align}
Once again, the continuum-energy domain is selected by $\mathcal{A}$ in \eqref{eq:AbulkStates} as
\begin{equation}
	\intop^{\infty}_{-\infty}dE\,\mathcal{A}(E,m)=\intop^{gm-\lvert m\rvert}_{-\infty}dE+\intop^{\infty}_{gm+\lvert m\rvert}dE
\end{equation}
but note that the integral is now on the initial state energy. 

\subsubsection{Bulk-bulk contribution}
Finally, the bulk-bulk term contains two energy integrals and no restriction over the initial and final momenta. It reads
\begin{align}\label{eq:PiBB}
	&\Pi^{BB}(q;z,z';\omega)   \\
	& =  2\intop^{\infty}_{0}\frac{k\,dk}{(2\pi)^{2}}\intop^{2\pi}_{0}d\phi\intop^{\infty}_{-\infty}\frac{dE}{2\pi}\intop^{\infty}_{-\infty}\frac{dE'}{2\pi}\frac{f_{B}(E)-f_{B}(E')}{\hbar\omega+E-E'+i0^{+}}\nonumber \\
	&  \qquad\qquad
	\times\mathcal{A}(E,m)\mathcal{A}(E',m')\mathcal{K}_{BB}(E',m';E,m;z,z')\nonumber
\end{align}
For the sake of bookkeeping, we note that the factor $2$ in front accounts for spin degeneracy.

\subsection{Absorptive parts}\label{subsec:Absorptive-parts}
For later reference, we now consider the imaginary parts \cite{Giuliani,Jishi} of the expression above. For positive energy transfer $\hbar\omega>0$, we use the identity
\begin{equation}
	\text{Im}\frac{1}{x+i0^{+}}=-\pi\delta(x)
\end{equation}
for $x$ real to eliminate one continuum-energy integration, where present.

\subsubsection{Surface-to-bulk absorption}
In this sector, the energy-conservation delta function fixes $E'=E_{S}(k)+\hbar\omega$. Therefore 
\begin{equation}
	\begin{aligned}
		\Im\Pi^{BS}&=-\intop^{k_{\mathrm{NL}}}_{0}\frac{k\,dk}{(2\pi)^{2}}
		\intop^{2\pi}_{0}d\phi
		\Theta\left(E_{S}(k)+\hbar\omega-\mu\right) \\
		&\quad
		\times\Theta\left(\mu-E_{S}(k)\right)\mathcal{A}(E_{S}(k)+\hbar\omega,m')
		\\
		&\quad
		\times\mathcal{K}_{M}\left(E_{S}(k)+\hbar\omega,m';\kappa_{\boldsymbol{k}};z,z'\right).
	\end{aligned}\label{eq:ImPiBS}
\end{equation}
At fixed surface momentum $k<k_{\mathrm{NL}}$, and $m=m_{\boldsymbol{k}}<0$, the upper continuum
edge is
\begin{equation}
	E_{+}(m)=gm+\lvert m\rvert\label{eq:Eplus},
\end{equation}
see App. \ref{app:Eigenstates}, so upward surface-to-continuum vertical energy separation at fixed in-plane momentum is 
\begin{equation}
	E_{+}(m)-E_{S}(m)
	=2\lvert m\rvert\cos^{2}\frac{ \vartheta }{2}\label{eq:E+-ES}
\end{equation}
as given in App. \ref{app:Eigenstates}.

\subsubsection{Bulk-to-surface absorption}
In this sector, the energy-conservation delta function fixes $E=E_{S}(k')-\omega$, from which
\begin{align}
	\Im\Pi^{SB} & = -\int^{\infty}_{0}\frac{k\,dk}{(2\pi)^{2}}\int^{2\pi}_{0}d\phi\,
	\Theta(k_{\mathrm{NL}}-k')\Theta\left(E_{S}(k')-\mu\right) \nonumber\\
	& \quad \times\Theta\left(\mu-E_{S}(k')+\hbar\omega\right)
	\mathcal{A}(E_{S}(k')-\hbar\omega,m)\nonumber \\
	&  \quad\times\mathcal{K}_{M}\left(E_{S}(k')-\omega,m;\kappa_{\boldsymbol{k}'};z,z'\right)\label{eq:ImPiSB}
\end{align}
At fixed surface momentum $k<k_{\mathrm{NL}}$, and $m=m_{\boldsymbol{k}}<0$, the lower $E_{-}$
continuum edges is 
\begin{equation}
	E_{-}(m)=gm-\lvert m\rvert\label{eq:Eminus}
\end{equation}
from which the energy separation at fixed in-plane momentum between the lower continuum and the surface band is
\begin{equation}\label{eq:ES-E-}
	E_{S}(m)-E_{-}(m)=
	2\lvert m\rvert\sin^{2}\frac{ \vartheta }{2}.
\end{equation}

At zero temperature and $q=0$, denote by $\Delta_{S\rightarrow B}$ and
$\Delta_{B\rightarrow S}$  the minimum excitation energies
for surface-to-bulk and bulk-to-surface absorption, respectively.
For a partially filled drumhead, these follow by minimizing the
excitation energy subject to the occupation constraints in
\eqref{eq:ImPiBS} and \eqref{eq:ImPiSB}.

For $C>0$ and $\mu<0$, the surface-to-bulk minimum occurs at the surface Fermi contour, whereas the bulk-to-surface minimum occurs where the lower bulk edge reaches $\mu$. Thus
\begin{equation}
	\Delta_{S\rightarrow B}=\frac{|\mu|(1+\cos \vartheta )}{g+\cos \vartheta },
	\label{eq:DeltaSB-C>0}
\end{equation}
\begin{equation}
	\Delta_{B\rightarrow S}=\frac{|\mu|(1-\cos \vartheta )}{1+g}.
	\label{eq:DeltaBS-C>0}
\end{equation}

For $C<0$ and $\mu>0$, the surface-to-bulk minimum occurs
where the upper bulk edge reaches $\mu$, whereas the
bulk-to-surface minimum occurs at the surface Fermi contour.
Hence
\begin{equation}
	\Delta_{S\rightarrow B}=\frac{|\mu|(1+\cos \vartheta )}{1-g},
	\label{eq:DeltaSB-C<0}
\end{equation}
\begin{equation}
	\Delta_{B\rightarrow S}=\frac{|\mu|(1-\cos \vartheta )}{|g+\cos \vartheta |}.
	\label{eq:DeltaBS-C<0}
\end{equation}
These zero-momentum thresholds vanish linearly with $|\mu|$. At fixed nonzero $\mu$, the mixed absorption thresholds remain finite as $q\to0$, while the plasmon energy vanishes; mixed Landau damping is therefore absent at sufficiently small $q$.

\subsubsection{Bulk-bulk absorption}\label{subsec:Bulk-bulk-absorption}

In the bulk-bulk sector, the energy-conservation delta function fixes
$E'=E+\omega$. At zero temperature, 
\begin{equation}
	f_{B}(E)-f_{B}(E+\omega)=\Theta\left(E-\mu+\hbar\omega\right)\Theta\left(\mu-E\right).\label{eq:bulkabsinterval}
\end{equation}
Thus, having already accounted for the spin degeneracy, we write
\begin{align}
	\text{Im}\Pi^{BB}&=-\intop^{\infty}_{0}\frac{k\,dk}{(2\pi)^{2}}\intop^{2\pi}_{0}d\phi\intop^{\mu}_{\mu-\hbar\omega}\frac{dE}{2\pi}\mathcal{A}(E,m)
	\nonumber\\
	&\mathcal{A}(E+\hbar\omega,m')\mathcal{K}_{BB}\left(E+\hbar\omega,m';E,m;z,z'\right).\label{eq:ImBB}
\end{align}
We still have one finite energy integral in this sector.

\subsubsection{Surface-surface imaginary part}
Assuming $C\neq0$ (not flat band), the energy-conservation condition is 
\begin{equation}
	\omega-E_{\mathrm{S}}(|\boldsymbol{k}+\boldsymbol{q}|)+E_{\mathrm{S}}(k)
	=\hbar\omega-C(q^{2}+2kq\cos\phi)=0
	\label{eq:energyconservationsurface}
\end{equation}
see\eqref{eq:ES(kpq)-ES(k)}. Define $x_{k}=\frac{\hbar\omega/C-q^{2}}{2kq}$ and $k^{2}_{\omega}=k^{2}+\frac{\hbar\omega}{C}$.
The angular delta function gives 
\begin{equation}
	\int^{2\pi}_{0}d\phi\,\delta\left[\hbar\omega-C(q^{2}+2kq\cos\phi)\right]
	=\frac{\Theta(1-\lvert x_{k}\rvert)}{\lvert C\rvert kq\sqrt{1-x^{2}_{k}}}
\end{equation}
On the energy shell $k'=k_{\omega}$ is selected by the angular frequency and
\begin{equation}
	\begin{aligned}
		&\text{Im}\Pi^{SS}(q;z,z';\omega)  =  -\frac{1}{\lvert C\rvert q}\intop^{k_{\mathrm{NL}}}_{0}\frac{dk}{2\pi}\frac{\Theta\left(1-\lvert x_{k}\rvert\right)}{\sqrt{1-x^{2}_{k}}}\Theta(k^{2}_{\omega})
		\\
		&\qquad\times
		\Theta(k_{\mathrm{NL}}^2-k^{2}_{\omega})\Theta\left(\mu-E_{S}(k)\right)
		\Theta\left(E_{S}(k)+\hbar\omega-\mu\right)
		\\
		&\qquad\times
		4\kappa_{k}\kappa_{k_{\omega}}e^{-(\kappa_{k}+\kappa_{k_{\omega}})(z+z')}\label{eq:ImPiSS}
	\end{aligned}
\end{equation}
where spin degeneracy has been taken into account and 
the inverse penetration depth $\kappa_{k}$  ($\kappa_{k_{\omega}}$) follow from
\eqref{eq:kappa}, evaluated at the initial momentum
$k$ (the final momentum $k_\omega$). The exact surface spectral function has been reduced to one integral over $k$.

Equations \eqref{eq:ImPiBS}, \eqref{eq:ImPiSB}, \eqref{eq:ImBB},
and \eqref{eq:ImPiSS} determine the absorptive parts of the four
polarization sectors.
We now derive directly the long-wavelength drumhead polarization required for the surface-plasmon pole equation.

\subsection{Long-wavelength surface contribution}
\label{subsec:Long-wavelength-surface}
\noindent
We expand the surface bubble directly at fixed phase velocity
$\upsilon_\omega=\omega/q>v_S$, where $\omega$ is the angular frequency and
$v_S=2|C|k_F/\hbar$. The surface Fermi contour satisfies
$\mu=C(k_F^2-k_{\mathrm{NL}}^2)$, giving
\begin{equation}
	m_F=\frac{\mu}{g+\cos \vartheta }.
	\label{eq:mFdef}
\end{equation}
The inverse penetration depth at the surface Fermi
contour is given in \eqref{eq:kappaF}. 	We assume $0<k_F<k_{\mathrm{NL}}$ and
$q\ll\min(k_F,q_k)$, with
$q_k=(k_{\mathrm{NL}}^2-k_F^2)/(2k_F)$. 
Introducing the notation 
$\boldsymbol{k}_{\pm}=\boldsymbol{k}\pm\boldsymbol{q}/2$,
with $E_{S,\pm}=E_S(|\boldsymbol{k}_{\pm}|)$ and
$\kappa_{\pm}=\kappa_{\boldsymbol{k}_{\pm}}$, 
\eqref{eq:PiSS0} is written as
\begin{equation}
	\begin{aligned}
		&\Pi^{SS}(q;z,z';\omega)
		=2\int\frac{d^2\boldsymbol{k}}{(2\pi)^2}
		\frac{f(E_{S,-})-f_S(E_{S,+})}
		{\hbar\omega+E_{S,-}-E_{S,+}+i0^+}\\
		&\qquad\times\rho_{+-}(z)\rho_{+-}(z')\Theta(k_{\mathrm{NL}}-|\boldsymbol{k}_-|)
		\Theta(k_{\mathrm{NL}}-|\boldsymbol{k}_+|),
	\end{aligned}
	\label{eq:PiSSsymmetric}
\end{equation}
where the factor two accounts for spin and
\mbox{$\rho_{+-}(z)=2\sqrt{\kappa_-\kappa_+}
	e^{-(\kappa_-+\kappa_+)z}$}.
At zero temperature, \mbox{$f(E)=\Theta(\mu-E)$}. For the parabolic band, the shifted quantities are
\begin{equation}
	\begin{aligned}
		E_{S,\pm}
		&=E_S(k)+\frac{Cq^2}{4}
		\pm C\boldsymbol{k}\cdot\boldsymbol{q},\\
		\kappa_{\pm}
		&=\kappa_{\boldsymbol{k}}
		\mp\frac{D\sin \vartheta }{\hbar v}
		\boldsymbol{k}\cdot\boldsymbol{q}
		-\frac{D\sin \vartheta }{4\hbar v}q^2.
	\end{aligned}
	\label{eq:kappaSymmetricExpansion}
\end{equation}
The common quadratic energy shift cancels from the denominator and the occupation difference and transition density have the expansion
\begin{equation}
	\begin{aligned}
		f(E_{S,-})-f(E_{S,+})
		&=2C\boldsymbol{k}\cdot\boldsymbol{q}\,
		\delta(E_S(k)-\mu)+\mathcal{O}(q^3),\\
		\rho_{+-}(z)
		&=2\kappa_{\boldsymbol{k}}
		e^{-2\kappa_{\boldsymbol{k}}z}
		+\mathcal{O}(q^2).
	\end{aligned}
	\label{eq:fsymmetricExpansion}
\end{equation}
Since the denominator is of order $q$ at fixed $\upsilon_\omega$, both
remainders contribute only at order $q^2$ to the polarization.
The support functions equal one in the selected Fermi shell.

Using $\delta(E_S(k)-\mu)=\delta(k-k_F)/(2|C|k_F)$
in \eqref{eq:PiSSsymmetric}, the radial integration
produces the surface DOS and selects
$\kappa_{\boldsymbol k}=\kappa_F$.
The angular integration then gives
\begin{equation}
	\begin{aligned}
		\chi_S(\upsilon_\omega)
		&=\frac{\nu_S}{2\pi}\int_0^{2\pi}
		\frac{v_S\cos\phi}{\upsilon_\omega-v_S\cos\phi}\,d\phi\\
		&=\nu_S\left[
		\frac{\upsilon_\omega}{\sqrt{\upsilon_\omega^2-v_S^2}}-1
		\right].
	\end{aligned}
	\label{eq:chiSurf}
\end{equation}
For $C<0$, a shift by $\pi$ gives the same integral. Consequently,
\begin{equation}
	\Pi^{SS}(q;z,z';\upsilon_\omega q)
	=w_F(z)w_F(z')\chi_S(\upsilon_\omega)+\mathcal{O}(q^2).
	\label{eq:PiSSnoLinearCorrection}
\end{equation}
Thus neither the penetration profiles nor the surface Lindhard
function produces a linear momentum correction at fixed
$\upsilon_\omega>v_S$, away from the continuum threshold. Their corrections
first affect the plasmon frequency at order $q^3$.

\section{Static regular polarization}
\label{app:StaticRegular}

As seen in App. \ref{app:Polarization}, the regular polarization contains a bulk contribution $\Pi^{BB}$ and a mixed contribution
\begin{equation}
	\Pi_M=\Pi^{BS}+\Pi^{SB}.
\end{equation}
Using their static forms computed in App. \ref{app:Polarization},
this technical Appendix supplies the static response needed to calculate the screened interaction. App.  \ref{app:finite-green} uses that response to obtain the effective interaction between drumhead electrons.

\subsection{Static uniform limit of the bulk polarization}\label{sec:Static-uniform-limit}

Consider now the half-space bulk Lindhard function \eqref{eq:PiBB} in the limit $q\to0$, $\omega\to0$. Its homogeneous component is
\begin{align}
	&\Pi^{BB}_{bulk}\left(0,0\right) =  \lim_{L\to\infty}\frac{1}{L}\intop^{L}_{0}dz\intop^{L}_{0}dz'\Pi^{BB}\left(0;z,z';0\right)\\
	& \quad =  \lim_{L\to\infty}\frac{2}{L}\intop\frac{d^{2}\boldsymbol{k}}{(2\pi)^{2}}\intop^{\infty}_{-\infty}\frac{dE}{2\pi}\intop^{\infty}_{-\infty}\frac{dE'}{2\pi}\frac{f_{B}(E)-f_{B}(E')}{E-E'+i0^{+}}\nonumber\\
	&\qquad
	\times\mathcal{A}(E,m)\mathcal{A}(E',m')\intop^{L}_{0}dz\rho^{*}_{E',E}(z)\intop^{L}_{0}dz'\rho_{E',E}(z')\nonumber 
\end{align}
As the states are normalized to the energy, see \eqref{eq:normalizationbulk}, 
one has
\begin{equation}
	\begin{aligned}
		&	\Pi^{BB}_{bulk}\left(0,0\right) \\
		&\qquad  = \lim_{L\to\infty}\frac{2}{L}\intop\frac{d^{2}\boldsymbol{k}}{(2\pi)^{2}}\frac{\partial f_{B}(E)}{\partial E}\mathcal{A}(E,m)\intop^{L}_{0}dz\rho^{*}_{E,E}(z)
	\end{aligned}
\end{equation}
Using the eigenstate \eqref{eq:psibulk},
\begin{equation}
	\begin{aligned}
		&\lim_{L\to\infty}\frac{1}{L}
		\int_0^L dz\,\rho_{E,E}(z)\\
		&=\frac{\eta p}{\varepsilon-m\cos \vartheta }
		\left[
		1+\frac{\varepsilon^2+m^2
			-2\varepsilon m\cos \vartheta }{\hbar^2v^2p^2}
		\right]\\
		&=\frac{2|\varepsilon|}{\hbar^2v^2p}
		=2J(E,m).
	\end{aligned}
\end{equation}
Substitution gives
\begin{equation}
	\Pi^{BB}_{bulk}\left(0,0\right)=4\intop\frac{d^{2}\boldsymbol{k}}{(2\pi)^{2}}\intop^{\infty}_{-\infty}\frac{dE}{2\pi}\frac{\partial f_{B}(E)}{\partial E}\mathcal{A}(E,m)J(E,m).
\end{equation}
In the limit $T\to0$, the derivative of the distribution function is $\frac{\partial f_{B}(E)}{\partial E}\to-\delta\left[E-\mu\right]$ and
\begin{equation}
	\Pi^{BB}_{bulk}\left(0,0\right)=-4\intop\frac{d^{2}\boldsymbol{k}}{(2\pi)^{3}}\mathcal{A}(\mu,m)J(\mu,m).
\end{equation}
Using the Jacobian \eqref{eq:Jacobian}, the two normal-momentum
roots $k_z=\pm p(\mu,m)$ each contribute $J(\mu,m)$ to the
bulk energy-delta integral. Including the continuum support
\eqref{eq:AbulkStates}, we obtain
\begin{align}
	\Pi^{BB}_{bulk}\left(0,0\right)&=-2\intop\frac{d^{2}\boldsymbol{k}}{(2\pi)^{3}}\int dk_{z}\delta\left(\mu-E_{bulk}(\boldsymbol{k},k_{z})\right) \nonumber\\
	&=-2\nu_{B}(\mu),  \label{eq:bulkDOS}
\end{align}
the factor $2$ in front appears because of spin degeneracy.

We also note that the first derivative of the projected static mixed response in $q$ at $q=0$ vanishes, provided it is differentiable. Indeed, in a function of the form $\mathcal{F}(k,m';z,z_{1})$ integrated in momentum, the angular dependence enters only through $m'$, hence
\begin{equation}
	\begin{aligned}
		&\left.\frac{d}{dq}\int^{2\pi}_{0}d\varphi\,\mathcal{F}(k,m+D[2kq\cos\varphi+q^{2}];z,z_{1})\right|_{q=0} \nonumber\\
		&\qquad=2Dk\,\partial_{m'}\mathcal{F}(k,m;z,z_{1})\int^{2\pi}_{0}\cos\varphi\,d\varphi=0.\label{eq:ivc-angular-linear}
	\end{aligned}
\end{equation}
After projection onto the drumhead states, $\Pi_{M}(q;z,z_{1};0)=\Pi_{M,0}(z,z_{1})+o(q)$, where $o(q)/q\to0$ as $q\to0^{+}$.

\subsection{Static mixed polarization}
We now combine the two surface-bulk transition channels \eqref{eq:PiBS} and \eqref{eq:PiSB}, into a single expression for the static mixed polarization.
To this aim, set $\omega=0$ in \eqref{eq:PiBS} and \eqref{eq:PiSB}.
In the latter, change the integration variable to the surface
momentum $\boldsymbol{k}'=\boldsymbol{k}+\boldsymbol{q}$ and
then invert that momentum. Rotational invariance brings both
contributions to the same surface-momentum domain, with the
bulk mass given by \eqref{eq:mprime}. Their static occupation
quotients are equal, and their transition kernels agree by
\eqref{eq:KM}. Hence
\begin{equation}
	\begin{aligned}
		&\Pi_M(q;z,z';0)
		=-4\int_0^{k_{\mathrm{NL}}}
		\frac{k\,dk}{(2\pi)^2}\int_0^{2\pi}d\phi
		\int_{-\infty}^{\infty}\frac{dE}{2\pi}\\
		&\qquad\times\mathcal A(E,m')
		\frac{f_S(k)-f_B(E)}{E-E_S(k)}
		\mathcal K_M(E,m';\kappa_{\boldsymbol{k}};z,z').
	\end{aligned}
	\label{eq:mixed-static}
\end{equation}
The factor four counts the two spin projections and the two
transition directions. The static bulk contribution follows
by setting $\omega=0$ in \eqref{eq:PiBB}.

\subsection{Zero-momentum mixed kernel}
In order to evaluate the screening matrix elements in \eqref{eq:two-potential-mixed-entry} below,
we now derive the zero-momentum mixed polarization $\Pi_{M,0}$ in \eqref{eq:mixed-zero}.

At zero in-plane momentum transfer, the surface and bulk states entering each mixed transition have the same in-plane momentum $\boldsymbol{k}$, so \mbox{$m'=m=m_{\boldsymbol{k}}\in(-Dk_{\mathrm{NL}}^2,0)$}.
In this notation, \mbox{$E_S(m')=(g+\cos \vartheta )m'$} and
\mbox{$\kappa'=-m'\sin \vartheta /(\hbar v)$}.
Equations~\eqref{eq:E+-ES} and \eqref{eq:ES-E-} place the surface energy between the bulk thresholds.
At zero temperature, an occupied surface state contributes
only through unoccupied upper-band states, and an unoccupied
surface state only through occupied lower-band states.
The contributing branch is therefore
\begin{equation}
	\eta'(m')
	=\mathrm{sgn}\!\left[\mu-(g+\cos \vartheta )m'\right].
	\label{eq:mixed-selected-branch}
\end{equation}
On this branch, introduce the dimensionless continuum variable
\begin{equation}
	x=\frac{\hbar vp'}{|\varepsilon'|+|m'|},
	\qquad 0\leq x<1,
	\label{eq:mixed-x-definition}
\end{equation}
where $\varepsilon'=E-gm'$ and $p'=p(E,m')$ is defined by
\eqref{eq:p}. Inverting gives
\begin{equation}
	\begin{aligned}
		E_{\eta'}(x,m')
		&=gm'+\eta'|m'|\frac{1+x^2}{1-x^2},\\
		p'(x,m')
		&=\frac{2|m'|x}{\hbar v(1-x^2)},\\
		\left|\frac{dE_{\eta'}}{dx}\right|
		&=\frac{4|m'|x}{(1-x^2)^2}.
	\end{aligned}
	\label{eq:mixed-parametrization}
\end{equation}
Opposite occupation requires $E_{\eta'}>\mu$ on the upper
branch and $E_{\eta'}<\mu$ on the lower branch.
These conditions give $x\geq x_0(m')$, with
\begin{equation}
	x_0(m')=
	\begin{cases}
		0,
		&\eta'(\mu-gm')\leq |m'|,\\[2pt]
		\displaystyle
		\sqrt{\frac{\eta'(\mu-gm')-|m'|}
			{\eta'(\mu-gm')+|m'|}},
		&\eta'(\mu-gm')>|m'|.
	\end{cases}
	\label{eq:mixed-x-endpoint}
\end{equation}

Evaluate \eqref{eq:rhoBS} at
$E=E_{\eta'}(x,m')$, $m=m'$, and $\kappa_s=\kappa'$.
Removing its normalization factor defines the dimensionless
depth function
\begin{equation}
	h_{m',x}(z)
	=\sqrt{\frac{\varepsilon'-m'\cos \vartheta }
		{4\kappa'\eta'p'}}\,\rho_{BS}(z).
	\label{eq:TransitionShape}
\end{equation}
On the allowed occupation domain, the quotient in
\eqref{eq:mixed-static} equals $1/|E-E_S(m')|$.
Combining this factor with the energy Jacobian and the
normalization in \eqref{eq:rhoBS} gives
\begin{equation}
	\begin{aligned}
		&	\mathcal W(x,m')
		=\left|\frac{dE_{\eta'}}{dx}\right|
		\frac{4\kappa'\eta'p'}
		{(\varepsilon'-m'\cos \vartheta )
			|E_{\eta'}-E_S(m')|}\\
		&\qquad=\frac{32\kappa'x^2}
		{\hbar v(1-x^2)
			\left[(1+\eta'\cos \vartheta )
			+(1-\eta'\cos \vartheta )x^2\right]^2},
	\end{aligned}
	\label{eq:mixed-weight}
\end{equation}
where $E_{\eta'}-E_S(m')=\varepsilon'-m'\cos \vartheta $.

The angular integral gives $2\pi$ and
$k\,dk=dm'/(2D)$. Including the continuum measure
$dE/(2\pi)$, we obtain
\begin{align}
	&\Pi_{M,0}(z,z')
	\equiv\Pi_M(0;z,z';0)  		\label{eq:mixed-zero} \\
	&=-
	\intop_{-Dk_{\mathrm{NL}}^2}^{0}\frac{dm'}{2\pi^2D}
	\intop_{x_0(m')}^1dx\,
	\mathcal W(x,m')h_{m',x}(z)h_{m',x}(z').\nonumber
\end{align}
The kernel is negative semidefinite because
$\mathcal W\geq0$.
The surface-bulk orthogonality established in App. \ref{app:Eigenstates} also implies $\int_0^\infty h_{m',x}(z)\,dz=0$, hence, the zero-momentum mixed response annihilates a potential independent of depth.

\subsection{Local bulk approximation and screening scales}
We now use the uniform bulk response established in App. \ref{sec:Static-uniform-limit} to specify a local approximation to bulk screening for the toroidal Fermi surface and evaluate the bulk density of states and Thomas-Fermi wave number. This will later enter the screening equation \eqref{eq:finite-static-green}.
We begin by approximating the depth dependence of the bulk polarization by
\begin{equation}
	\Pi^{BB}(q;z,z';0)
	\simeq-2\nu_B(\mu)\delta(z-z').
	\label{eq:adopt-local}
\end{equation}
The uniform response fixes the coefficient; locality in depth,
including the treatment of boundary corrections, is an
additional approximation.

At the chemical potential, the continuum support
\eqref{eq:AbulkStates} restricts the mass to the interval
with endpoints
\begin{equation}
	m_{\pm}(\mu)
	=\frac{-g\mu\pm|\mu|}{1-g^2}.
	\label{eq:bulk-mass-endpoints}
\end{equation}
The physical radial domain also requires
$m\geq-Dk_{\mathrm{NL}}^2$.
When the chemical potential intersects the drumhead band,
$\mathrm{sgn}(\mu)=-\mathrm{sgn}(g+\cos \vartheta )$, so
\begin{equation}
	\begin{aligned}
		1-\mathrm{sgn}(\mu)g-|g+\cos \vartheta |
		&=1+\mathrm{sgn}(\mu)\cos \vartheta >0,\\
		m_-(\mu)
		=-\frac{|\mu|}{1-\mathrm{sgn}(\mu)g}
		&>m_F>-Dk_{\mathrm{NL}}^2,
	\end{aligned}
	\label{eq:interior-untruncated}
\end{equation}
where $m_F$ is defined by \eqref{eq:mFdef}.
The full bulk mass interval therefore contributes.

Evaluating the bulk density of states over this interval gives
\begin{equation}
	\begin{aligned}
		\nu_B(\mu)&=\frac{|\mu|}{4\pi D\hbar v(1-g^2)^{3/2}},\\
		q_{\mathrm{TF}}^2&=\frac{2e^2\nu_B(\mu)}{\varepsilon_0\epsilon_\perp}=\frac{e^2|\mu|}{2\pi\varepsilon_0\epsilon_\perp D\hbar v(1-g^2)^{3/2}}.
	\end{aligned}
	\label{eq:qTF}
\end{equation}
The inverse length entering the local bulk differential
operator is
\begin{equation}
	Q(q)=\sqrt{q_{\mathrm{TF}}^2+
		\frac{\epsilon_{\parallel}}{\epsilon_{\perp}}q^2}.
	\label{eq:Q(q)}
\end{equation}
The mixed contribution remains nonlocal and is retained in the regular screened interaction, constructed in App. \ref{app:finite-green}.

\section{Joint static screening in a space of two potentials}	
\label{app:finite-green}
As the chemical potential approaches the nodal line, the
separations between the surface energy and the bulk continua,
given by \eqref{eq:E+-ES} and \eqref{eq:ES-E-}, decrease at
the surface Fermi momentum. This motivates retaining the
mixed surface-bulk polarization together with the bulk
polarization when constructing the regular screened
interaction.

Using the static response specified in App. \ref{app:StaticRegular}, we formulate the screening
equation in weak form and project it onto a space spanned by two trial potentials. These potentials are chosen from the depth dependence of the solution with local bulk screening. The mixed polarization is included in the projected screening matrix before inversion.
The screened potential satisfies \cite{Giuliani}
\begin{equation}
	\begin{aligned}
		\delta(z-z')=&\frac{\varepsilon_0\epsilon_\perp}{e^2}\left[-\partial_z^2+Q(q)^2\right]W_r(q;z,z')
		\\&-\int_0^\infty dz_1\,\Pi_{M,0}(z,z_1)W_r(q;z_1,z').
	\end{aligned}
	\label{eq:finite-static-green}
\end{equation}
It is continuous at the source, has derivative jump  $-e^2/(\varepsilon_0\epsilon_\perp)$ there, and decays at infinity. It agrees with the retained static problem through first order in $q$, while the $O(q^{2})$ polarization correction is omitted. The retained terms determine the plasmon frequency
through order $q^2$.

Let $U,V$ be real, decaying, finite-energy potentials. 
If the Robin condition $V'(0)=\epsilon_{2}qV(0)/\epsilon_{\perp}$ holds, projecting \eqref{eq:finite-static-green} onto $U$ yields the symmetric form \cite{BrennerScott2008}
\begin{align}
	a_q(U,V)=&\frac{\varepsilon_0\epsilon_\perp}{e^2}\int_0^\infty dz\left[U'V'+Q(q)^2UV\right] \nonumber \\
	&+\frac{\varepsilon_0\epsilon_2 q}{e^2}U(0)V(0)         \label{eq:finite-form}\\
	&-\int_0^\infty dz\int_0^\infty dz_1\,U(z)\Pi_{M,0}(z,z_1)V(z_1) \nonumber
\end{align}
The Green function therefore obeys the weak condition \cite{ChoiKim2014}
\begin{equation}
	\mathfrak{a}_{q}(U,W_{r}(q;\cdot,z'))=U(z'),\label{eq:ivc-weak-green-equation}
\end{equation}
where the dot marks the $z$ (depth) argument acted on by the form.

As 
$
Q(q)=q_{\mathrm{TF}}+\frac{\epsilon_{\parallel}}{2\epsilon_{\perp}q_{\mathrm{TF}}}q^{2}+O(q^{4})
$
in \eqref{eq:finite-form}, the first momentum correction to the interaction comes from the interface
term. The change of $Q$ first affects the frequency at cubic order.
Depth profiles chosen at $q=0$ therefore suffice through quadratic order.

We need the screened potential generated by $w_{F}$, and its projection back onto $w_{F}$. In general, the potential generated by a density profile $\varrho$ is
\begin{equation}
	F_{\varrho}(z)=\int^{\infty}_{0}dz'\,W_{r}(q;z,z';0)\varrho(z'),\label{eq:two-potential-source-map}
\end{equation}
which is equivalent to the weak equation 
\begin{equation}
	\mathfrak{a}_{q}(U,F_{\varrho})=\int^{\infty}_{0}dz\,U(z)\varrho(z).\label{eq:two-potential-weak-source}
\end{equation}
for every admissible $U$.

\subsection{Bulk screening and the trial subspace}
\label{app:bulk-only}

We drop the mixed kernel for now. The potential of $w_{F}$ in this reference problem is denoted $F_{\mathrm{TF}}$. At $q=0$, \eqref{eq:finite-static-green} becomes
\begin{equation}
	\begin{aligned} & \left(-\partial^{2}_{z}+q^{2}_{\mathrm{TF}}\right)F_{\mathrm{TF}}(z)
		=\frac{2\kappa_{F}e^2}{\varepsilon_0\epsilon_\perp}e^{-2\kappa_{F}z},\\
		& F_{\mathrm{TF}}'(0)=0,
	\end{aligned}
	\label{eq:two-potential-BB-reference}
\end{equation}
We also require that the potential is associated with a surface plasmon 
by requiring $F_{\mathrm{TF}}(z\to\infty)\to0$. 
The source suggests a particular solution $\propto e^{-2\kappa_{F}z}$. 
Adding the decaying homogeneous solution, 
\begin{equation}
	F_{\mathrm{TF}}(z)=\frac{2\kappa_F e^2/(\varepsilon_0\epsilon_\perp)}{q^{2}_{\mathrm{TF}}-(2\kappa_{F})^{2}}e^{-2\kappa_{F}z}+d_{B}e^{-q_{\mathrm{TF}}z}.
	\label{eq:two-potential-reference-general}
\end{equation}
The Neumann condition in \eqref{eq:two-potential-BB-reference} fixes $d_{B}$, so that the solution reads
\begin{equation}
	F_{\mathrm{TF}}(z)=\frac{2\kappa_F e^2/(\varepsilon_0\epsilon_\perp)}{q^{2}_{\mathrm{TF}}-(2\kappa_{F})^{2}}\left(e^{-2\kappa_{F}z}-\frac{2\kappa_{F}}{q_{\mathrm{TF}}}e^{-q_{\mathrm{TF}}z}\right).\label{eq:two-potential-reference-solution}
\end{equation}
Projecting the bulk-screened potential back onto the drumhead profile gives the coupling with bulk screening alone. Using the definition of $\zeta=\nu_S\mathcal{V}_0$ and the drumhead profile $w_{F}(z)=2\kappa_{F}\exp\{-2\kappa_{F}z\}$ in the main body of the paper, one arrives at
\begin{equation}
	\begin{aligned}
		\zeta _B
		&=\nu_S\int_0^\infty dz\,w_F(z)F_{\mathrm{TF}}(z)\\
		&  =\frac{\nu_S e^2\kappa_F(q_{\mathrm{TF}}+4\kappa_F)}{\varepsilon_0\epsilon_\perp q_{\mathrm{TF}}(q_{\mathrm{TF}}+2\kappa_F)^2}
	\end{aligned}\label{eq:bulk-only-projection}
\end{equation}
The factor $q_{\mathrm{TF}}-2\kappa_F$ cancels, so the projected coupling remains finite when the screening and source exponents coincide. Using \eqref{eq:qTF} and the compressibility ratio $ \cratio $ gives $\zeta _B= \cratio \frac{1+4\rho}{(1+2\rho)^2}$, see the main text.

We now restore the mixed kernel. Based on the result \eqref{eq:two-potential-reference-solution},
we approximate the potential as a function in the span of 
\begin{equation}
	\phi_{i}(z)=e^{-s_{i}z},\qquad s_{1}=q_{\mathrm{TF}},\qquad s_{2}=2\kappa_{F}\;.
	\label{eq:phi}
\end{equation}
For any source, we write the screened potential resulting from projecting \eqref{eq:finite-static-green}
onto the trial subspace as
\begin{equation}
	F^{(2)}_{\varrho}(z)=\sum^{2}_{i=1}b_{\varrho,i}(q)\phi_{i}(z),\label{eq:two-potential-trial-source}
\end{equation}
where the coefficients are fixed by projecting the full screening
equation onto this space using \eqref{eq:finite-form},
which includes the dielectric boundary condition.

\subsection{Screening matrix}
\noindent
Choosing $U=\phi_{j}$ in \eqref{eq:two-potential-weak-source} and
using linearity gives 
\begin{equation}
	\mathfrak{a}_{q}(\phi_{j},F^{(2)}_{\varrho})=
	\sum^{2}_{i=1}\mathfrak{a}_{q}(\phi_{j},\phi_{i})b_{\varrho,i}(q)=\int^{\infty}_{0}dz\,\phi_{j}(z)\varrho(z)
	\label{eq:two-potential-tested-equation}
\end{equation}
This fixes the screening matrix $A(q)_{ji}=\mathfrak{a}_{q}(\phi_{j},\phi_{i})$ in the trial subspace.
From \eqref{eq:finite-form},
\begin{align}
	A(q)_{ji}= & \frac{\varepsilon_0\epsilon_\perp}{e^2}\int^{\infty}_{0}dz\,\left[\phi_{j}'\phi_{i}'+\left(q^{2}_{\mathrm{TF}}
	+\frac{\epsilon_{\parallel}}{\epsilon_{\perp}}q^{2}\right)\phi_{j}\phi_{i}\right]
	\nonumber
	\\
	& +\frac{\varepsilon_0\epsilon_2 q}{e^2}\phi_{j}(0)\phi_{i}(0)
	\label{eq:Aexplicit}
	\\
	&
	-\int^{\infty}_{0}dz\int^{\infty}_{0}dz_{1}\,\phi_{j}(z)\Pi_{M,0}(z,z_{1})\phi_{i}(z_{1})\nonumber 
\end{align}
Its $q$ dependence sits in the interior $q^{2}$ term and the interface $q$ term.
The local integrals follow from 
\begin{equation}
	\int^{\infty}_{0}dz\,\phi_{i}\phi_{j}=\frac{1}{s_{i}+s_{j}}\;,\qquad
	\int^{\infty}_{0}dz\,\phi_{i}'\phi_{j}'=\frac{s_{i}s_{j}}{s_{i}+s_{j}}
	\label{eq:two-potential-local-integrals}
\end{equation}
having used $\phi_{i}'=-s_{i}\phi_{i}$ for the base \eqref{eq:phi}. 
The mixed contribution, which contains both channels $\Pi^{BS}_{0}+\Pi^{SB}_{0}$, is 
\begin{equation}\label{eq:two-potential-mixed-entry}
	M_{ij}=-\int^{\infty}_{0}dz\int^{\infty}_{0}dz_{1}\,\phi_{i}(z)\Pi_{M,0}(z,z_{1})\phi_{j}(z_{1}).
\end{equation}
Its minus sign comes from the inverse screening operator, so $M$
is positive semidefinite. Writing $A=A(q=0)$, we have
\begin{equation}
	A_{ij}=\mathfrak{a}_{0}(\phi_{i},\phi_{j})=\frac{\varepsilon_0\epsilon_\perp(s_is_j+q_{\mathrm{TF}}^2)}{e^2(s_i+s_j)}+M_{ij},
	\label{eq:finiteA}
\end{equation}
that is, 
\begin{equation}
	\begin{aligned}
		&A_{1j}=\frac{\varepsilon_0\epsilon_\perp q_{\mathrm{TF}}}{e^2}+M_{1j},\qquad j=1,2, \\
		& A_{22}=\frac{\varepsilon_0\epsilon_\perp\left[(2\kappa_F)^2+q_{\mathrm{TF}}^2\right]}{4e^2\kappa_F}+M_{22}.
		\label{eq:two-potential-explicit-A}
	\end{aligned}
\end{equation}
The mixed entries are evaluated in closed form in App. \ref{sec:finite-mixed}. It can be checked that the matrix $A$ is positive definite and \mbox{$\det A>0$}.

\subsection{Two-potential Green function}
\noindent
For a unit point source at depth $z'$, the approximate Green function
is 
\begin{equation}
	W^{(2)}_{r}(q;z,z';0)=d_{1}(q;z')\phi_{1}(z)+d_{2}(q;z')\phi_{2}(z).\label{eq:W(2)}
\end{equation}

Testing with $\phi_{j}$ and using linearity, 
\begin{equation}
	\mathfrak{a}_{q}\!\left(\phi_{j},W^{(2)}_{r}(q;\cdot,z';0)\right)=\sum^{2}_{i=1}A(q)_{ji}d_{i}(q;z').\label{eq:two-potential-point-test}
\end{equation}
For the exact Green function $W_r$ with an interior source, \eqref{eq:ivc-weak-green-equation}
gives 
\begin{equation}
	\mathfrak{a}_{q}\!\left(\phi_{j},W_{r}(q;\cdot,z';0)\right)=\int^{\infty}_{0}dz\,\phi_{j}(z)\delta(z-z')=\phi_{j}(z').\label{eq:revision-point-source-test}
\end{equation}
Requiring the same in the trial space, 
\begin{equation}
	\sum^{2}_{i=1}A(q)_{ji}d_{i}(q;z')=\phi_{j}(z'),\qquad j=1,2.\label{eq:two-potential-point-source}
\end{equation}
With $\boldsymbol{d}=(d_{1},d_{2})^{T}$ and $\boldsymbol{\phi}=(\phi_{1},\phi_{2})^{T}$,
this gives $\boldsymbol{d}(q;z')=A(q)^{-1}\boldsymbol{\phi}(z')$, from which
\begin{equation}
	\begin{aligned}
		W^{(2)}_{r}(q;z,z';0)
		&=\boldsymbol{\phi}(z)^{T}A(q)^{-1}\boldsymbol{\phi}(z').\label{eq:two-potential-Green-derived}
	\end{aligned}
\end{equation}
This kernel solves the weak equation for every test potential in the span.

Using the argument at the end of App. \ref{app:Polarization}, the projected mixed response has no term linear in $q$. 
This leaves the interface term of \eqref{eq:Aexplicit} as the only linear in $q$ correction,
so that
\begin{equation}
	A(q)_{ij}=A_{ij}+\frac{\varepsilon_0\epsilon_2 q}{e^2}\phi_{i}(0)\phi_{j}(0)+o(q).\label{eq:two-potential-boundary-entry}
\end{equation}
The boundary-value vector has components $v_i=\phi_i(0)$. In particular, for the exponential basis \eqref{eq:phi}, $\boldsymbol v=(1,1)^T$, giving
\begin{equation}
	\begin{aligned}A(q) & =A+\frac{\varepsilon_{0}\epsilon_{2}}{e^{2}}q\boldsymbol{v}\boldsymbol{v}^{T}+o(q).
	\end{aligned}
	\label{eq:finite-W}
\end{equation}
The momentum dependence displayed in \eqref{eq:Aexplicit}
comes from the interface term proportional to $q$ and the
interior term proportional to $q^2$.
Their size relative to the static screening form can be
bounded directly.
Writing $\mathfrak{a}_0=\mathfrak{a}_{q=0}$, the negative
semidefiniteness of $\Pi_{M,0}$ established in
\eqref{eq:mixed-zero} implies, for any real, decaying,
finite-energy test potential $U$,
\begin{equation}
	\begin{aligned}
		q_{\mathrm{TF}}U(0)^2
		\leq
		\int_0^\infty dz\,\left[(U')^2+q_{\mathrm{TF}}^2U^2\right]
		&\leq\frac{e^2\mathfrak{a}_0(U,U)}{\varepsilon_0\epsilon_\perp},
		\\
		q_{\mathrm{TF}}^2\int_0^\infty dz\,U^2
		& \leq\frac{e^2\mathfrak{a}_0(U,U)}{\varepsilon_0\epsilon_\perp}.
	\end{aligned}
	\label{eq:finite-electrostatic-bounds}
\end{equation}
The first inequality follows by integrating
$(U^2)'$ and applying the Cauchy-Schwarz inequality. 
Using these bounds in \eqref{eq:finite-form}, the
interface and interior corrections are bounded,
respectively, by
$\epsilon_2q/(\epsilon_\perp q_{\mathrm{TF}})$ and
$\epsilon_\parallel q^2/(\epsilon_\perp q_{\mathrm{TF}}^2)$
times $\mathfrak{a}_0(U,U)$.
Sufficient conditions for treating both electrostatic
corrections as small when expanding $A(q)^{-1}$ are
therefore
\begin{equation}
	\frac{\epsilon_2q}{\epsilon_\perp q_{\mathrm{TF}}}\ll1,
	\qquad
	\frac{\epsilon_\parallel q^2}
	{\epsilon_\perp q_{\mathrm{TF}}^2}\ll1.
	\label{eq:finite-electrostatic-window}
\end{equation}

\subsection{The drumhead source}

For $\varrho=w_{F}$, the right-hand side of \eqref{eq:two-potential-tested-equation}
is the vector of source overlaps 
\begin{equation}
	t_{i}=\int^{\infty}_{0}dz\,w_{F}(z)\phi_{i}(z)=\frac{2\kappa_{F}}{2\kappa_{F}+s_{i}}
	\label{eq:two-potential-source-overlaps}
\end{equation}
At $q=0$, the screened potential projected onto the drumhead
is $F^{(2)}_{0}=b_{1}\phi_{1}+b_{2}\phi_{2}$,
with 
$\boldsymbol{b}=A^{-1}\boldsymbol{t}$, see \eqref{eq:two-potential-Green-derived}.
Projecting the potential back onto $w_{F}$ gives the static interaction
\begin{equation}
	\begin{aligned}
		\mathcal{V}_{0}&=\int^{\infty}_{0}dz\,w_{F}F^{(2)}_{0}=t_{1}b_{1}+t_{2}b_{2}\\
		&=\frac{A_{22}t^{2}_{1}-2A_{12}t_{1}t_{2}+A_{11}t^{2}_{2}}{\det A},\label{eq:two-potential-static-interaction}
	\end{aligned}
\end{equation}
in which the matrix elements are known from \eqref{eq:finiteA}, \eqref{eq:two-potential-explicit-A}. Multiplying by the surface DOS,
this expression determines the coupling $ \zeta $, see definition in the main text.	The same coefficients give the interface potential $F^{(2)}_{0}(0)=b_{1}+b_{2}$, which enters the momentum correction.


\subsection{Surface charge and the parameter $ \zeta $}
\label{app:beta-capacitance}
\noindent
The parameter  $ \zeta $ measures the residual repulsion between surface electrons after screening by the regular sectors. Its relation to the ASP velocity follows the screened-interaction description of metallic ASPs \cite{Pitarke2004}.   In the notation of Ref.~\cite{Pitarke2004}, $\zeta=\nu_S I(z_d)$. Here we connect $ \zeta $ to the chemical-potential response to a variation of surface charge and interpret it as a ratio of capacitances.
A variation in the surface charge density  can be induced by two distinct mechanisms: either by a chemical potential shift or by an electrostatic potential \cite{Christen1996}. At fixed drumhead band energies, a chemical potential shift produces a surface charge variation $\delta\sigma_{\mathrm{fill}}$. Instead, an electric potential changes the electron energies at fixed chemical potential and therefore the surface states occupation by a value $\delta\sigma_{\mathrm{pot}}$.
To quantify the band energy change, consider a small variation $\delta n$ of the surface electron density,
with depth profile $w_F(z)\delta n$. 
From this, \eqref{eq:two-potential-source-map} gives the screened electron potential-energy shift $F_0^{(2)}(z)\delta n$ in the static, long-wavelength limit and within the two-potential approximation.
Its expectation value in a drumhead state at the Fermi contour gives the first-order shift of that state's energy.
Using \eqref{eq:two-potential-static-interaction}, this shift is written as
\begin{equation}
	\delta n\int_0^\infty dz\,w_F(z)F_0^{(2)}(z)
	=\mathcal V_0\delta n
	=-\frac{\mathcal V_0}{e}\,\delta\sigma,   \label{eq:electron-energy-shift}
\end{equation}
where $\delta\sigma=-e\,\delta n$.
The corresponding charge contribution therefore obeys
\begin{equation}
	\delta\sigma_{\mathrm{pot}}
	=e\nu_S\mathcal V_0\delta n
	=- \zeta \,\delta\sigma,
\end{equation}
The total surface charge variation is the sum of both contributions
$\delta\sigma=\delta\sigma_{\mathrm{fill}} +\delta\sigma_{\mathrm{pot}}$.
These equations determine the total drumhead charge variation
\begin{equation}
	\delta\sigma
	=\frac{\delta\sigma_{\mathrm{fill}}}{1+ \zeta },
	\qquad
	\delta\sigma_{\mathrm{pot}}
	=-\frac{ \zeta }{1+ \zeta }\,
	\delta\sigma_{\mathrm{fill}}.
	\label{eq:beta-charge-response}
\end{equation}
For $ \zeta \ll1$, the charge response approaches the value obtained at fixed band energies, while for $ \zeta \gg1$, the induced potential strongly opposes the occupation change.

The band compressibility is conventionally expressed through the quantum capacitance per unit area, \mbox{$c_Q=e^2\nu_S$} \cite{Luryi1988}. In analogy, one can define a screening capacitance per unit area \cite{Scalfi2020}. Dividing the electron potential energy shift \eqref{eq:electron-energy-shift} by its charge $-e$ gives the corresponding electric potential change relative to the bulk. The ratio between the surface charge variation and this potential change defines the effective capacitance associated with regular-sector screening 
\begin{equation}
	c_{\mathrm{scr}}
	=\frac{-e\,\delta n}{-\mathcal V_0\delta n/e}
	=\frac{e^2}{\mathcal V_0}.
\end{equation}
In the weak-screening regime of App. \ref{sec:weak-mixed},
\eqref{eq:bulk-only-projection} and \eqref{eq:weak-hierarchy} give
$
\mathcal V_0\simeq
\frac{e^2}{\varepsilon_0\epsilon_\perp q_{\mathrm{TF}}}
$, from which $c_{\mathrm{scr}}\simeq\varepsilon_0\epsilon_\perp q_{\mathrm{TF}}$.
The screening capacitance then has the parallel-plate form, with the Thomas-Fermi length acting as an effective separation.
From the definitions above, we see that $ \zeta =c_Q/c_{\mathrm{scr}}$.

\subsection{Surface charge sheet}
\label{sec:surface-sheet}
\noindent
We now extend the formalism developed in this section to describe a NLS
in which the finite-width drumhead charge profile is replaced  by a sheet at
the interface with the dielectric.
This idealization allows one to describe floating surface states and can be directly modeled by the source equation
\eqref{eq:two-potential-weak-source}.

A charge sheet is represented by the distribution
$\delta_+(z)$, defined on the material half-space by
\begin{equation}
	\int_0^\infty dz\,\delta_+(z)U(z)=U(0),
	\label{eq:sheet-normalization}
\end{equation}
where $U$ is a test potential.
The weak source equation \eqref{eq:ivc-weak-green-equation} therefore becomes
$\mathfrak{a}_0(U,F_{\mathrm{sh}})=U(0)$, where
$F_{\mathrm{sh}}$ is the static screened potential
generated by the sheet at zero in-plane momentum.
Using the expansion
\eqref{eq:two-potential-trial-source} and testing with
$U=\phi_j$, as in \eqref{eq:two-potential-tested-equation}, gives
\begin{equation}
	F_{\mathrm{sh}}^{(2)}(z)
	=\boldsymbol\phi(z)^T\boldsymbol b_{\mathrm{sh}}
	\;,\qquad
	A\boldsymbol b_{\mathrm{sh}}=\boldsymbol{v}
	\label{eq:sheet-screened-potential}
\end{equation}
Here $\boldsymbol b_{\mathrm{sh}}$ contains the potential
coefficients, $A=A(0)$ is the screening matrix
\eqref{eq:finiteA}, and $\boldsymbol v$ is the
boundary-value vector introduced in \eqref{eq:finite-W}.
Thus a sheet source replaces the drumhead overlap
vector $\boldsymbol t$ by the boundary value vector $\boldsymbol v$.
Projecting the potential back onto the sheet then gives
\begin{equation}
	\mathcal V_{\mathrm{sh},0}
	=F_{\mathrm{sh}}^{(2)}(0)
	=\boldsymbol v^T A^{-1}\boldsymbol v.
	\label{eq:sheet-static-interaction}
\end{equation}
\noindent
To obtain a bulk-screening estimate, we treat the FSS
and bulk as separate electronic subsystems coupled
through the Coulomb interaction.
Within this approximation, the density operator acts
separately in the two subsystems.
The density vertices $F_{\lambda'\lambda}$ defined
below \eqref{eq:PiRbubble} therefore vanish between a
surface state and a bulk state.
The corresponding mixed sectors of
\eqref{eq:PiRbubble} give
\begin{equation}
	\Pi^{BS}=\Pi^{SB}=0,
	\qquad
	M_{ij}=0,
	\label{eq:sheet-independent-subsystems}
\end{equation}
where the second equality follows from
\eqref{eq:two-potential-mixed-entry}.
This specifies the screening approximation in addition
to the sheet approximation for the source. 
With $M_{ij}=0$,
\eqref{eq:two-potential-explicit-A} gives
$A_{11}=A_{21}=\varepsilon_0\epsilon_\perp q_{\mathrm{TF}}/e^2$.
Consequently, the source equation
\eqref{eq:sheet-screened-potential} is solved by
\begin{equation}
	A
	\begin{pmatrix}e^2/(\varepsilon_0\epsilon_\perp q_{\mathrm{TF}})\\0\end{pmatrix}
	=
	\begin{pmatrix}
		1\\
		1
	\end{pmatrix}
	=\boldsymbol v.
	\label{eq:sheet-bulk-source}
\end{equation}
The potential and its sheet projection are therefore
\begin{equation}
	\begin{aligned}
		&	F^{(2)}_{\mathrm{sh}}(z)=\frac{e^2}{\varepsilon_0\epsilon_\perp q_{\mathrm{TF}}}e^{-q_{\mathrm{TF}}z}\;,
		\qquad
		V_{sh,0}=\frac{e^2}{\varepsilon_0\epsilon_\perp q_{\mathrm{TF}}}.
	\end{aligned}
	\label{eq:sheet-bulk-interaction}
\end{equation}
\noindent
For an isotropic surface Fermi contour, the response
\eqref{eq:chiSurf} gives the leading pole condition
$\mathcal V_{\mathrm{sh},0}\chi_S(\aspv )=1$.
Applying $\zeta=\nu_S\mathcal{V}_0$ and \eqref{eq:u0} yields
\begin{equation}
	\begin{aligned}
		\zeta _{\mathrm{sh}}
		&=\nu_S\mathcal V_{\mathrm{sh},0}
		=\frac{\nu_S e^2}{\varepsilon_0\epsilon_\perp q_{\mathrm{TF}}},
		\\
		\aspv 
		&=v_S\frac{1+ \zeta _{\mathrm{sh}}}{\sqrt{1+2 \zeta _{\mathrm{sh}}}}.
	\end{aligned}
	\label{eq:sheet-bulk-velocity}
\end{equation}
Here $ \zeta _{\mathrm{sh}}$ is the coupling for the sheet source in the independent-subsystem approximation.

\subsection{Explicit evaluation of the static mixed matrix}
\label{sec:finite-mixed}
\noindent
We now provide integral expressions for the mixed polarization represented in the trial subspace, see \eqref{eq:two-potential-mixed-entry}.

\subsubsection{Depth overlaps}
\noindent
Substituting \eqref{eq:mixed-zero} into \eqref{eq:two-potential-mixed-entry},
the minus signs cancel and the depth integrals factorize: 
\begin{equation}
	\begin{aligned}M_{ij}= & \frac{1}{2\pi^{2}D}\int^{0}_{-Dk^{2}_{\mathrm{NL}}}dm'\int^{1}_{x_{0}(m')}dx\,\mathcal{W}(x,m')\\
		& \times\left[\int^{\infty}_{0}dz\,e^{-s_{i}z}h_{m',x}(z)\right]\left[\int^{\infty}_{0}dz'\,e^{-s_{j}z'}h_{m',x}(z')\right].
	\end{aligned}
	\label{eq:depth-projection-factorized}
\end{equation}
For an inverse length $s>0$, the transition overlap is formally a Laplace transform, which we denote as
\begin{equation}
	\begin{aligned}H_{s}(m',x) & =\int^{\infty}_{0}dz\,e^{-(\kappa'+s)z}\left[\cos(p'z)-\frac{\kappa'}{p'}\sin(p'z)\right]\\
		& 
		=\frac{s}{(\kappa'+s)^{2}+p'^{2}}
	\end{aligned}
	\label{eq:finite-H}
\end{equation}
This expression vanishes at $s=0$, which is simply the orthogonality between surface and bulk states. Hence 
\begin{equation}
	M_{ij}=\intop^{0}_{-Dk^{2}_{\mathrm{NL}}}\frac{dm'}{2\pi^{2}D}\intop^{1}_{x_{0}(m')}dx\,\mathcal{W}(x,m')H_{s_{i}}(m',x)H_{s_{j}}(m',x).
	\label{eq:finite-M-integral}
\end{equation}
We reverse the integration order in \eqref{eq:finite-M-integral}.
The branch selection \eqref{eq:mixed-selected-branch} and the
occupation constraint \eqref{eq:mixed-x-endpoint} give
\begin{equation}
	\begin{aligned}
		&\eta'=\mathrm{sgn}(\mu):\quad
		m(k_F)<m'<\frac{-|\mu|(1-x^2)}
		{1+x^2-g(1-x^2)\mathrm{sgn}(\mu)},\\
		&\eta'=-\mathrm{sgn}(\mu):\quad
		-Dk_{\mathrm{NL}}^2<m'<m(k_F),\\
		&\qquad\qquad\qquad\qquad0<x<1.
	\end{aligned}
	\label{eq:finite-domains}
\end{equation}
The upper limit in the first interval follows from
$E_{\mathrm{sgn}(\mu)}(x,m')=\mu$, with the bulk energy
given by \eqref{eq:mixed-parametrization}.
In the second interval, the selected bulk branch and the
surface state have opposite occupations for every $x$.

\subsubsection{Further manipulation of the integrand}
\noindent
The denominator of $H_{s}$ factorizes into two conjugate factors as
\begin{equation}
	\begin{aligned}
		& H_{s}(m',x)=\frac{s(\hbar v)^{2}}{(\hbar vs-m'u_{+})(\hbar vs-m'u_{-})}
	\end{aligned}
	\label{eq:depth-factored-overlap}
\end{equation}
where $u_{\pm}(x) =\sin \vartheta \pm\frac{2ix}{1-x^{2}}$, $u_{-} =u^{*}_{+}$.
Define now the dimensionless function
\begin{equation}
	P_\eta(x)=
	\frac{x^2}
	{(1-x^2)[1+x^2+\eta\cos \vartheta (1-x^2)]^2},
	\label{eq:depth-branch-weight}
\end{equation}
where $\eta=\pm1$ and $0<x<1$.
Using $\hbar v\kappa'=-m'\sin \vartheta $, the weight
\eqref{eq:mixed-weight} is rewritten as
\begin{equation}
	\mathcal W(x,m')
	=-\frac{32m'\sin \vartheta }{(\hbar v)^2}P_{\eta'}(x),
	\label{eq:depth-weight-mass}
\end{equation}
Together with \eqref{eq:depth-factored-overlap}, this leaves
the rational dependence on $m'$ in
\begin{equation}
	f_{ij}(m',x)=
	\frac{-m'}
	{\displaystyle\prod_{\sigma=\pm}
		(\hbar vs_i-m'u_\sigma)(\hbar vs_j-m'u_\sigma)},
	\label{eq:finite-fij}
\end{equation}
where the product runs over $\sigma=+$ and $\sigma=-$.
Substitution into \eqref{eq:finite-M-integral}, with the
domains \eqref{eq:finite-domains}, gives
\begin{equation}
	\begin{aligned}
		M_{ij}={}&
		\frac{16\sin \vartheta (\hbar v)^2s_is_j}{\pi^2D}
		\int_0^1dx\,\Bigg[
		P_{\mathrm{sgn}(\mu)}(x)
		\\
		&\times
		\int_{m(k_F)}^{
			-\frac{|\mu|(1-x^2)}
			{1+x^2-g(1-x^2)\mathrm{sgn}(\mu)}}
		f_{ij}(m',x)\,dm'
		\\
		&+
		P_{-\mathrm{sgn}(\mu)}(x)
		\int_{-Dk_{\mathrm{NL}}^2}^{m(k_F)}
		f_{ij}(m',x)\,dm'
		\Bigg].
	\end{aligned}
	\label{eq:finite-M-rational}
\end{equation}

\section{Bulk linewidth and quality factor}
\label{app:Landau-damping}

We calculate the bulk contribution to the plasmon linewidth using the static screening matrix of
App. \ref{app:finite-green}, with the mixed entries evaluated in App.  \ref{sec:finite-mixed}. 
The calculation is at zero temperature.
We consider sufficiently small $q$ that the plasmon lies above the surface particle-hole continuum and below both mixed absorption thresholds, see App. \ref{subsec:Absorptive-parts}.

\subsection{Low-frequency bulk absorption}

For the trial potentials $\phi_i$ of \eqref{eq:phi}, with
$i,j=1,2$, define the bulk absorption matrix by
\begin{equation}
	\begin{aligned}
		\mathsf C_{ij}
		={}&-\lim_{\omega\to0^+}\frac{1}{\omega}
		\int_0^\infty dz\int_0^\infty dz'\,
		\phi_i(z)
		\\
		&\qquad\times
		\Im\Pi^{BB}(0;z,z';\omega)\phi_j(z').
	\end{aligned}
	\label{eq:qf-projected-loss}
\end{equation}
At $q=0$, the angular integration in \eqref{eq:ImBB} gives
$2\pi$, and $k\,dk=dm/(2D)$.
The energy integral extends from $\mu-\hbar\omega$ to $\mu$.
To first order in $\omega$, its integrand is evaluated at
$E=E'=\mu$.
The bulk transition kernel then factorizes into the product of
the equal-energy densities at $z$ and $z'$.
Consequently,
\begin{equation}
	\begin{aligned}
		\mathsf C_{ij}
		&=\frac{\hbar}{8\pi^2D}
		\int_{m_-(\mu)}^{m_+(\mu)}
		dm\,\mathcal L_{s_i}(m)\mathcal L_{s_j}(m),
		\\
		\mathcal L_s(m)
		&=\int_0^\infty dz\,e^{-sz}
		\Phi_{\mu,m}^\dagger(z)\Phi_{\mu,m}(z).
	\end{aligned}
	\label{eq:scaling-dimensional-absorption}
\end{equation}
The limits $m_\pm(\mu)$ are defined in
\eqref{eq:bulk-mass-endpoints}.
The full interval contributes in the surface-band regime,
as established in \eqref{eq:interior-untruncated}.
The spin degeneracy is already included in \eqref{eq:ImBB}.
The factor $\hbar$ in
\eqref{eq:scaling-dimensional-absorption} comes from the
energy interval of width $\hbar\omega$.

To evaluate the depth integral, write
$\varepsilon_m=\mu-gm$ and $p_m=p(\mu,m)>0$, with $p$ defined
in \eqref{eq:p}.
On the bulk Fermi surface,
$\mathrm{sgn}(\varepsilon_m)=\mathrm{sgn}(\mu)$.
For $s>0$, the elementary integrals are
\begin{equation}
	\begin{aligned}
		\int_0^\infty e^{-sz}\cos^2(p_mz)\,dz
		&=\frac{s^2+2p_m^2}{s(s^2+4p_m^2)},
		\\
		\int_0^\infty e^{-sz}\sin^2(p_mz)\,dz
		&=\frac{2p_m^2}{s(s^2+4p_m^2)},
		\\
		\int_0^\infty e^{-sz}\sin(p_mz)\cos(p_mz)\,dz
		&=\frac{p_m}{s^2+4p_m^2}.
	\end{aligned}
	\label{eq:qf-depth-integrals}
\end{equation}
Substituting the bulk state \eqref{eq:psibulk} gives
\begin{equation}
	\begin{aligned}
		\mathcal L_s(m)
		={}&
		\frac{2\,\mathrm{sgn}(\mu)p_m}
		{s(\varepsilon_m-m\cos \vartheta )(s^2+4p_m^2)}
		\\
		&\times
		\left[
		s^2+\frac{2sm\sin \vartheta }{\hbar v}
		+\frac{4\varepsilon_m(\varepsilon_m-m\cos \vartheta )}
		{(\hbar v)^2}
		\right].
	\end{aligned}
	\label{eq:qf-density-projection}
\end{equation}
Equations~\eqref{eq:scaling-dimensional-absorption} and
\eqref{eq:qf-density-projection} determine all entries of
$\mathsf C$ through one integral over $m$.
Each integrand in
\eqref{eq:scaling-dimensional-absorption} is an outer product
of a real vector with itself, multiplied by a positive measure.
Thus $\mathsf C$ is positive semidefinite.

\subsection{Absorption of the screened drumhead potential}
\label{app:absorption-drumhead-potential}

We retain the leading low-frequency bulk loss,
with $\mathsf C$ evaluated at zero in-plane momentum.
We assume that this projected coefficient approaches
\eqref{eq:qf-projected-loss} continuously along the acoustic
limit $\omega/q=\aspv $ as $q\to0$.
For the static momentum dependence, we use
\eqref{eq:finite-W}.
Its absence of a linear mixed contribution requires the
projected static mixed response to be differentiable with
respect to the in-plane momentum vector at the origin.
Rotational invariance then makes that derivative vanish.

The inverse screening operator contains $-\Pi_r$, so the negative imaginary bulk response produces a positive
$i\omega\mathsf C$ term in the projected matrix.
Adding the leading bulk absorption to \eqref{eq:finite-W} gives
\begin{equation}
	\begin{aligned}
		A(q,\omega)
		={}&A+\frac{\varepsilon_0\epsilon_2}{e^2}
		q\,\boldsymbol v\boldsymbol v^T
		+i\omega\mathsf C +o(q,\omega).
	\end{aligned}
	\label{eq:qf-dynamic-matrix}
\end{equation}
The surface source vector $\boldsymbol t$ is defined in
\eqref{eq:two-potential-source-overlaps}, and its static
potential coefficients introduced in
App. \ref{app:finite-green} obey $\boldsymbol b=A^{-1}\boldsymbol t$.
Expanding the inverse of \eqref{eq:qf-dynamic-matrix} and
projecting onto $\boldsymbol t$ gives
\begin{equation}
	\begin{aligned}
		\mathcal V(q,\omega)
		={}&\mathcal V_0
		-\frac{\varepsilon_0\epsilon_2}{e^2}
		(b_1+b_2)^2q
		-i\omega\mathcal C_r
		\\
		&+o(q)+o(\omega),
		\\
		\mathcal C_r
		={}&\boldsymbol b^T\mathsf C\boldsymbol b\geq0.
	\end{aligned}
	\label{eq:qf-interaction}
\end{equation}
Here $\mathcal V_0$ is the static interaction in
\eqref{eq:two-potential-static-interaction}.
The matrix $A$ includes the local bulk term and both mixed
channels before inversion.
The contraction $\mathcal C_r$ therefore weights the bulk
absorption with the jointly screened potential.

\subsection{Complex pole and linewidth}

The surface response is given by \eqref{eq:chiSurf}.
Together with \eqref{eq:PiSSnoLinearCorrection}, it yields the
pole condition
\begin{equation}
	1=\mathcal V(q,\omega_{\mathrm{SP}})
	\chi_S\!\left(\frac{\omega_{\mathrm{SP}}}{q}\right)
\end{equation}
to the order needed below.
Write
\begin{equation}
	\omega_{\mathrm{SP}}(q)
	=\aspv q+(c_2-i\Gamma_2)q^2+o(q^2),
	\qquad
	\Gamma_2\geq0.
	\label{eq:qf-pole-expansion}
\end{equation}
The leading real pole satisfies $\mathcal V_0\chi_S(\aspv )=1$, is characterized by the slope 
\begin{equation}
	\aspv =v_S\frac{1+ \zeta }{\sqrt{1+2 \zeta }}
\end{equation}
quoted in the main body of the paper.
A prime denotes differentiation of $\chi_S$ with respect
to its phase-velocity argument.
From \eqref{eq:chiSurf},
\begin{equation}
	\chi_S'(\upsilon_\omega)
	=-\frac{\nu_Sv_S^2}{(\upsilon_\omega^2-v_S^2)^{3/2}}.
	\label{eq:qf-surface-derivative}
\end{equation}
Expanding the surface response at the complex pole,
\begin{equation}
	\chi_S\!\left(\frac{\omega_{\mathrm{SP}}}{q}\right)
	=\chi_S(\aspv )
	+\chi_S'(\aspv )(c_2-i\Gamma_2)q+o(q).
	\label{eq:qf-surface-expansion}
\end{equation}
In the absorptive term of \eqref{eq:qf-interaction},
$\omega_{\mathrm{SP}}$ can be replaced by $\aspv q$ at this order.
The coefficient of $q$ in the pole condition is therefore
\begin{equation}
	\begin{aligned}
		0={}&\mathcal V_0\chi_S'(\aspv )(c_2-i\Gamma_2)
		\\
		&-
		\left[
		\frac{\varepsilon_0\epsilon_2}{e^2}(b_1+b_2)^2
		+i \aspv \mathcal C_r
		\right]\chi_S(\aspv ).
	\end{aligned}
	\label{eq:qf-pole-first-order}
\end{equation}
Using $ \zeta =\nu_S\mathcal V_0$ and the ASP velocity given in the main body of the paper, one has
\begin{equation}
	\begin{aligned}
		\aspv ^2-v_S^2
		&=\frac{v_S^2 \zeta ^2}{1+2 \zeta },
		\\
		-\frac{1}{\mathcal V_0^2\chi_S'(\aspv )}
		&=\frac{\nu_Sv_S \zeta }{(1+2 \zeta )^{3/2}}.
	\end{aligned}
	\label{eq:qf-pole-prefactor}
\end{equation}
Separating real and imaginary parts of
\eqref{eq:qf-pole-first-order} gives
\begin{equation}
	\begin{aligned}
		c_2
		&=-\frac{\varepsilon_0\epsilon_2}{e^2}
		\frac{\nu_Sv_S \zeta }{(1+2 \zeta )^{3/2}}
		(b_1+b_2)^2,
		\\
		\Gamma_2
		&=\frac{\nu_Sv_S\aspv  \zeta }{(1+2 \zeta )^{3/2}}
		\mathcal C_r
		\\
		&=\frac{\nu_Sv_S^2 \zeta (1+ \zeta )}
		{(1+2 \zeta )^2}\mathcal C_r .
	\end{aligned}
	\label{eq:qf-pole-coefficients}
\end{equation}
In particular, $c_2\leq0$ and $\Gamma_2\geq0$.
The bulk-induced half-width is $\Gamma_B(q)=-\Im\omega_{\mathrm{SP}}(q)=\Gamma_2q^2+o(q^2)$ in the regime considered.

\subsection{Quality factor}

The quality factor defined in the main text is
\begin{equation}
	\begin{aligned}
		Q_{\mathrm{SP}}(q)
		&=\frac{\aspv }{2\Gamma_2q}[1+o(1)]
		\\
		&=\frac{(1+2 \zeta )^{3/2}}
		{2\nu_Sv_S \zeta \mathcal C_r}
		\frac{1}{q}[1+o(1)].
	\end{aligned}
	\label{eq:qf-quality-factor}
\end{equation}
This expression applies when $\mathcal C_r>0$.
The leading coefficient depends on the static interaction through $ \zeta $ and on the screened depth profile through $\mathcal C_r$.
The constant term in the momentum expansion of $Q_{\mathrm{SP}}$ requires the next linewidth coefficient.
The upper edge of the drumhead particle-hole continuum is $\omega=v_Sq+|C|q^2/\hbar$. The ratio of the separation of the resonance from this edge to the bulk-induced half-width is therefore
\begin{equation}
	\frac{\Re\omega_{\mathrm{SP}}(q)-v_Sq}{\Gamma_B(q)}
	=2\left(1-\frac{\sqrt{1+2\zeta}}{1+\zeta}\right)Q_{\mathrm{SP}}(q)\,[1+o(1)],
	\label{eq:qf-resolution}
\end{equation}
where the ASP slope $\aspv$ given in the main text has been used. The prefactor is of order $\zeta^2$ for $\zeta\ll1$.

At $q_{\mathrm{TF}}=2\kappa_F$, the two trial exponentials coincide. One uses instead the regular basis
\begin{equation}
	\phi_1(z)=e^{-q_{\mathrm{TF}}z},
	\qquad
	\phi_2(z)=\kappa_F z e^{-q_{\mathrm{TF}}z}.
	\label{eq:qf-regular-basis}
\end{equation}
to evaluate matrix and source integrals.
The quality factor retains \eqref{eq:qf-quality-factor}, with $\mathcal V_0$ and $\mathcal C_r$ evaluated in the same basis.


\subsection{Surface response and loss function}
\label{sec:surface-loss}

We now determine the exterior response $g_{\mathrm{surf}}(q,\omega)$ defined from the screened potential
\begin{align}\label{eq:WRgsurf}
	&W^{R}\left(q;z<0,z'<0;\omega\right)\\
	&\qquad=\frac{e^{2}}{2\epsilon_{0}\epsilon_{2}q}\left[e^{-q\left|z-z'\right|}-g_{surf}\left(q,\omega\right)e^{q\left(z+z'\right)}\right]
	\nonumber
\end{align}
in the main text.
Throughout this subsection, the frequency remains below the mixed absorption thresholds, and $A(q,\omega)$ is evaluated with the terms displayed in \eqref{eq:qf-dynamic-matrix}: the static mixed response and the leading bulk absorption are thus retained in the same matrix inverse.

Let $r_r(q,\omega)$ denote the reflection coefficient defined as in \eqref{eq:WRgsurf}, with the full interaction replaced
by the regular interaction $W_r$.
For an exterior source at $z'<0$, continuity of the potential and of the normal dielectric displacement gives the interface condition
\begin{equation}
	\left[	\partial_z-\frac{\epsilon_2q}{\epsilon_\perp}	\right]
	W_r(q;z,z';\omega)\bigg|_{z=0^+}
	=-\frac{e^2}{\varepsilon_0\epsilon_\perp}e^{qz'}.
	\label{eq:qf-exterior-boundary}
\end{equation}
In the weak formulation \eqref{eq:finite-form}, this inhomogeneous boundary condition supplies the source $e^{qz'}U(0)$ for a test potential $U$.
Projecting onto the two trial potentials yields the source vector $e^{qz'}\boldsymbol v$, where $v_i=\phi_i(0)$.
The projected interaction across the interface and its boundary trace are consequently
\begin{equation}
	\begin{aligned}
		&W_r^{(2)}(q;z>0,z'<0;\omega)
		=e^{qz'}\boldsymbol\phi(z)^T
		A(q,\omega)^{-1}\boldsymbol v,
		\\
		&r_r(q,\omega)
		=1-\frac{2\varepsilon_0\epsilon_2q}{e^2}
		\boldsymbol v^T A(q,\omega)^{-1}\boldsymbol v.
	\end{aligned}
	\label{eq:qf-regular-exterior}
\end{equation}
As discussed, the dielectric boundary condition is incorporated through the interface term in the weak formulation.
For the loss spectrum, we evaluate the surface response \eqref{eq:chiSurf} at $\upsilon_\omega=(\omega+i0^+)/q$, where $0^+$ is a positive infinitesimal.
The square root is analytic in the upper half of the complex $u$ plane and approaches $u$ at large modulus.
We denote this retarded boundary value by $\chi_S(q,\omega)$.
For positive real frequency, this prescription gives
\begin{equation}
	-\Im\chi_S(q,\omega)
	=
	\begin{cases}
		\displaystyle
		\frac{\nu_S\omega}{\sqrt{v_S^2q^2-\omega^2}},
		&0<\omega<v_Sq,
		\\[6pt]
		0,
		&\omega>v_Sq.
	\end{cases}
	\label{eq:qf-surface-absorption}
\end{equation}

The full interaction is obtained by including the surface polarization in the Dyson equation for the regular interaction:
\begin{equation}
	W^R=W_r+W_r\Pi^{SS}W^R.
	\label{eq:qf-surface-dyson}
\end{equation}
where $W_r$ already includes BB, BS, and SB screening.
We solve \eqref{eq:qf-surface-dyson} using the factorized
surface polarization in
\eqref{eq:PiSSnoLinearCorrection} and evaluating the
regular interaction in the two-potential approximation.

The interaction projected onto the surface charge
profile is
\begin{equation}
	\begin{aligned}
		\mathcal V(q,\omega)
		&=
		\int_0^\infty dz\int_0^\infty dz'\,
		w_F(z)W_r^{(2)}(q;z,z';\omega)w_F(z')
		\\
		&=
		\boldsymbol t^T A(q,\omega)^{-1}\boldsymbol t.
	\end{aligned}
	\label{eq:qf-loss-interaction}
\end{equation}
Its expansion is \eqref{eq:qf-interaction}.
Multiplying \eqref{eq:qf-surface-dyson} by $w_F(z)$ and integrating over the observation depth gives a closed scalar equation for the surface projection of $W^R$.
For an exterior source, \eqref{eq:qf-regular-exterior} then gives
\begin{equation}
	\begin{aligned}
		&\left[1-\chi_S(q,\omega)\mathcal V(q,\omega)\right]
		\int_0^\infty d z_1 \,w_F(z_1)
		W^R(q; z_1 ,z';\omega)
		\\
		&\qquad
		=e^{qz'}\boldsymbol t^T
		A(q,\omega)^{-1}\boldsymbol v,
		\qquad z'<0,
	\end{aligned}
	\label{eq:qf-surface-projection}
\end{equation}
where $z_1$ is an interior depth. Substituting the solution of \eqref{eq:qf-surface-projection} into \eqref{eq:qf-surface-dyson}, and using reciprocity for
the coupling to the exterior probe, yields
\begin{equation}
	\begin{aligned}
		&W^R(q;z,z';\omega)-W_r(q;z,z';\omega)
		\\
		&\qquad
		=e^{q(z+z')}
		\frac{
			\chi_S(q,\omega)
			\left[
			\boldsymbol v^T A(q,\omega)^{-1}\boldsymbol t
			\right]^2
		}{
			1-\chi_S(q,\omega)\mathcal V(q,\omega)
		},
	\end{aligned}
	\label{eq:qf-exterior-surface-correction}
\end{equation}
for $z,z'<0$.
Comparison with \eqref{eq:WRgsurf}, using the regular
reflection coefficient in
\eqref{eq:qf-regular-exterior}, then yields
\begin{equation}
	\begin{aligned}
		g_{\mathrm{surf}}(q,\omega)
		={}&1-\frac{2\varepsilon_0\epsilon_2q}{e^2}
		\Bigg[
		\boldsymbol v^T A(q,\omega)^{-1}\boldsymbol v
		\\
		&\qquad+
		\frac{
			\chi_S(q,\omega)
			\left[
			\boldsymbol v^T A(q,\omega)^{-1}\boldsymbol t
			\right]^2
		}{
			1-\chi_S(q,\omega)\mathcal V(q,\omega)
		}
		\Bigg].
	\end{aligned}
	\label{eq:qf-surface-response-int}
\end{equation}
Bulk absorption enters the regular reflection, the coupling
to the surface charge, and the collective denominator
through the same $A(q,\omega)$.

Considering projected interaction \eqref{eq:qf-loss-interaction} and using the symmetry of the screening matrix $A(q,\omega)$, one writes
\begin{equation}
	\mathcal V(q,\omega)
	=\frac{A_{22}t_1^2-2A_{12}t_1t_2+A_{11}t_2^2}
	{\det A(q,\omega)}.
\end{equation}
Expanding the determinant after the surface contribution is included gives
\begin{equation}
	\begin{aligned}
		&\det\!\left[A(q,\omega)
		-\chi_S(q,\omega)\boldsymbol t\boldsymbol t^T\right]\\
		&=\det A(q,\omega)
		-\chi_S(q,\omega)
		\left[A_{22}t_1^2-2A_{12}t_1t_2+A_{11}t_2^2\right],
	\end{aligned}
\end{equation}
whereas the terms quadratic in the surface response cancel. Substituting the expression for the projected interaction, one sees that the denominator in \eqref{eq:qf-surface-response} obeys
\begin{equation}
	\begin{aligned}
		&\det\left[
		A(q,\omega)-\chi_S(q,\omega)
		\boldsymbol t\boldsymbol t^T
		\right]
		\\
		&\qquad
		=\det A(q,\omega)
		\left[
		1-\chi_S(q,\omega)\mathcal V(q,\omega)
		\right].
	\end{aligned}
	\label{eq:qf-surface-determinant}
\end{equation}
Using \eqref{eq:qf-surface-determinant}, the surface response function \eqref{eq:qf-surface-response-int} becomes
\begin{equation}
	g_{\mathrm{surf}}(q,\omega)
	=1-\frac{2\varepsilon_0\epsilon_2q}{e^2}
	\boldsymbol v^T
	\left[
	A(q,\omega)
	-\chi_S(q,\omega)\boldsymbol t\boldsymbol t^T
	\right]^{-1}
	\boldsymbol v.
	\label{eq:qf-surface-response}
\end{equation}
\noindent
To separate the bulk and surface contributions to the
loss function, we first express the induced surface
density in terms of the fully screened interior potential.
For an exterior source, we introduce its coefficients
in the trial basis through
\begin{equation}
	W^{R,(2)}(q;z>0,z'<0;\omega)
	=e^{qz'}\boldsymbol\phi(z)^T\boldsymbol d^R(q,\omega).
\end{equation}
Using the source overlaps
\eqref{eq:two-potential-source-overlaps}, the drumhead
average of this potential is
\begin{equation}
	\int_0^\infty dz\,w_F(z)
	W^{R,(2)}(q;z,z';\omega)
	=e^{qz'}\boldsymbol t^T\boldsymbol d^R.
\end{equation}
At the retained order, the factorized surface polarization
\eqref{eq:PiSSnoLinearCorrection} converts this averaged
potential into the induced surface density response:
\begin{equation}
	\begin{aligned}
		&\int_0^\infty dz_1\,
		\Pi^{SS}(q;z,z_1;\omega)
		W^{R,(2)}(q;z_1,z';\omega)\\
		&\qquad=
		e^{qz'}w_F(z)\chi_S(q,\omega)
		\left(\boldsymbol t^T\boldsymbol d^R\right).
	\end{aligned}
\end{equation}
Thus $\chi_S(q,\omega)$ multiplies the averaged potential
to give the induced surface density amplitude, whose
depth profile is $w_F(z)$. 
This induced density supplies an additional source for
the potential through the regular screened interaction
in the Dyson equation \eqref{eq:qf-surface-dyson}.
Projecting the density onto each trial potential gives
the overlap $t_j$.
Combining this contribution with the exterior source
in \eqref{eq:qf-regular-exterior}, and cancelling the
common factor $e^{qz'}$, yields
\begin{equation}
	A(q,\omega)\boldsymbol d^R
	=\boldsymbol v
	+\chi_S(q,\omega)\boldsymbol t
	\left(\boldsymbol t^T\boldsymbol d^R\right).
	\label{eq:qf-screened-exterior-source}
\end{equation}
Here $\boldsymbol v$ represents the exterior source,
while the second term represents the projected induced
surface density.
Because that density responds to the fully screened potential itself, this equation determines the potential coefficients self-consistently. From \eqref{eq:qf-screened-exterior-source}
\begin{equation}
	\boldsymbol d^R =
	\left[A(q,\omega)-\chi_S(q,\omega)\boldsymbol t\boldsymbol t^T\right]^{-1}\boldsymbol v
	\label{eq:dR}
\end{equation}
so that the surface response function \eqref{eq:qf-surface-response} is equivalently written as
\begin{equation}
	g_{\mathrm{surf}}(q,\omega)
	=1-\frac{2\varepsilon_0\epsilon_2q}{e^2}
	\boldsymbol v^T\boldsymbol d^R.
	\label{eq:gsurf-usable}
\end{equation}
Equation~\eqref{eq:gsurf-usable} allows to express the loss function
through the imaginary part of $\boldsymbol v^T\boldsymbol d^R$.
\noindent
To separate in this combination the contributions from bulk and surface particle-hole creation,
we multiply the coefficient equation \eqref{eq:qf-screened-exterior-source} from the left by $(\boldsymbol d^R)^\dagger$, obtaining
\begin{equation}
	\left(\boldsymbol d^R\right)^\dagger\boldsymbol v
	=\left(\boldsymbol d^R\right)^\dagger
	A(q,\omega)\boldsymbol d^R-\chi_S(q,\omega)
	\left|\boldsymbol t^T\boldsymbol d^R\right|^2.
	\label{eq:qf-loss-source-balance}
\end{equation}
In the retained approximation \eqref{eq:qf-dynamic-matrix} for $A(q,\omega)$,
the static screening matrix and the interface correction are real and symmetric,
so their quadratic forms in the equation above are real.
The term $i\omega\mathsf C$, which accounts for bulk
particle-hole creation through
\eqref{eq:qf-projected-loss}, therefore gives
\begin{equation}
	\Im\!\left[
	\left(\boldsymbol d^R\right)^\dagger
	A(q,\omega)\boldsymbol d^R
	\right]
	=\omega
	\left(\boldsymbol d^R\right)^\dagger
	\mathsf C\boldsymbol d^R.
\end{equation}
Taking the imaginary part of \eqref{eq:qf-loss-source-balance} and using the above identity gives
\begin{equation}
	\begin{aligned}
		-\Im\!\left[\boldsymbol v^T\boldsymbol d^R\right]
		={}&\omega
		\left(\boldsymbol d^R\right)^\dagger
		\mathsf C\boldsymbol d^R-\Im\chi_S(q,\omega)
		\left|\boldsymbol t^T\boldsymbol d^R\right|^2.
	\end{aligned}
\end{equation}
\noindent
The loss function $L(q,\omega)$ defined in the main text can therefore be written as
\begin{equation}
	\begin{aligned}
		&L(q,\omega)
		=-\frac{2\varepsilon_0\epsilon_2q}{e^2}
		\Im\!\left[\boldsymbol v^T\boldsymbol d^R\right]\\
		&\qquad=\frac{2\varepsilon_0\epsilon_2q}{e^2}
		\Bigg[
		\omega\left(\boldsymbol d^R\right)^\dagger
		\mathsf C\boldsymbol d^R
		-\Im\chi_S(q,\omega)
		\left|\boldsymbol t^T\boldsymbol d^R\right|^2
		\Bigg]
	\end{aligned}
	\label{eq:qf-surface-loss}
\end{equation}
The bulk absorption matrix
\eqref{eq:qf-projected-loss} weights the potential
coefficients in the first term, which therefore describes bulk particle-hole
creation in the fully screened interior potential.
The second term describes instead surface particle-hole creation:
the imaginary part of the surface response is weighted
by the squared drumhead average of that potential.
Both contributions are nonnegative because
$\mathsf C$ is positive semidefinite and
\eqref{eq:qf-surface-absorption} gives a nonpositive
imaginary part of the retarded surface response.

Above the surface particle-hole continuum, the surface
absorption term vanishes and the retained bulk term alone supplies the loss.
The mixed sectors affect the screened potential through the static screening matrix. Their absorption thresholds
lie however above the frequencies considered here.


\section{Screening and sensitivity regimes}
\label{app:scaling}

We express the static interaction of App. \ref{app:finite-green} in dimensionless form and identify the weak-screening regime and the regime of large chemical-potential sensitivity.

\subsection{Static coupling and material parameters}

Define the ratio of the surface velocity to the bulk velocity
\begin{equation}
	\mathcal{U}=\frac{v_S}{v}.
	\label{eq:scaling-original-parameters}
\end{equation}
At fixed material parameters, the chemical-potential dependence of this velocity ratio and of the penetration and screening lengths can be written as
\begin{equation}
	\rho=\rho_0\sqrt{\mathcal Y},\qquad
	\mathcal{U}=\mathcal{U}_0 \sqrt{1-\mathcal Y},
	\label{eq:scaling-path}
\end{equation}
with $\rho_0$ in the main text and
\begin{equation}\label{rq:U0}
	\mathcal{U}_0=\frac{2|C|k_{\mathrm{NL}}}{\hbar v}
\end{equation}
representing a surface-bulk velocity anisotropy. 
It is convenient to formulate our solution in terms of the dimensionless quantities
\begin{equation}
	\overline A=\frac{A}{\nu_S},\qquad
	\overline M=\frac{M}{\nu_S},\qquad
	\overline{\boldsymbol b}=\nu_S\boldsymbol b.
	\label{eq:scaling-matrices}
\end{equation}
Using \eqref{eq:qTF} in \eqref{eq:two-potential-explicit-A} gives
\begin{equation}
	\begin{aligned}
		\overline A&=\overline B+\overline M,\\
		\overline B
		&=\frac{1}{ \cratio }
		\begin{pmatrix}
			\rho&\rho\\
			\rho&\rho^2+\tfrac14
		\end{pmatrix},\\
		\boldsymbol t
		&=\begin{pmatrix}
			2\rho/(1+2\rho)\\
			1/2
		\end{pmatrix},\qquad
		\overline A\,\overline{\boldsymbol b}=\boldsymbol t.
	\end{aligned}
	\label{eq:scaling-static-matrix}
\end{equation}
The band-dependent surface–bulk compressibility ratio $\cratio$ is defined in the main text,
$\overline B$ is the local bulk contribution and $\boldsymbol{t}$ comes from \eqref{eq:two-potential-source-overlaps}.
The entries of $M$ are given by \eqref{eq:finite-M-rational}, with the trial exponents
in \eqref{eq:phi}.
Measuring inverse lengths in units of $\kappa_F$ and the
mass integration variable in units of
$|m_F|=Dk_{\mathrm{NL}}^2\mathcal Y$ shows that
$\overline M$ depends on $\rho,\mathcal Y,g, \vartheta $.
The resulting static coupling is
\begin{equation}
	\zeta 
	=\boldsymbol t^T\overline A^{-1}\boldsymbol t
	=\boldsymbol t^T\overline{\boldsymbol b}.
	\label{eq:scaling-beta}
\end{equation}
Its component expression follows from
\eqref{eq:two-potential-static-interaction}.

\subsection{Weak-screening regime}
\label{sec:weak-mixed}
\noindent
The surface state depth profile is of paramount importance in defining the impact of bulk screening \cite{Silkin2005}.
Weak bulk screening on the surface penetration scale corresponds to $\rho\gg1$.
We take $\rho\to\infty$ at fixed $\mathcal Y,g, \vartheta $.
Let $\overline{\boldsymbol b}_B$ denote the potential coefficients obtained from the local bulk block $\overline B$.
Inverting this block in \eqref{eq:scaling-static-matrix}, and taking the large-$\rho$ limit of $\zeta_B$, gives
\begin{equation}
	\begin{aligned}
		\overline{\boldsymbol b}_B
		&=\frac{ \cratio }{(1+2\rho)(2\rho-1)}
		\begin{pmatrix}4\rho\\-2\end{pmatrix},\\
		\zeta _B
		&\sim\frac{ \cratio }{\rho}.
	\end{aligned}
	\label{eq:weak-beta-B}
\end{equation}
Thus $\overline b_{B,1}=O(\rho^{-1})$ and
$\overline b_{B,2}=O(\rho^{-2})$.

The factor of $s$ in the transition overlap
\eqref{eq:finite-H}, with
$s_1=\kappa_F/\rho$ and $s_2=2\kappa_F$,
gives the mixed-matrix orders
\begin{equation}
	\overline M_{11}=O(\rho^{-2}),\quad
	\overline M_{12}=O(\rho^{-1}),\quad
	\overline M_{22}=O(1).
	\label{eq:weak-matrix-orders}
\end{equation}
Expanding the inverse of $\overline B+\overline M$
then gives
\begin{equation}
	\zeta 
	= \zeta _B-
	\overline{\boldsymbol b}_B^T\overline M\,
	\overline{\boldsymbol b}_B+O(\rho^{-6}).
	\label{eq:weak-coupling-expansion}
\end{equation}
The displayed contraction is $O(\rho^{-4})$.
Positive semidefiniteness of $\overline M$ also gives
the bound
\begin{equation}
	0\leq\frac{ \zeta _B- \zeta }{ \zeta _B}
	\leq
	\frac{\overline{\boldsymbol b}_B^T\overline M\,
		\overline{\boldsymbol b}_B}{ \zeta _B}
	=O(\rho^{-3}).
	\label{eq:weak-hierarchy}
\end{equation}
Mixed screening therefore lowers the coupling, with a relative correction bounded by the displayed contraction.
The latter can be evaluated directly from the mixed matrix of App. \ref{sec:finite-mixed}.

The additional condition $ \cratio /\rho\ll1$ gives weak effective coupling.
In that case the expression of $\aspv$ yields $\aspv /v_S-1= \zeta ^2/2+O( \zeta ^3)$.
Along the material path \eqref{eq:scaling-path}, the weak-screening window is $\rho_0^{-2}\ll\mathcal Y<1$ and exists for $\rho_0\gg1$.
At fixed dielectric ratios, the electrostatic conditions \eqref{eq:finite-electrostatic-window} require $q/q_{\mathrm{TF}}\ll1$ throughout the weak-screening limit.

\subsection{Around the drumhead band edge}
\label{app:band-endpoint}
\noindent
We now approach the drumhead band edge by varying the chemical potential at fixed material parameters.
Writing $\delta=1-\mathcal Y$, the surface dispersion in App. \ref{app:Eigenstates} and the fixed-material parameterization \eqref{eq:scaling-path} behave as
\begin{equation}
	\begin{aligned}
		k_F&=k_{\mathrm{NL}}\sqrt{\delta},\\
		\rho&=\rho_0[1-\delta/2+O(\delta^2)],\\
		v_S&=v\mathcal{U}=v\mathcal{U}_0\sqrt{\delta}
	\end{aligned}
	\label{eq:article-y1-band}
\end{equation}
to leading order.
In this limit $\delta\to0^+$, the surface Fermi contour contracts to a point and the surface Fermi velocity vanishes, 
while the screening ratio approaches a material-dependent endpoint value  $\rho_0$. 
The constants $\rho_0$ and $\mathcal{U}_0$ are given in the main text and in  \eqref{rq:U0}, respectively.

The mass term at the Fermi surface \eqref{eq:mFdef} can be written as $m_F=-Dk_{\mathrm{NL}}^2\mathcal Y$:
the interval 
in \eqref{eq:finite-domains} in which mixed transitions are active
therefore shrinks to zero as $Dk_{\mathrm{NL}}^2\delta$.
The interval associated with $\eta'=\mathrm{sgn}(\mu)$ in Eq. \eqref{eq:finite-domains}
remains instead finite and contributes to the limiting value of the mixed screening matrix.
After the $m'$ integration in \eqref{eq:finite-M-rational},
the remaining $x$-integrands are $O(x^2)$ as $x\to0$
and $O(1-x)$ as $x\to1$.
Differentiation with respect to $\mathcal Y$ along
\eqref{eq:scaling-path}, including the dependence of
the mass-integration limits and trial exponents,
preserves these endpoint bounds.
The integrals defining $d\overline M_{ij}/d\mathcal Y$
therefore converge, so the mixed-matrix entries have
finite limits and finite first derivatives at the band edge.
The positive static screening form gives a positive-definite
limiting matrix, its inverse remains therefore finite.

Denoting with a star the value at the band edge,
$(\rho,\mathcal Y)=(\rho_0,1)$, 
\begin{equation}
	\begin{aligned}
		\overline A_*&=\overline A(\rho_0,1),\\
		\boldsymbol t_*
		&=\begin{pmatrix}
			2\rho_0/(1+2\rho_0)\\
			1/2
		\end{pmatrix},\\
		\zeta _*
		&=\boldsymbol t_*^T\overline A_*^{-1}\boldsymbol t_*,
		\qquad
		\zeta = \zeta _*+O(\delta).
	\end{aligned}
	\label{eq:article-y1-coupling}
\end{equation}
where the limiting matrix follows from \eqref{eq:scaling-static-matrix},
and the coupling from \eqref{eq:scaling-beta}.

Differentiating the ASP velocity $\aspv$ at fixed material parameters, see \eqref{eq:scaling-path},
separates the variation of the surface Fermi velocity from that of the screened coupling
\begin{equation}
	\begin{aligned}
		\frac{d\aspv }{d\mathcal Y}
		={}&-\frac{v\mathcal{U}_0}{2\sqrt{1-\mathcal Y}}\,
		\frac{1+ \zeta }{\sqrt{1+2 \zeta }}\\
		&+v\mathcal{U}_0\sqrt{1-\mathcal Y}\,
		\frac{ \zeta }{(1+2 \zeta )^{3/2}}\,
		\frac{d \zeta }{d\mathcal Y},\\
		\frac{d \zeta }{d\mathcal Y}
		={}&\frac{\rho}{2\mathcal Y}\,
		\partial_\rho \zeta +\partial_{\mathcal Y} \zeta .
	\end{aligned}
	\label{eq:scaling-speed-sweep}
\end{equation}
Here $\partial_\rho \zeta $ holds $\mathcal Y,g, \vartheta $
fixed, while $\partial_{\mathcal Y} \zeta $ holds
$\rho,g, \vartheta $ fixed.
The screening matrix and source have finite first derivatives at the endpoint, so $d \zeta /d\mathcal Y$ remains finite.

The contribution to  \eqref{eq:scaling-speed-sweep} containing the coupling derivative vanishes as $\delta^{1/2}$, whereas the surface-velocity contribution diverges as $\delta^{-1/2}$. 
Using \eqref{eq:article-y1-coupling} in the expression for $\aspv$ in the main text, together with \eqref{eq:scaling-speed-sweep} gives
\begin{equation}
	\begin{aligned}
		\frac{\aspv }{v}
		&=\mathcal{U}_0
		\frac{1+ \zeta _*}{\sqrt{1+2 \zeta _*}}\,
		\sqrt{\delta}\,[1+O(\delta)],\\
		\frac{1}{v}\frac{d\aspv }{d\mathcal Y}
		&=-\frac{\mathcal{U}_0(1+ \zeta _*)}
		{2\sqrt{1+2 \zeta _*}}\,
		\frac{1}{\sqrt{\delta}}\,[1+O(\delta)].
	\end{aligned}
	\label{eq:band-edge-speed}
\end{equation}
The ASP velocity therefore vanishes at the band edge, while its dependence on chemical potential becomes increasingly steep.

The change in surface electron density follows from the change in occupation of the drumhead band: for positive curvature, an expanding Fermi contour adds electrons, whereas for negative curvature, it removes electrons.
Including the spin degeneracy used in $\nu_S$ gives
\begin{equation}
	\begin{aligned}
		\delta n
		&=\frac{\mathrm{sgn}(C)}{2\pi}\,\delta(k_F^2)
		=-\frac{\mathrm{sgn}(C)k_{\mathrm{NL}}^2}{2\pi}\,
		\delta\mathcal Y,\\
		\frac{d\sigma}{d\mathcal Y}
		&=\frac{e\,\mathrm{sgn}(C)k_{\mathrm{NL}}^2}{2\pi},\\
		\frac{d\aspv }{d\sigma}
		&=\frac{2\pi\,\mathrm{sgn}(C)}{e k_{\mathrm{NL}}^2}
		\frac{d\aspv }{d\mathcal Y}.
	\end{aligned}
	\label{eq:band-edge-charge-conversion}
\end{equation}
Combining this conversion with \eqref{eq:band-edge-speed}
gives the charge derivative 
\begin{equation}
	\left|\frac{d\aspv }{d\sigma}\right| = \frac{\pi v\mathcal{U}_0}{e k_{\mathrm{NL}}^2} \frac{1+ \zeta _*}{\sqrt{1+2 \zeta _*}} \frac{1}{\sqrt{1-\mathcal Y}}\left[1+O(1-\mathcal Y)\right],
	\label{eq:band-edge-speed-speed}
\end{equation}
which is rewritten in the main text.
Its magnitude diverges as the inverse square root of the distance to the band edge.
Near that edge, adding electrons increases the ASP velocity for positive curvature; for negative curvature, it reduces the hole population and decreases the velocity. 
For charge variations satisfying $|\delta\sigma|\ll e k_F^2/(2\pi)$,
the dispersion $\omega_{\mathrm{SP}}$ in the main text gives the resonance displacement to first order in the charge variation.
From the quality factor $Q_{\mathrm{SP}}$, also in the main text, its magnitude relative to the bulk-induced half-width is, 
\begin{equation}
	\begin{aligned}
		\delta\Re\omega_{\mathrm{SP}}(q)
		&\simeq q\,\frac{d\aspv }{d\sigma}\,\delta\sigma,\\
		\frac{|\delta\Re\omega_{\mathrm{SP}}(q)|}{\Gamma_B(q)}
		&\simeq
		\frac{2\pi Q_{\mathrm{SP}}(q)}
		{e k_{\mathrm{NL}}^2\delta}\,|\delta\sigma|,
	\end{aligned}
	\label{eq:band-edge-charge-resolution}
\end{equation}
to leading order in momentum and near the band edge. 
The small carrier population enhances the fractional resonance displacement, while the quality factor determines
its size relative to the intrinsic linewidth.

The band-edge regime overlaps with weak screening
when $\rho_0\gg1$, because the screening ratio remains
close to its endpoint value. The relative mixed correction is then controlled by
\eqref{eq:weak-hierarchy}. 
Finally, we note that the surface-response expansion in
App. \ref{subsec:Long-wavelength-surface} requires $q\ll\min(k_F,q_k)$.
Near the band edge, the penetration-variation scale exceeds the surface Fermi wavevector.
Using \eqref{eq:article-y1-band}, their ratio and the
resulting momentum restriction are
\begin{equation}
	\frac{q_k}{k_F}
	=\frac{1-\delta}{2\delta}\gg1 \;, \qquad
	\left(\frac{q}{k_{\mathrm{NL}}}\right)^2
	\ll\delta\ll1.
	\label{eq:band-edge-momentum}
\end{equation}
Thus the restrictive surface-response condition is
that the momentum remain small compared with the
shrinking surface Fermi wavevector.
At fixed momentum, this sets a lower bound on the
distance of the Fermi energy from the band edge.

\section{Material fit}\label{app:Materials}

The estimates below specify how the bulk and surface bands are mapped onto the Hamiltonian and surface dispersion in the main text.
We take $\epsilon_\perp=10$ unless otherwise stated.
The couplings reported in the main text include both bulk and mixed screening in the two-potential approximation of App. \ref{app:finite-green}.
Bulk-only estimates are identified separately below. Velocities are quoted through $\hbar v_S$ and $\hbar \aspv $, in eV\AA.

\subsection{Ca$_3$P$_2$}\label{app:Ca3P2}

\paragraph{Bulk and surface bands.}
The spin-rotation-invariant, two-orbital fit in \cite{Chan2016a}, describes a single nodal ring in the $k_z=0$ plane. Its scalar term is $f(\boldsymbol{k})=\nu_0(k_\parallel^2-k_0^2)+V_0$, where $k_0$ is the ring radius, $\nu_0 $ is the scalar curvature, and $V_0$ is the energy of the crossing relative to the energy origin of the fit. Denoting the
fitted mass curvature and normal dispersion coefficient by
$\nu_\parallel$ and $\nu_z$, respectively, the reduction to first order in $k_z$ gives
\begin{equation}
	\begin{aligned}
		k_{\mathrm{NL}}=k_0&=0.206\ \text{\AA}^{-1},&D=\nu_\parallel&=4.34\ \mathrm{eV}\text{\AA}^2,\\
		\hbar v=\nu_z&=2.50\ \mathrm{eV}\text{\AA},&g=\nu_0/\nu_\parallel&=-0.2288,
	\end{aligned}
\end{equation}
with $\nu_0=-0.993$ eV\AA$^2$ and $V_0=0.095$ eV. The fit does not take
spin-orbit coupling into account \cite{Chan2016a}.

The calculated drumhead effective mass $m^*\simeq4m_e$, where $m_e$ is
the free-electron mass, gives $C=\hbar^2/(2m^*)=0.9525$ eV\AA$^2$ and
$ \vartheta =\arccos(C/D-g)=1.106$~\cite{Chan2016a}. Thus $C/D=0.219$.
The inferred boundary angle applies to the termination used for that
surface dispersion. Surface relaxation and band bending depend on
termination and can obscure the drumhead among other weakly dispersing
bands~\cite{Alassaf2024}.

\paragraph{Chemical potential and screening.}
The partially filled drumhead requires $-Ck_{\mathrm{NL}}^2<\mu<0$, or $-40.42<\mu<0$ meV. We choose its midpoint, $\mu=-Ck_{\mathrm{NL}}^2/2=-20.21$ meV, so that $\mathcal Y=1/2$. Interpreting the fitted $V_0$ as the crossing energy above the reference Fermi level gives $\mu_{\mathrm{ref}}=-95$ meV; the chosen occupation therefore requires raising the Fermi level by $74.8$ meV. 
This only specifies the required energy shift, not a particular doping mechanism.

The dispersion $E_S$ in the main body of the paper,  \eqref{eq:kappaF}, and \eqref{eq:qTF} 
\begin{equation}
	\begin{aligned}
		k_F&=0.1457\ \text{\AA}^{-1},&\kappa_F&=0.03293\ \text{\AA}^{-1},\\
		q_{\mathrm{TF}}&=0.0763\ \text{\AA}^{-1},&\kappa_F/q_{\mathrm{TF}}&=0.432.
	\end{aligned}
\end{equation}
The densities of states are $\nu_S=0.1671$ eV$^{-1}$\AA$^{-2}$ and
$\nu_B=1.607\times10^{-4}$ eV$^{-1}$\AA$^{-3}$.
The amplitude decay length is $\kappa_F^{-1}=30.4$ \AA, 
$\hbar v_S=0.277$ eV\AA\ and $\hbar \aspv =0.568$ eV\AA.
The compressibility ratio $ \cratio =\frac{\nu_S\kappa_F}{2\nu_B}$ is $17.1$.

Joint bulk and mixed screening produces an effective coupling $ \zeta =6.84$
(reduced from bulk-only screening) and a velocity ratio $\aspv /v_S=2.05$ in the two-potential
approximation, as reported in the Table.

\subsection{BaSn$_2$}
\label{app:BaSn2}

\paragraph{Band mapping.}
Without spin-orbit coupling, there is a nonplanar calculated nodal loop, with crossing energies between $0.064$ and $0.117$ eV above the Fermi level $E_F$ and a drumhead bandwidth of approximately $0.18$ eV for the Sn termination~\cite{Huang2016}.
Spin-orbit coupling gaps the loop, so the planar circular NLS model only provides an approximate local description of the spin-orbit-free bands.

The estimates used here are $k_{\mathrm{NL}}=0.136(2\pi/a)=0.184$ \AA$^{-1}$, with the lattice
constants $a=4.652$ \AA and $c=5.546$ \AA ~\cite{Fumega2021,Kim2008}, and a radial bulk-band slope $s_r\simeq2.0\pm0.6$ eV\AA. 
Identifying $D=s_r/(2k_{\mathrm{NL}})$ gives $D\simeq5.44$ eV\AA$^2$, with the range $3.8$-$7.1$ eV\AA$^2$.
The estimated nodal coordinates $(k,E_{\mathrm{node}}-E_F)=(0.184,0.117)$, $(0.150,0.090)$,
and $(0.122,0.064)$, in \AA$^{-1}$ and eV, give $dE_{\mathrm{node}}/d(k^2)\simeq2.76$ eV\AA$^2$~\cite{Huang2016};
here $E_{\mathrm{node}}$ denotes the crossing energy.
Assigning this slope to $gD$ gives $g\simeq0.51$ (this identification is approximate, because the crossing energy also varies with $k_z$ along the loop).
Mapping the surface bandwidth to \eqref{eq:ES} gives $C\simeq0.18/k_{\mathrm{NL}}^2=5.32$ eV\AA$^2$,
$ \vartheta \simeq1.08$, and $C/D\simeq0.98$. 
For lack of data, we assume $\hbar v=1.00$ eV\AA.

\paragraph{Occupation and acoustic slope.}
Taking the highest crossing as the local energy reference gives $\mu=-0.117$ eV at the calculated Fermi level and $\mathcal Y=0.65$.
The mapped model gives
\begin{equation}
	\begin{aligned}
		k_F&\simeq0.109\ \text{\AA}^{-1},&\kappa_F&\simeq0.1054\ \text{\AA}^{-1},\\
		q_{\mathrm{TF}}&=0.312\ \text{\AA}^{-1},&\kappa_F/q_{\mathrm{TF}}&=0.338.
	\end{aligned}
\end{equation}
The densities of states are $\nu_S=0.02992$ eV$^{-1}$\AA$^{-2}$ and $\nu_B=2.694\times10^{-3}$ eV$^{-1}$\AA$^{-3}$.
The surface DOS assumes a complete circular pocket, and the bulk DOS includes only the modeled ring. The amplitude decay length is $\kappa_F^{-1}\simeq9.5$ \AA. 
Including mixed screening in the two-potential approximation gives $ \zeta =0.445$ and $\aspv /v_S=1.05$, as in the main body of the paper.
Additional bulk pockets would strengthen screening and reduce the velocity ratio.

\subsection{CaAgAs}
\label{app:CaAgAs}

\paragraph{Model and parameter source.}
The estimates refer to the spin-orbit-free fit in \cite{Zhao2023}.
The calculated spin-orbit gap \cite{Wang2017,Takane2018} has not been observed, to our knowledge.
Photoemission finds instead hole-doped samples, with a Fermi-level downshift of about $0.5$ eV relative to the calculation~\cite{Wang2017}, while bulk-sensitive measurements place the sample Fermi level about $0.1$ eV below the ungapped nodes and report a Hall hole density of approximately $1.6\times10^{20}$ cm$^{-3}$~\cite{Takane2018}.
For the estimates below we choose the spin-orbit-free model and a Fermi level inside its drumhead window.

Up to an orbital rotation, the fitted Hamiltonian is
\begin{equation}
	\begin{aligned}
		H_{\mathrm{fit}}={}&(C_0+C_\perp k^2+C_zk_z^2)\sigma_0\\
		&+(M_0-M_\perp k^2-M_zk_z^2)\sigma_z+Ak_z\sigma_y,
	\end{aligned}
\end{equation}
where $k=|\boldsymbol{k}_\parallel|$, $k_z$ is normal to the (0001) surface, $\sigma_0$ is the orbital identity matrix, and $\sigma_y,\sigma_z$ are orbital Pauli matrices.
The coefficients $C_\perp,C_z$ describe the scalar curvatures, $M_0,M_\perp,M_z$ the orbital mass, and $A$ the normal linear dispersion. 
The values are $C_\perp=154$, $C_z=650$, $M_\perp=-386$, and $M_z=-1170$ eV\AA$^2$, $M_0=-0.30$ eV, and $A=8.0$ eV\AA\, see \cite{Zhao2023}, Fig.~2(b). The reference fits the coefficients to the photoemission results of \cite{Wang2017}. We choose the common energy offset $C_0=-0.08$ eV.

\paragraph{Reduction and surface dispersion.}
At $k_z=0$, the ring radius is $k_{\mathrm{NL}}=\sqrt{M_0/M_\perp}=0.02788$ \AA$^{-1}$ and its energy in the chosen convention is $E_{\mathrm{NL}}=C_0+C_\perp k_{\mathrm{NL}}^2=39.69$ meV.
Subtracting $E_{\mathrm{NL}}$ and retaining terms to first order in $k_z$ produces the minimal Hamiltonian used in this study with $D=-M_\perp=386$ eV\AA$^2$, $g=C_\perp/D=0.3990$, and $\hbar v=A=8.0$ eV\AA.
The omitted normal-curvature terms are $C_zk_z^2\sigma_0-M_zk_z^2\sigma_z$.

The open-boundary surface dispersion is $E_S^{\mathrm{fit}}(k)=C_0+C_zM_0/M_z+
(C_\perp-C_zM_\perp/M_z)k^2$, for $k^2<M_0/M_\perp$, see \cite{Zhao2023}, Eq.~(3).
Thus $C=-60.44$ eV\AA$^2$ and matching to \eqref{eq:ES} fixes $\cos \vartheta =(C-C_\perp)/D=C_z/M_z=-5/9$, or $ \vartheta =2.160$.
The complete two-spin surface contour has $\nu_S=2.633\times10^{-3}$ eV$^{-1}$\AA$^{-2}$.
Partial filling requires $0<\mu<|C|k_{\mathrm{NL}}^2=46.98$ meV, or $39.7<E_F<86.7$ meV in the chosen energy convention.

\paragraph{Occupation and surface profile.}
We take $E_F=42$ meV, corresponding to $122$ meV in the energy
convention of \cite{Zhao2023}, Fig.~3(b). Then
$\mu=E_F-E_{\mathrm{NL}}=2.311$ meV and $\mathcal Y=0.049$.
Equations~\eqref{eq:ES} and \eqref{eq:kappaF} give then 
$k_F=0.027184$ \AA$^{-1}$ and
$\kappa_F=1.534\times10^{-3}$ \AA$^{-1}$, with amplitude decay
length $\kappa_F^{-1}=652$ \AA.

The single exponential $w_F$, see App. \ref{app:Eigenstates}, belongs to the linearized model. The quadratic fit instead has two normal decay constants.
Writing $\Delta_F=E_F-C_0-C_\perp k_F^2$ and
$m_F=M_0-M_\perp k_F^2$, the equation
$\det[E_F-H_{\mathrm{fit}}(k_F,i\kappa)]=0$ is, with
$t=\kappa^2$,
\begin{equation}
	\begin{aligned}
		0={}&(C_z^2-M_z^2)t^2\\
		&+(2C_z\Delta_F-2M_zm_F+A^2)t+\Delta_F^2-m_F^2.
	\end{aligned}
\end{equation}
Its roots give inverse lengths $2.04\times10^{-3}$ and
$6.18\times10^{-3}$ \AA$^{-1}$. The screening estimates below use
the single-exponential profile of the linearized model.

\paragraph{Bulk screening and acoustic slope.}
Equation~\eqref{eq:qTF} gives
$\nu_B=7.724\times10^{-8}$ eV$^{-1}$\AA$^{-3}$ per spin.
With the assumed $\epsilon_\perp=10$, $q_{\mathrm{TF}}=1.672\times10^{-3}$ \AA$^{-1}$
and $\kappa_F/q_{\mathrm{TF}}=0.917$, $\hbar v_S=3.286$ eV\AA.
Joint screening in the two-potential approximation gives $ \zeta =8.42$
and $\aspv /v_S=2.23$, as in the Table in the main text.

\subsection{ZrSiSe}
\label{app:ZrSiSe}

\paragraph{Surface band.}
ZrSiSe hosts no drumhead states at the Fermi level. Its low-energy surface
states, observed in photoemission and quasiparticle interference, are instead a
FSS. These are tightly localized at the surface~\cite{Topp2017,Zhu2018}.
We model them by the charge sheet of App. \ref{sec:surface-sheet},
which corresponds to the limit $\kappa_F\to\infty$, i.e.\ $\rho=\infty$.

The surface DOS can be extracted from quantum oscillations data.
In a magnetic field $B$ perpendicular to the surface, 
electrons at the Fermi energy perform closed cyclotron orbits along the Fermi contour, whose Landau quantization makes the magnetoresistance oscillate periodically in $1/B$.
The oscillation frequency $F$ (in tesla) is fixed by the $k$-space area $A_k$ enclosed by the orbit through the Onsager relation $F=\hbar A_k/(2\pi e)$~\cite{Onsager1952,GirvinYang2019}.
The thermal damping of the oscillation amplitude, described by the Lifshitz-Kosevich
formula~\cite{LifshitzKosevich1956}, determines the cyclotron mass
$m_c=(\hbar^2/2\pi)\,\partial A_k/\partial E$ at the Fermi energy~\cite{Shoenberg1984}.
For ZrSiSe, \cite{Liu2021} attributes an oscillation of frequency with $F=445$ T (of period $\Delta(1/B)=1/F$) 
to an electron pocket of the floating surface band near $\bar X$,
with cross-section $A_k=4.25$ nm$^{-2}$ and cyclotron mass $m_c\simeq0.39\,m_e$.
The DOS of a two-dimensional pocket per spin is $(2\pi)^{-2}\,\partial A_k/\partial E=m_c/(2\pi\hbar^2)$, independently of the shape of the pocket. The floating band forms one electron pocket around each edge midpoint of the
square surface Brillouin zone~\cite{Topp2017,Liu2021}. Since each of the four edge midpoints is shared by two neighboring zones, the zone contains two pockets, 
related by a fourfold rotation  \cite{Topp2017,Zhu2018}. Together with both spin projections, this gives
\begin{equation}
	\nu_S=\frac{2m_c}{\pi\hbar^2}=0.03258\ \mathrm{eV}^{-1}\text{\AA}^{-2}.
	\label{eq:ZrSiSe-nuS}
\end{equation}
Using for simplicity a circular parabolic pocket of the same area, we estimate
$k_F=\sqrt{A_k/\pi}=0.116$ \AA$^{-1}$ and $\hbar v_S=\hbar^2k_F/m_c=2.27$ eV\AA.

\paragraph{Bulk density of states and permittivity.}
The calculated bulk density of states at the Fermi level is
$N(E_F)=0.47$ states/eV per unit cell, summed over spins, with lattice
constants $a=3.623$ \AA\ and $c=8.365$ \AA \cite{Yen2020}. Dividing by twice
the cell volume $a^2c$ gives the DOS per spin entering \eqref{eq:qTF},
\begin{equation}
	\nu_B=\frac{N(E_F)}{2a^2c}=2.140\times10^{-3}\ \mathrm{eV}^{-1}\text{\AA}^{-3}.
	\label{eq:ZrSiSe-nuB}
\end{equation}

The background permittivity $\epsilon_\perp$ is the permittivity along the
surface normal, i.e. along the $c$ axis for the cleavage plane of ZrSiSe.
It can be estimated from optical data through the high-frequency permittivity $\epsilon_{\infty}$, which accounts for the optical transitions above the Drude response.
In a Drude description, $\epsilon_1(\omega)$ changes sign at the screened plasma frequency
$\omega_{\mathrm{pl}}^{\mathrm{scr}}=\omega_{\mathrm{pl}}/\sqrt{\epsilon_\infty}$,
where $\omega_{\mathrm{pl}}$ is the unscreened Drude plasma
frequency~\cite{EbadAllah2022,DresselGruner2002,Grosso}. Therefore
$\epsilon_{\infty,c}=(\omega_{\mathrm{pl},c}/\omega^{\mathrm{scr}}_{\mathrm{pl},c})^2$
for light polarized along $c$, and we identify $\epsilon_\perp=\epsilon_{\infty,c}$.
For ZrSiSe, only the screened $c$-axis plasma frequency,
$\simeq0.37$ eV, is reported~\cite{Shao2022}, which does not determine
$\epsilon_{\infty,c}$. For the isostructural compounds ZrGeSe and ZrGeS,
Table~I of \cite{EbadAllah2022} reports
$\omega_{\mathrm{pl},c}=1.06$ and $0.95$ eV and
$\omega^{\mathrm{scr}}_{\mathrm{pl},c}=0.44$ and $0.36$ eV, hence
$\epsilon_{\infty,c}=5.8$ and $7.0$, respectively.
We take $\epsilon_\perp=6$ as an estimate.

\paragraph{Screening and acoustic slope.}
In the independent-subsystem approximation \eqref{eq:sheet-independent-subsystems}, the mixed sectors do not contribute, and the screened sheet potential \eqref{eq:sheet-bulk-interaction} lies in the trial subspace \eqref{eq:phi}.
The two-potential projection is therefore exact for this source.
For $\epsilon_\perp=6$, $e^2/(\varepsilon_0\epsilon_\perp)=30.16$ eV\AA, and \eqref{eq:qTF} with \eqref{eq:ZrSiSe-nuB} gives
\begin{equation}
	\begin{aligned}
		&q_{\mathrm{TF}}=\sqrt{\frac{2e^2\nu_B}{\varepsilon_0\epsilon_\perp}}=0.359\ \text{\AA}^{-1},\\
		& 	\mathcal V_{\mathrm{sh},0}=\frac{e^2}{\varepsilon_0\epsilon_\perp q_{\mathrm{TF}}}=83.9\ \mathrm{eV}\text{\AA}^2 .
	\end{aligned}
\end{equation}
With \eqref{eq:ZrSiSe-nuS}, the coupling and velocity \eqref{eq:sheet-bulk-velocity} are
\begin{equation}
	\zeta _{\mathrm{sh}}=\sqrt{\frac{e^2\nu_S^2}{2\varepsilon_0\epsilon_\perp\nu_B}}=2.73,
	\qquad
	\frac{\aspv}{v_S}=\frac{1+ \zeta _{\mathrm{sh}}}{\sqrt{1+2 \zeta _{\mathrm{sh}}}}=1.47,
	\label{eq:ZrSiSe-beta}
\end{equation}
as reported in the main body of this study. The ratio depends on the surface
band only through $\nu_S$, i.e.\ through $m_c$ and the number of pockets,
while $v_S$ sets the scale of the absolute slope, $\hbar\aspv=3.33$ eV\AA\ with
the orbit velocity above.

The sheet coupling bounds the bulk-screened coupling of a finite-depth source from above.
Using \eqref{eq:qTF} and \eqref{eq:sheet-bulk-velocity}, the bulk-only coupling $\zeta_B$ 
reads
\begin{equation}
	\zeta _B= \zeta _{\mathrm{sh}}\,
	\frac{\rho(1+4\rho)}{(1+2\rho)^2}.
\end{equation}
The last factor increases monotonically with $\rho$ and tends to $1$ for
$\rho\to\infty$, where $ \zeta _B\to \zeta _{\mathrm{sh}}$. At fixed $\nu_S$,
$\nu_B$ and $\epsilon_\perp$, a finite penetration length of the surface charge therefore
lowers $ \zeta $ and $\aspv/v_S$.
It is worth noting that the estimate assumes an isotropic response \eqref{eq:chiSurf},
whereas the floating pockets are in general anisotropic.

\paragraph{Comparison with experiment.}
Electron energy loss spectroscopy on ZrSiSe gives an acoustic slope $\aspv=8.24\times10^5$ m/s along $\bar\Gamma\bar M$, which  \cite{Xue2026} compares with a surface Fermi velocity $\hbar v_{\mathrm{F2D}}=3.30$-$4.15$ eV\AA, obtaining $\aspv/v_{\mathrm{F2D}}=1.3$-$1.6$.
The ratio $\aspv/v_S=1.47$ obtained above lies within this interval.
Combined with the same surface velocities, it gives $u = (7.4-9.3)\times 10^{5}$~m/s, which includes the measured slope.
The orbit velocity $v_S=3.45\times10^5$ m/s, considered as an average over the Fermi contour, is smaller than $v_{\mathrm{F2D}}$, measured along $\bar\Gamma\bar M$; the difference should indicate an anisotropic pocket. 
Since the velocity ratio does not depend on $v_S$, the comparison is made for $\aspv/v_S$.

\newpage

\bibliography{NLSASP}

\begin{thebibliography}{94}%
\makeatletter
\providecommand \@ifxundefined [1]{%
 \@ifx{#1\undefined}
}%
\providecommand \@ifnum [1]{%
 \ifnum #1\expandafter \@firstoftwo
 \else \expandafter \@secondoftwo
 \fi
}%
\providecommand \@ifx [1]{%
 \ifx #1\expandafter \@firstoftwo
 \else \expandafter \@secondoftwo
 \fi
}%
\providecommand \natexlab [1]{#1}%
\providecommand \enquote  [1]{``#1''}%
\providecommand \bibnamefont  [1]{#1}%
\providecommand \bibfnamefont [1]{#1}%
\providecommand \citenamefont [1]{#1}%
\providecommand \href@noop [0]{\@secondoftwo}%
\providecommand \href [0]{\begingroup \@sanitize@url \@href}%
\providecommand \@href[1]{\@@startlink{#1}\@@href}%
\providecommand \@@href[1]{\endgroup#1\@@endlink}%
\providecommand \@sanitize@url [0]{\catcode `\\12\catcode `\$12\catcode
  `\&12\catcode `\#12\catcode `\^12\catcode `\_12\catcode `\%12\relax}%
\providecommand \@@startlink[1]{}%
\providecommand \@@endlink[0]{}%
\providecommand \url  [0]{\begingroup\@sanitize@url \@url }%
\providecommand \@url [1]{\endgroup\@href {#1}{\urlprefix }}%
\providecommand \urlprefix  [0]{URL }%
\providecommand \Eprint [0]{\href }%
\providecommand \doibase [0]{https://doi.org/}%
\providecommand \selectlanguage [0]{\@gobble}%
\providecommand \bibinfo  [0]{\@secondoftwo}%
\providecommand \bibfield  [0]{\@secondoftwo}%
\providecommand \translation [1]{[#1]}%
\providecommand \BibitemOpen [0]{}%
\providecommand \bibitemStop [0]{}%
\providecommand \bibitemNoStop [0]{.\EOS\space}%
\providecommand \EOS [0]{\spacefactor3000\relax}%
\providecommand \BibitemShut  [1]{\csname bibitem#1\endcsname}%
\let\auto@bib@innerbib\@empty
\bibitem [{\citenamefont {Burkov}\ \emph {et~al.}(2011)\citenamefont {Burkov},
  \citenamefont {Hook},\ and\ \citenamefont {Balents}}]{Burkov2011}%
  \BibitemOpen
  \bibfield  {author} {\bibinfo {author} {\bibfnamefont {A.~A.}\ \bibnamefont
  {Burkov}}, \bibinfo {author} {\bibfnamefont {M.~D.}\ \bibnamefont {Hook}},\
  and\ \bibinfo {author} {\bibfnamefont {L.}~\bibnamefont {Balents}},\
  }\bibfield  {title} {\bibinfo {title} {Topological nodal semimetals},\ }\href
  {https://doi.org/10.1103/PhysRevB.84.235126} {\bibfield  {journal} {\bibinfo
  {journal} {Phys. Rev. B}\ }\textbf {\bibinfo {volume} {84}},\ \bibinfo
  {pages} {235126} (\bibinfo {year} {2011})}\BibitemShut {NoStop}%
\bibitem [{\citenamefont {Fang}\ \emph {et~al.}(2015)\citenamefont {Fang},
  \citenamefont {Chen}, \citenamefont {Kee},\ and\ \citenamefont
  {Fu}}]{Fang2015}%
  \BibitemOpen
  \bibfield  {author} {\bibinfo {author} {\bibfnamefont {C.}~\bibnamefont
  {Fang}}, \bibinfo {author} {\bibfnamefont {Y.}~\bibnamefont {Chen}}, \bibinfo
  {author} {\bibfnamefont {H.-Y.}\ \bibnamefont {Kee}},\ and\ \bibinfo {author}
  {\bibfnamefont {L.}~\bibnamefont {Fu}},\ }\bibfield  {title} {\bibinfo
  {title} {Topological nodal line semimetals with and without spin-orbital
  coupling},\ }\href {https://doi.org/10.1103/physrevb.92.081201} {\bibfield
  {journal} {\bibinfo  {journal} {Phys. Rev. B}\ }\textbf {\bibinfo {volume}
  {92}},\ \bibinfo {pages} {081201(R)} (\bibinfo {year} {2015})}\BibitemShut
  {NoStop}%
\bibitem [{\citenamefont {Fang}\ \emph {et~al.}(2016)\citenamefont {Fang},
  \citenamefont {Weng}, \citenamefont {Dai},\ and\ \citenamefont
  {Fang}}]{Fang2016}%
  \BibitemOpen
  \bibfield  {author} {\bibinfo {author} {\bibfnamefont {C.}~\bibnamefont
  {Fang}}, \bibinfo {author} {\bibfnamefont {H.}~\bibnamefont {Weng}}, \bibinfo
  {author} {\bibfnamefont {X.}~\bibnamefont {Dai}},\ and\ \bibinfo {author}
  {\bibfnamefont {Z.}~\bibnamefont {Fang}},\ }\bibfield  {title} {\bibinfo
  {title} {Topological nodal line semimetals},\ }\href
  {https://doi.org/10.1088/1674-1056/25/11/117106} {\bibfield  {journal}
  {\bibinfo  {journal} {Chin. Phys. B}\ }\textbf {\bibinfo {volume} {25}},\
  \bibinfo {pages} {117106} (\bibinfo {year} {2016})}\BibitemShut {NoStop}%
\bibitem [{\citenamefont {Chiu}\ and\ \citenamefont
  {Schnyder}(2014)}]{Chiu2014}%
  \BibitemOpen
  \bibfield  {author} {\bibinfo {author} {\bibfnamefont {C.-K.}\ \bibnamefont
  {Chiu}}\ and\ \bibinfo {author} {\bibfnamefont {A.~P.}\ \bibnamefont
  {Schnyder}},\ }\bibfield  {title} {\bibinfo {title} {Classification of
  reflection-symmetry-protected topological semimetals and nodal
  superconductors},\ }\href {https://doi.org/10.1103/PhysRevB.90.205136}
  {\bibfield  {journal} {\bibinfo  {journal} {Phys. Rev. B}\ }\textbf {\bibinfo
  {volume} {90}},\ \bibinfo {pages} {205136} (\bibinfo {year}
  {2014})}\BibitemShut {NoStop}%
\bibitem [{\citenamefont {Syzranov}\ and\ \citenamefont
  {Skinner}(2017)}]{Syzranov2017}%
  \BibitemOpen
  \bibfield  {author} {\bibinfo {author} {\bibfnamefont {S.~V.}\ \bibnamefont
  {Syzranov}}\ and\ \bibinfo {author} {\bibfnamefont {B.}~\bibnamefont
  {Skinner}},\ }\bibfield  {title} {\bibinfo {title} {Electron transport in
  nodal-line semimetals},\ }\href {https://doi.org/10.1103/PhysRevB.96.161105}
  {\bibfield  {journal} {\bibinfo  {journal} {Phys. Rev. B}\ }\textbf {\bibinfo
  {volume} {96}},\ \bibinfo {pages} {161105} (\bibinfo {year}
  {2017})}\BibitemShut {NoStop}%
\bibitem [{\citenamefont {Rui}\ \emph {et~al.}(2018)\citenamefont {Rui},
  \citenamefont {Zhao},\ and\ \citenamefont {Schnyder}}]{Rui2018}%
  \BibitemOpen
  \bibfield  {author} {\bibinfo {author} {\bibfnamefont {W.~B.}\ \bibnamefont
  {Rui}}, \bibinfo {author} {\bibfnamefont {Y.~X.}\ \bibnamefont {Zhao}},\ and\
  \bibinfo {author} {\bibfnamefont {A.~P.}\ \bibnamefont {Schnyder}},\
  }\bibfield  {title} {\bibinfo {title} {Topological transport in {Dirac}
  nodal-line semimetals},\ }\href {https://doi.org/10.1103/PhysRevB.97.161113}
  {\bibfield  {journal} {\bibinfo  {journal} {Phys. Rev. B}\ }\textbf {\bibinfo
  {volume} {97}},\ \bibinfo {pages} {161113} (\bibinfo {year}
  {2018})}\BibitemShut {NoStop}%
\bibitem [{\citenamefont {Biderang}\ \emph {et~al.}(2018)\citenamefont
  {Biderang}, \citenamefont {Leonhardt}, \citenamefont {Raghuvanshi},
  \citenamefont {Schnyder},\ and\ \citenamefont {Akbari}}]{Biderang2018}%
  \BibitemOpen
  \bibfield  {author} {\bibinfo {author} {\bibfnamefont {M.}~\bibnamefont
  {Biderang}}, \bibinfo {author} {\bibfnamefont {A.}~\bibnamefont {Leonhardt}},
  \bibinfo {author} {\bibfnamefont {N.}~\bibnamefont {Raghuvanshi}}, \bibinfo
  {author} {\bibfnamefont {A.~P.}\ \bibnamefont {Schnyder}},\ and\ \bibinfo
  {author} {\bibfnamefont {A.}~\bibnamefont {Akbari}},\ }\bibfield  {title}
  {\bibinfo {title} {Drumhead surface states and their signatures in
  quasiparticle scattering interference},\ }\href
  {https://doi.org/10.1103/PhysRevB.98.075115} {\bibfield  {journal} {\bibinfo
  {journal} {Phys. Rev. B}\ }\textbf {\bibinfo {volume} {98}},\ \bibinfo
  {pages} {075115} (\bibinfo {year} {2018})}\BibitemShut {NoStop}%
\bibitem [{\citenamefont {Yang}\ \emph {et~al.}(2022)\citenamefont {Yang},
  \citenamefont {Luo},\ and\ \citenamefont {Chen}}]{Yang2022}%
  \BibitemOpen
  \bibfield  {author} {\bibinfo {author} {\bibfnamefont {M.-X.}\ \bibnamefont
  {Yang}}, \bibinfo {author} {\bibfnamefont {W.}~\bibnamefont {Luo}},\ and\
  \bibinfo {author} {\bibfnamefont {W.}~\bibnamefont {Chen}},\ }\bibfield
  {title} {\bibinfo {title} {Quantum transport in topological nodal-line
  semimetals},\ }\href {https://doi.org/10.1080/23746149.2022.2065216}
  {\bibfield  {journal} {\bibinfo  {journal} {Advances in Physics: X}\ }\textbf
  {\bibinfo {volume} {7}},\ \bibinfo {pages} {2065216} (\bibinfo {year}
  {2022})}\BibitemShut {NoStop}%
\bibitem [{\citenamefont {Mukherjee}\ and\ \citenamefont
  {Carbotte}(2017)}]{Mukherjee2017}%
  \BibitemOpen
  \bibfield  {author} {\bibinfo {author} {\bibfnamefont {S.~P.}\ \bibnamefont
  {Mukherjee}}\ and\ \bibinfo {author} {\bibfnamefont {J.~P.}\ \bibnamefont
  {Carbotte}},\ }\bibfield  {title} {\bibinfo {title} {Transport and optics at
  the node in a nodal loop semimetal},\ }\href
  {https://doi.org/10.1103/PhysRevB.95.214203} {\bibfield  {journal} {\bibinfo
  {journal} {Phys. Rev. B}\ }\textbf {\bibinfo {volume} {95}},\ \bibinfo
  {pages} {214203} (\bibinfo {year} {2017})}\BibitemShut {NoStop}%
\bibitem [{\citenamefont {Pronin}\ and\ \citenamefont
  {Dressel}(2021)}]{Pronin2021}%
  \BibitemOpen
  \bibfield  {author} {\bibinfo {author} {\bibfnamefont {A.~V.}\ \bibnamefont
  {Pronin}}\ and\ \bibinfo {author} {\bibfnamefont {M.}~\bibnamefont
  {Dressel}},\ }\bibfield  {title} {\bibinfo {title} {Nodal semimetals: A
  survey on optical conductivity},\ }\href
  {https://doi.org/10.1002/pssb.202000027} {\bibfield  {journal} {\bibinfo
  {journal} {Phys. Status Solidi B}\ }\textbf {\bibinfo {volume} {258}},\
  \bibinfo {pages} {2000027} (\bibinfo {year} {2021})}\BibitemShut {NoStop}%
\bibitem [{\citenamefont {Anand}\ and\ \citenamefont
  {De~Martino}(2026)}]{Anand2026}%
  \BibitemOpen
  \bibfield  {author} {\bibinfo {author} {\bibfnamefont {A.}~\bibnamefont
  {Anand}}\ and\ \bibinfo {author} {\bibfnamefont {A.}~\bibnamefont
  {De~Martino}},\ }\bibfield  {title} {\bibinfo {title} {Toroidal {Fermi}
  surface geometry and phonon-limited transport in nodal-line semimetals},\
  }\href {https://doi.org/10.1103/3tl2-n2qd} {\bibfield  {journal} {\bibinfo
  {journal} {Phys. Rev. B}\ }\textbf {\bibinfo {volume} {113}},\ \bibinfo
  {pages} {235432} (\bibinfo {year} {2026})}\BibitemShut {NoStop}%
\bibitem [{\citenamefont {Rhim}\ and\ \citenamefont {Kim}(2016)}]{Rhim2016}%
  \BibitemOpen
  \bibfield  {author} {\bibinfo {author} {\bibfnamefont {J.-W.}\ \bibnamefont
  {Rhim}}\ and\ \bibinfo {author} {\bibfnamefont {Y.~B.}\ \bibnamefont {Kim}},\
  }\bibfield  {title} {\bibinfo {title} {Anisotropic density fluctuations,
  plasmons, and {Friedel} oscillations in nodal line semimetal},\ }\href
  {https://doi.org/10.1088/1367-2630/18/4/043010} {\bibfield  {journal}
  {\bibinfo  {journal} {New J. Phys.}\ }\textbf {\bibinfo {volume} {18}},\
  \bibinfo {pages} {043010} (\bibinfo {year} {2016})}\BibitemShut {NoStop}%
\bibitem [{\citenamefont {Singh}\ \emph {et~al.}(2023)\citenamefont {Singh},
  \citenamefont {Sebastian}, \citenamefont {Chen}, \citenamefont {Chang},\ and\
  \citenamefont {Belyanin}}]{Singh2023}%
  \BibitemOpen
  \bibfield  {author} {\bibinfo {author} {\bibfnamefont {A.}~\bibnamefont
  {Singh}}, \bibinfo {author} {\bibfnamefont {M.}~\bibnamefont {Sebastian}},
  \bibinfo {author} {\bibfnamefont {Y.}~\bibnamefont {Chen}}, \bibinfo {author}
  {\bibfnamefont {P.-Y.}\ \bibnamefont {Chang}},\ and\ \bibinfo {author}
  {\bibfnamefont {A.}~\bibnamefont {Belyanin}},\ }\bibfield  {title} {\bibinfo
  {title} {Hyperbolic polaritons in topological nodal ring semimetals},\ }\href
  {https://doi.org/10.1103/PhysRevLett.131.096902} {\bibfield  {journal}
  {\bibinfo  {journal} {Phys. Rev. Lett.}\ }\textbf {\bibinfo {volume} {131}},\
  \bibinfo {pages} {096902} (\bibinfo {year} {2023})},\ \Eprint
  {https://arxiv.org/abs/2303.12893} {arXiv:2303.12893} \BibitemShut {NoStop}%
\bibitem [{\citenamefont {Shao}\ \emph {et~al.}(2022)\citenamefont {Shao},
  \citenamefont {Sternbach}, \citenamefont {Kim}, \citenamefont {Rikhter},
  \citenamefont {Xu}, \citenamefont {De~Giovannini}, \citenamefont {Jing},
  \citenamefont {Chae}, \citenamefont {Sun}, \citenamefont {Lee}, \citenamefont
  {Zhu}, \citenamefont {Mao}, \citenamefont {Hone}, \citenamefont {Queiroz},
  \citenamefont {Millis}, \citenamefont {Schuck}, \citenamefont {Rubio},
  \citenamefont {Fogler},\ and\ \citenamefont {Basov}}]{Shao2022}%
  \BibitemOpen
  \bibfield  {author} {\bibinfo {author} {\bibfnamefont {Y.}~\bibnamefont
  {Shao}}, \bibinfo {author} {\bibfnamefont {A.~J.}\ \bibnamefont {Sternbach}},
  \bibinfo {author} {\bibfnamefont {B.~S.~Y.}\ \bibnamefont {Kim}}, \bibinfo
  {author} {\bibfnamefont {A.~A.}\ \bibnamefont {Rikhter}}, \bibinfo {author}
  {\bibfnamefont {X.}~\bibnamefont {Xu}}, \bibinfo {author} {\bibfnamefont
  {U.}~\bibnamefont {De~Giovannini}}, \bibinfo {author} {\bibfnamefont
  {R.}~\bibnamefont {Jing}}, \bibinfo {author} {\bibfnamefont {S.~H.}\
  \bibnamefont {Chae}}, \bibinfo {author} {\bibfnamefont {Z.}~\bibnamefont
  {Sun}}, \bibinfo {author} {\bibfnamefont {S.~H.}\ \bibnamefont {Lee}},
  \bibinfo {author} {\bibfnamefont {Y.}~\bibnamefont {Zhu}}, \bibinfo {author}
  {\bibfnamefont {Z.}~\bibnamefont {Mao}}, \bibinfo {author} {\bibfnamefont
  {J.~C.}\ \bibnamefont {Hone}}, \bibinfo {author} {\bibfnamefont
  {R.}~\bibnamefont {Queiroz}}, \bibinfo {author} {\bibfnamefont {A.~J.}\
  \bibnamefont {Millis}}, \bibinfo {author} {\bibfnamefont {P.~J.}\
  \bibnamefont {Schuck}}, \bibinfo {author} {\bibfnamefont {A.}~\bibnamefont
  {Rubio}}, \bibinfo {author} {\bibfnamefont {M.~M.}\ \bibnamefont {Fogler}},\
  and\ \bibinfo {author} {\bibfnamefont {D.~N.}\ \bibnamefont {Basov}},\
  }\bibfield  {title} {\bibinfo {title} {Infrared plasmons propagate through a
  hyperbolic nodal metal},\ }\href {https://doi.org/10.1126/sciadv.add6169}
  {\bibfield  {journal} {\bibinfo  {journal} {Sci. Adv.}\ }\textbf {\bibinfo
  {volume} {8}},\ \bibinfo {pages} {eadd6169} (\bibinfo {year} {2022})},\
  \Eprint {https://arxiv.org/abs/2206.01828} {arXiv:2206.01828} \BibitemShut
  {NoStop}%
\bibitem [{\citenamefont {Santamaria-Garcia}\ \emph {et~al.}(2026)\citenamefont
  {Santamaria-Garcia}, \citenamefont {Blevins}, \citenamefont {Pajovic},
  \citenamefont {Orona-Navar},\ and\ \citenamefont
  {Boriskina}}]{Santamariagarcia2026}%
  \BibitemOpen
  \bibfield  {author} {\bibinfo {author} {\bibfnamefont {V.~J.}\ \bibnamefont
  {Santamaria-Garcia}}, \bibinfo {author} {\bibfnamefont {M.~G.}\ \bibnamefont
  {Blevins}}, \bibinfo {author} {\bibfnamefont {S.}~\bibnamefont {Pajovic}},
  \bibinfo {author} {\bibfnamefont {C.}~\bibnamefont {Orona-Navar}},\ and\
  \bibinfo {author} {\bibfnamefont {S.~V.}\ \bibnamefont {Boriskina}},\ }\href
  {https://arxiv.org/abs/2605.12611} {\bibinfo {title} {{Visible-NIR}-frequency
  hyperbolic response in nodal-line semimetal {PbTaSe$_2$}}} (\bibinfo {year}
  {2026}),\ \Eprint {https://arxiv.org/abs/2605.12611} {arXiv:2605.12611
  [physics.optics]} \BibitemShut {NoStop}%
\bibitem [{\citenamefont {Chang}(2025)}]{Chang2025}%
  \BibitemOpen
  \bibfield  {author} {\bibinfo {author} {\bibfnamefont {P.-Y.}\ \bibnamefont
  {Chang}},\ }\bibfield  {title} {\bibinfo {title} {Nodal-line semimetals and
  their variance},\ }\href
  {https://doi.org/https://doi.org/10.1016/j.mtquan.2025.100057} {\bibfield
  {journal} {\bibinfo  {journal} {Materials Today Quantum}\ }\textbf {\bibinfo
  {volume} {8}},\ \bibinfo {pages} {100057} (\bibinfo {year}
  {2025})}\BibitemShut {NoStop}%
\bibitem [{\citenamefont {Nagao}\ \emph {et~al.}(2001)\citenamefont {Nagao},
  \citenamefont {Hildebrandt}, \citenamefont {Henzler},\ and\ \citenamefont
  {Hasegawa}}]{Nagao2001}%
  \BibitemOpen
  \bibfield  {author} {\bibinfo {author} {\bibfnamefont {T.}~\bibnamefont
  {Nagao}}, \bibinfo {author} {\bibfnamefont {T.}~\bibnamefont {Hildebrandt}},
  \bibinfo {author} {\bibfnamefont {M.}~\bibnamefont {Henzler}},\ and\ \bibinfo
  {author} {\bibfnamefont {S.}~\bibnamefont {Hasegawa}},\ }\bibfield  {title}
  {\bibinfo {title} {Dispersion and damping of a two-dimensional plasmon in a
  metallic surface-state band},\ }\href
  {https://doi.org/10.1103/PhysRevLett.86.5747} {\bibfield  {journal} {\bibinfo
   {journal} {Phys. Rev. Lett.}\ }\textbf {\bibinfo {volume} {86}},\ \bibinfo
  {pages} {5747} (\bibinfo {year} {2001})}\BibitemShut {NoStop}%
\bibitem [{\citenamefont {Diaconescu}\ \emph {et~al.}(2007)\citenamefont
  {Diaconescu}, \citenamefont {Pohl}, \citenamefont {Vattuone}, \citenamefont
  {Savio}, \citenamefont {Hofmann}, \citenamefont {Silkin}, \citenamefont
  {Pitarke}, \citenamefont {Chulkov}, \citenamefont {Echenique}, \citenamefont
  {Far{\'\i}as},\ and\ \citenamefont {Rocca}}]{Diaconescu2007}%
  \BibitemOpen
  \bibfield  {author} {\bibinfo {author} {\bibfnamefont {B.}~\bibnamefont
  {Diaconescu}}, \bibinfo {author} {\bibfnamefont {K.}~\bibnamefont {Pohl}},
  \bibinfo {author} {\bibfnamefont {L.}~\bibnamefont {Vattuone}}, \bibinfo
  {author} {\bibfnamefont {L.}~\bibnamefont {Savio}}, \bibinfo {author}
  {\bibfnamefont {P.}~\bibnamefont {Hofmann}}, \bibinfo {author} {\bibfnamefont
  {V.~M.}\ \bibnamefont {Silkin}}, \bibinfo {author} {\bibfnamefont {J.~M.}\
  \bibnamefont {Pitarke}}, \bibinfo {author} {\bibfnamefont {E.~V.}\
  \bibnamefont {Chulkov}}, \bibinfo {author} {\bibfnamefont {P.~M.}\
  \bibnamefont {Echenique}}, \bibinfo {author} {\bibfnamefont {D.}~\bibnamefont
  {Far{\'\i}as}},\ and\ \bibinfo {author} {\bibfnamefont {M.}~\bibnamefont
  {Rocca}},\ }\bibfield  {title} {\bibinfo {title} {Low-energy acoustic
  plasmons at metal surfaces},\ }\href {https://doi.org/10.1038/nature05975}
  {\bibfield  {journal} {\bibinfo  {journal} {Nature}\ }\textbf {\bibinfo
  {volume} {448}},\ \bibinfo {pages} {57} (\bibinfo {year} {2007})}\BibitemShut
  {NoStop}%
\bibitem [{\citenamefont {Silkin}\ \emph {et~al.}(2008)\citenamefont {Silkin},
  \citenamefont {Pitarke}, \citenamefont {Chulkov}, \citenamefont {Diaconescu},
  \citenamefont {Pohl}, \citenamefont {Vattuone}, \citenamefont {Savio},
  \citenamefont {Hofmann}, \citenamefont {Far\'{\i}as}, \citenamefont {Rocca},\
  and\ \citenamefont {Echenique}}]{Silkin2008}%
  \BibitemOpen
  \bibfield  {author} {\bibinfo {author} {\bibfnamefont {V.~M.}\ \bibnamefont
  {Silkin}}, \bibinfo {author} {\bibfnamefont {J.~M.}\ \bibnamefont {Pitarke}},
  \bibinfo {author} {\bibfnamefont {E.~V.}\ \bibnamefont {Chulkov}}, \bibinfo
  {author} {\bibfnamefont {B.}~\bibnamefont {Diaconescu}}, \bibinfo {author}
  {\bibfnamefont {K.}~\bibnamefont {Pohl}}, \bibinfo {author} {\bibfnamefont
  {L.}~\bibnamefont {Vattuone}}, \bibinfo {author} {\bibfnamefont
  {L.}~\bibnamefont {Savio}}, \bibinfo {author} {\bibfnamefont
  {P.}~\bibnamefont {Hofmann}}, \bibinfo {author} {\bibfnamefont
  {D.}~\bibnamefont {Far\'{\i}as}}, \bibinfo {author} {\bibfnamefont
  {M.}~\bibnamefont {Rocca}},\ and\ \bibinfo {author} {\bibfnamefont {P.~M.}\
  \bibnamefont {Echenique}},\ }\bibfield  {title} {\bibinfo {title} {Band
  structure effects on the {Be(0001)} acoustic surface plasmon energy
  dispersion},\ }\href {https://doi.org/10.1002/pssa.200778164} {\bibfield
  {journal} {\bibinfo  {journal} {Phys. Status Solidi A}\ }\textbf {\bibinfo
  {volume} {205}},\ \bibinfo {pages} {1307} (\bibinfo {year}
  {2008})}\BibitemShut {NoStop}%
\bibitem [{\citenamefont {Stauber}\ \emph {et~al.}(2017)\citenamefont
  {Stauber}, \citenamefont {Gomez-Santos},\ and\ \citenamefont
  {Brey}}]{Stauber2017}%
  \BibitemOpen
  \bibfield  {author} {\bibinfo {author} {\bibfnamefont {T.}~\bibnamefont
  {Stauber}}, \bibinfo {author} {\bibfnamefont {G.}~\bibnamefont
  {Gomez-Santos}},\ and\ \bibinfo {author} {\bibfnamefont {L.}~\bibnamefont
  {Brey}},\ }\bibfield  {title} {\bibinfo {title} {Plasmonics in topological
  insulators: Spin–charge separation, the influence of the inversion layer,
  and phonon–plasmon coupling},\ }\href
  {https://doi.org/10.1021/acsphotonics.7b00524} {\bibfield  {journal}
  {\bibinfo  {journal} {ACS Photonics}\ }\textbf {\bibinfo {volume} {4}},\
  \bibinfo {pages} {2978} (\bibinfo {year} {2017})}\BibitemShut {NoStop}%
\bibitem [{\citenamefont {Di~Pietro}\ \emph {et~al.}(2020)\citenamefont
  {Di~Pietro}, \citenamefont {Adhlakha}, \citenamefont {Piccirilli},
  \citenamefont {Di~Gaspare}, \citenamefont {Moon}, \citenamefont {Oh},
  \citenamefont {Di~Mitri}, \citenamefont {Spampinati}, \citenamefont
  {Perucchi},\ and\ \citenamefont {Lupi}}]{DiPietro2020}%
  \BibitemOpen
  \bibfield  {author} {\bibinfo {author} {\bibfnamefont {P.}~\bibnamefont
  {Di~Pietro}}, \bibinfo {author} {\bibfnamefont {N.}~\bibnamefont {Adhlakha}},
  \bibinfo {author} {\bibfnamefont {F.}~\bibnamefont {Piccirilli}}, \bibinfo
  {author} {\bibfnamefont {A.}~\bibnamefont {Di~Gaspare}}, \bibinfo {author}
  {\bibfnamefont {J.}~\bibnamefont {Moon}}, \bibinfo {author} {\bibfnamefont
  {S.}~\bibnamefont {Oh}}, \bibinfo {author} {\bibfnamefont {S.}~\bibnamefont
  {Di~Mitri}}, \bibinfo {author} {\bibfnamefont {S.}~\bibnamefont
  {Spampinati}}, \bibinfo {author} {\bibfnamefont {A.}~\bibnamefont
  {Perucchi}},\ and\ \bibinfo {author} {\bibfnamefont {S.}~\bibnamefont
  {Lupi}},\ }\bibfield  {title} {\bibinfo {title} {Terahertz tuning of {Dirac}
  plasmons in {Bi$_2$Se$_3$} topological insulator},\ }\href
  {https://doi.org/10.1103/PhysRevLett.124.226403} {\bibfield  {journal}
  {\bibinfo  {journal} {Phys. Rev. Lett.}\ }\textbf {\bibinfo {volume} {124}},\
  \bibinfo {pages} {226403} (\bibinfo {year} {2020})}\BibitemShut {NoStop}%
\bibitem [{\citenamefont {Song}\ and\ \citenamefont {Rudner}(2017)}]{Song2017}%
  \BibitemOpen
  \bibfield  {author} {\bibinfo {author} {\bibfnamefont {J.~C.~W.}\
  \bibnamefont {Song}}\ and\ \bibinfo {author} {\bibfnamefont {M.~S.}\
  \bibnamefont {Rudner}},\ }\bibfield  {title} {\bibinfo {title} {{Fermi} arc
  plasmons in {Weyl} semimetals},\ }\href
  {https://doi.org/10.1103/PhysRevB.96.205443} {\bibfield  {journal} {\bibinfo
  {journal} {Phys. Rev. B}\ }\textbf {\bibinfo {volume} {96}},\ \bibinfo
  {pages} {205443} (\bibinfo {year} {2017})}\BibitemShut {NoStop}%
\bibitem [{\citenamefont {Andolina}\ \emph {et~al.}(2018)\citenamefont
  {Andolina}, \citenamefont {Pellegrino}, \citenamefont {Koppens},\ and\
  \citenamefont {Polini}}]{Andolina2018}%
  \BibitemOpen
  \bibfield  {author} {\bibinfo {author} {\bibfnamefont {G.~M.}\ \bibnamefont
  {Andolina}}, \bibinfo {author} {\bibfnamefont {F.~M.~D.}\ \bibnamefont
  {Pellegrino}}, \bibinfo {author} {\bibfnamefont {F.~H.~L.}\ \bibnamefont
  {Koppens}},\ and\ \bibinfo {author} {\bibfnamefont {M.}~\bibnamefont
  {Polini}},\ }\bibfield  {title} {\bibinfo {title} {Quantum nonlocal theory of
  topological {Fermi} arc plasmons in {Weyl} semimetals},\ }\href
  {https://doi.org/10.1103/PhysRevB.97.125431} {\bibfield  {journal} {\bibinfo
  {journal} {Phys. Rev. B}\ }\textbf {\bibinfo {volume} {97}},\ \bibinfo
  {pages} {125431} (\bibinfo {year} {2018})}\BibitemShut {NoStop}%
\bibitem [{\citenamefont {Peluso}\ \emph {et~al.}(2025)\citenamefont {Peluso},
  \citenamefont {De~Martino}, \citenamefont {Egger},\ and\ \citenamefont
  {Buccheri}}]{Weylcone}%
  \BibitemOpen
  \bibfield  {author} {\bibinfo {author} {\bibfnamefont {M.}~\bibnamefont
  {Peluso}}, \bibinfo {author} {\bibfnamefont {A.}~\bibnamefont {De~Martino}},
  \bibinfo {author} {\bibfnamefont {R.}~\bibnamefont {Egger}},\ and\ \bibinfo
  {author} {\bibfnamefont {F.}~\bibnamefont {Buccheri}},\ }\bibfield  {title}
  {\bibinfo {title} {Angular-momentum-selective nanofocusing with {Weyl}
  semimetals},\ }\href {https://doi.org/10.1103/8c99-rvmm} {\bibfield
  {journal} {\bibinfo  {journal} {Phys. Rev. Res.}\ }\textbf {\bibinfo {volume}
  {7}},\ \bibinfo {pages} {043121} (\bibinfo {year} {2025})}\BibitemShut
  {NoStop}%
\bibitem [{\citenamefont {Pellegrino}\ \emph {et~al.}(2025)\citenamefont
  {Pellegrino}, \citenamefont {Buccheri},\ and\ \citenamefont
  {Angilella}}]{Pellegrino2025}%
  \BibitemOpen
  \bibfield  {author} {\bibinfo {author} {\bibfnamefont {F.~M.~D.}\
  \bibnamefont {Pellegrino}}, \bibinfo {author} {\bibfnamefont
  {F.}~\bibnamefont {Buccheri}},\ and\ \bibinfo {author} {\bibfnamefont
  {G.~G.~N.}\ \bibnamefont {Angilella}},\ }\bibfield  {title} {\bibinfo {title}
  {Localized surface plasmons in a {Weyl} semimetal nanosphere},\ }\href
  {https://doi.org/10.1103/7b1l-gkjy} {\bibfield  {journal} {\bibinfo
  {journal} {Phys. Rev. B}\ }\textbf {\bibinfo {volume} {112}},\ \bibinfo
  {pages} {075431} (\bibinfo {year} {2025})}\BibitemShut {NoStop}%
\bibitem [{\citenamefont
  {{Bona{\v{c}}i{\'c}-Lo{\v{s}}i{\'c}}}(2022)}]{Losic2022}%
  \BibitemOpen
  \bibfield  {author} {\bibinfo {author} {\bibfnamefont {{\v{Z}}.}~\bibnamefont
  {{Bona{\v{c}}i{\'c}-Lo{\v{s}}i{\'c}}}},\ }\bibfield  {title} {\bibinfo
  {title} {Plasmon in dispersive nodal line semimetal},\ }\href
  {https://doi.org/https://doi.org/10.1139/cjp-2021-0371} {\bibfield  {journal}
  {\bibinfo  {journal} {Canadian Journal of Physics}\ }\textbf {\bibinfo
  {volume} {100}},\ \bibinfo {pages} {303} (\bibinfo {year}
  {2022})}\BibitemShut {NoStop}%
\bibitem [{\citenamefont {Muechler}\ \emph {et~al.}(2020)\citenamefont
  {Muechler}, \citenamefont {Topp}, \citenamefont {Queiroz}, \citenamefont
  {Krivenkov}, \citenamefont {Varykhalov}, \citenamefont {Cano}, \citenamefont
  {Ast},\ and\ \citenamefont {Schoop}}]{Muechler2020}%
  \BibitemOpen
  \bibfield  {author} {\bibinfo {author} {\bibfnamefont {L.}~\bibnamefont
  {Muechler}}, \bibinfo {author} {\bibfnamefont {A.}~\bibnamefont {Topp}},
  \bibinfo {author} {\bibfnamefont {R.}~\bibnamefont {Queiroz}}, \bibinfo
  {author} {\bibfnamefont {M.}~\bibnamefont {Krivenkov}}, \bibinfo {author}
  {\bibfnamefont {A.}~\bibnamefont {Varykhalov}}, \bibinfo {author}
  {\bibfnamefont {J.}~\bibnamefont {Cano}}, \bibinfo {author} {\bibfnamefont
  {C.~R.}\ \bibnamefont {Ast}},\ and\ \bibinfo {author} {\bibfnamefont {L.~M.}\
  \bibnamefont {Schoop}},\ }\bibfield  {title} {\bibinfo {title} {Modular
  arithmetic with nodal lines: Drumhead surface states in {ZrSiTe}},\ }\href
  {https://doi.org/10.1103/PhysRevX.10.011026} {\bibfield  {journal} {\bibinfo
  {journal} {Phys. Rev. X}\ }\textbf {\bibinfo {volume} {10}},\ \bibinfo
  {pages} {011026} (\bibinfo {year} {2020})}\BibitemShut {NoStop}%
\bibitem [{\citenamefont {Buccheri}\ \emph {et~al.}(2024)\citenamefont
  {Buccheri}, \citenamefont {Egger},\ and\ \citenamefont
  {De~Martino}}]{Buccheri2024}%
  \BibitemOpen
  \bibfield  {author} {\bibinfo {author} {\bibfnamefont {F.}~\bibnamefont
  {Buccheri}}, \bibinfo {author} {\bibfnamefont {R.}~\bibnamefont {Egger}},\
  and\ \bibinfo {author} {\bibfnamefont {A.}~\bibnamefont {De~Martino}},\
  }\bibfield  {title} {\bibinfo {title} {Dispersive drumhead states in
  nodal-line semimetal junctions},\ }\href
  {https://doi.org/10.1103/PhysRevResearch.6.013193} {\bibfield  {journal}
  {\bibinfo  {journal} {Phys. Rev. Res.}\ }\textbf {\bibinfo {volume} {6}},\
  \bibinfo {pages} {013193} (\bibinfo {year} {2024})}\BibitemShut {NoStop}%
\bibitem [{\citenamefont {Song}\ \emph {et~al.}(2018)\citenamefont {Song},
  \citenamefont {Zhang},\ and\ \citenamefont {Fang}}]{Song2018}%
  \BibitemOpen
  \bibfield  {author} {\bibinfo {author} {\bibfnamefont {Z.}~\bibnamefont
  {Song}}, \bibinfo {author} {\bibfnamefont {T.}~\bibnamefont {Zhang}},\ and\
  \bibinfo {author} {\bibfnamefont {C.}~\bibnamefont {Fang}},\ }\bibfield
  {title} {\bibinfo {title} {Diagnosis for nonmagnetic topological semimetals
  in the absence of spin-orbital coupling},\ }\href
  {https://doi.org/10.1103/PhysRevX.8.031069} {\bibfield  {journal} {\bibinfo
  {journal} {Phys. Rev. X}\ }\textbf {\bibinfo {volume} {8}},\ \bibinfo {pages}
  {031069} (\bibinfo {year} {2018})}\BibitemShut {NoStop}%
\bibitem [{\citenamefont {Zhang}\ \emph {et~al.}(2019)\citenamefont {Zhang},
  \citenamefont {Jiang}, \citenamefont {Song}, \citenamefont {Huang},
  \citenamefont {He}, \citenamefont {Fang}, \citenamefont {Weng},\ and\
  \citenamefont {Fang}}]{Zhang2019c}%
  \BibitemOpen
  \bibfield  {author} {\bibinfo {author} {\bibfnamefont {T.}~\bibnamefont
  {Zhang}}, \bibinfo {author} {\bibfnamefont {Y.}~\bibnamefont {Jiang}},
  \bibinfo {author} {\bibfnamefont {Z.}~\bibnamefont {Song}}, \bibinfo {author}
  {\bibfnamefont {H.}~\bibnamefont {Huang}}, \bibinfo {author} {\bibfnamefont
  {Y.}~\bibnamefont {He}}, \bibinfo {author} {\bibfnamefont {Z.}~\bibnamefont
  {Fang}}, \bibinfo {author} {\bibfnamefont {H.}~\bibnamefont {Weng}},\ and\
  \bibinfo {author} {\bibfnamefont {C.}~\bibnamefont {Fang}},\ }\bibfield
  {title} {\bibinfo {title} {Catalogue of topological electronic materials},\
  }\href {https://doi.org/10.1038/s41586-019-0944-6} {\bibfield  {journal}
  {\bibinfo  {journal} {Nature}\ }\textbf {\bibinfo {volume} {566}},\ \bibinfo
  {pages} {475} (\bibinfo {year} {2019})}\BibitemShut {NoStop}%
\bibitem [{\citenamefont {Vergniory}\ \emph {et~al.}(2022)\citenamefont
  {Vergniory}, \citenamefont {Wieder}, \citenamefont {Elcoro}, \citenamefont
  {Parkin}, \citenamefont {Felser}, \citenamefont {Bernevig},\ and\
  \citenamefont {Regnault}}]{Vergniory2022}%
  \BibitemOpen
  \bibfield  {author} {\bibinfo {author} {\bibfnamefont {M.~G.}\ \bibnamefont
  {Vergniory}}, \bibinfo {author} {\bibfnamefont {B.~J.}\ \bibnamefont
  {Wieder}}, \bibinfo {author} {\bibfnamefont {L.}~\bibnamefont {Elcoro}},
  \bibinfo {author} {\bibfnamefont {S.~S.~P.}\ \bibnamefont {Parkin}}, \bibinfo
  {author} {\bibfnamefont {C.}~\bibnamefont {Felser}}, \bibinfo {author}
  {\bibfnamefont {B.~A.}\ \bibnamefont {Bernevig}},\ and\ \bibinfo {author}
  {\bibfnamefont {N.}~\bibnamefont {Regnault}},\ }\bibfield  {title} {\bibinfo
  {title} {All topological bands of all nonmagnetic stoichiometric materials},\
  }\href {https://doi.org/10.1126/science.abg9094} {\bibfield  {journal}
  {\bibinfo  {journal} {Science}\ }\textbf {\bibinfo {volume} {376}},\ \bibinfo
  {pages} {eabg9094} (\bibinfo {year} {2022})}\BibitemShut {NoStop}%
\bibitem [{\citenamefont {Kim}\ \emph {et~al.}(2015)\citenamefont {Kim},
  \citenamefont {Wieder}, \citenamefont {Kane},\ and\ \citenamefont
  {Rappe}}]{Kim2015}%
  \BibitemOpen
  \bibfield  {author} {\bibinfo {author} {\bibfnamefont {Y.}~\bibnamefont
  {Kim}}, \bibinfo {author} {\bibfnamefont {B.~J.}\ \bibnamefont {Wieder}},
  \bibinfo {author} {\bibfnamefont {C.~L.}\ \bibnamefont {Kane}},\ and\
  \bibinfo {author} {\bibfnamefont {A.~M.}\ \bibnamefont {Rappe}},\ }\bibfield
  {title} {\bibinfo {title} {{Dirac} line nodes in inversion-symmetric
  crystals},\ }\href {https://doi.org/10.1103/PhysRevLett.115.036806}
  {\bibfield  {journal} {\bibinfo  {journal} {Phys. Rev. Lett.}\ }\textbf
  {\bibinfo {volume} {115}},\ \bibinfo {pages} {036806} (\bibinfo {year}
  {2015})}\BibitemShut {NoStop}%
\bibitem [{\citenamefont {Tiwari}\ and\ \citenamefont {Bzdu{\ifmmode
  \check{s}\else \v{s}\fi{}}ek}(2020)}]{Tiwari2020}%
  \BibitemOpen
  \bibfield  {author} {\bibinfo {author} {\bibfnamefont {A.}~\bibnamefont
  {Tiwari}}\ and\ \bibinfo {author} {\bibfnamefont {T.}~\bibnamefont
  {Bzdu{\ifmmode \check{s}\else \v{s}\fi{}}ek}},\ }\bibfield  {title} {\bibinfo
  {title} {Non-{Abelian} topology of nodal-line rings in
  $\mathcal{PT}$-symmetric systems},\ }\href
  {https://doi.org/10.1103/PhysRevB.101.195130} {\bibfield  {journal} {\bibinfo
   {journal} {Phys. Rev. B}\ }\textbf {\bibinfo {volume} {101}},\ \bibinfo
  {pages} {195130} (\bibinfo {year} {2020})}\BibitemShut {NoStop}%
\bibitem [{\citenamefont {Huang}\ \emph {et~al.}(2016)\citenamefont {Huang},
  \citenamefont {Liu}, \citenamefont {Vanderbilt},\ and\ \citenamefont
  {Duan}}]{Huang2016}%
  \BibitemOpen
  \bibfield  {author} {\bibinfo {author} {\bibfnamefont {H.}~\bibnamefont
  {Huang}}, \bibinfo {author} {\bibfnamefont {J.}~\bibnamefont {Liu}}, \bibinfo
  {author} {\bibfnamefont {D.}~\bibnamefont {Vanderbilt}},\ and\ \bibinfo
  {author} {\bibfnamefont {W.}~\bibnamefont {Duan}},\ }\bibfield  {title}
  {\bibinfo {title} {Topological nodal-line semimetals in alkaline-earth
  stannides, germanides, and silicides},\ }\href
  {https://doi.org/10.1103/PhysRevB.93.201114} {\bibfield  {journal} {\bibinfo
  {journal} {Phys. Rev. B}\ }\textbf {\bibinfo {volume} {93}},\ \bibinfo
  {pages} {201114} (\bibinfo {year} {2016})}\BibitemShut {NoStop}%
\bibitem [{\citenamefont {Weng}\ \emph {et~al.}(2016)\citenamefont {Weng},
  \citenamefont {Dai},\ and\ \citenamefont {Fang}}]{Weng2016}%
  \BibitemOpen
  \bibfield  {author} {\bibinfo {author} {\bibfnamefont {H.}~\bibnamefont
  {Weng}}, \bibinfo {author} {\bibfnamefont {X.}~\bibnamefont {Dai}},\ and\
  \bibinfo {author} {\bibfnamefont {Z.}~\bibnamefont {Fang}},\ }\bibfield
  {title} {\bibinfo {title} {Topological semimetals predicted from
  first-principles calculations},\ }\href
  {https://doi.org/10.1088/0953-8984/28/30/303001} {\bibfield  {journal}
  {\bibinfo  {journal} {Journal of Physics: Condensed Matter}\ }\textbf
  {\bibinfo {volume} {28}},\ \bibinfo {pages} {303001} (\bibinfo {year}
  {2016})}\BibitemShut {NoStop}%
\bibitem [{\citenamefont {Yang}\ \emph {et~al.}(2018)\citenamefont {Yang},
  \citenamefont {Yang}, \citenamefont {Derunova}, \citenamefont {Parkin},
  \citenamefont {Yan},\ and\ \citenamefont {Ali}}]{Yang2018s}%
  \BibitemOpen
  \bibfield  {author} {\bibinfo {author} {\bibfnamefont {S.-Y.}\ \bibnamefont
  {Yang}}, \bibinfo {author} {\bibfnamefont {H.}~\bibnamefont {Yang}}, \bibinfo
  {author} {\bibfnamefont {E.}~\bibnamefont {Derunova}}, \bibinfo {author}
  {\bibfnamefont {S.~S.~P.}\ \bibnamefont {Parkin}}, \bibinfo {author}
  {\bibfnamefont {B.}~\bibnamefont {Yan}},\ and\ \bibinfo {author}
  {\bibfnamefont {M.~N.}\ \bibnamefont {Ali}},\ }\bibfield  {title} {\bibinfo
  {title} {Symmetry demanded topological nodal-line materials},\ }\href
  {https://doi.org/10.1080/23746149.2017.1414631} {\bibfield  {journal}
  {\bibinfo  {journal} {Advances in Physics: X}\ }\textbf {\bibinfo {volume}
  {3}},\ \bibinfo {pages} {1414631} (\bibinfo {year} {2018})}\BibitemShut
  {NoStop}%
\bibitem [{\citenamefont {Chan}\ \emph {et~al.}(2016)\citenamefont {Chan},
  \citenamefont {Chiu}, \citenamefont {Chou},\ and\ \citenamefont
  {Schnyder}}]{Chan2016a}%
  \BibitemOpen
  \bibfield  {author} {\bibinfo {author} {\bibfnamefont {Y.-H.}\ \bibnamefont
  {Chan}}, \bibinfo {author} {\bibfnamefont {C.-K.}\ \bibnamefont {Chiu}},
  \bibinfo {author} {\bibfnamefont {M.~Y.}\ \bibnamefont {Chou}},\ and\
  \bibinfo {author} {\bibfnamefont {A.~P.}\ \bibnamefont {Schnyder}},\
  }\bibfield  {title} {\bibinfo {title} {{Ca$_3$P$_2$} and other topological
  semimetals with line nodes and drumhead surface states},\ }\href
  {https://doi.org/10.1103/PhysRevB.93.205132} {\bibfield  {journal} {\bibinfo
  {journal} {Phys. Rev. B}\ }\textbf {\bibinfo {volume} {93}},\ \bibinfo
  {pages} {205132} (\bibinfo {year} {2016})}\BibitemShut {NoStop}%
\bibitem [{\citenamefont {Hosen}\ \emph {et~al.}(2017)\citenamefont {Hosen},
  \citenamefont {Dimitri}, \citenamefont {Belopolski}, \citenamefont
  {Maldonado}, \citenamefont {Sankar}, \citenamefont {Dhakal}, \citenamefont
  {Dhakal}, \citenamefont {Cole}, \citenamefont {Oppeneer}, \citenamefont
  {Kaczorowski}, \citenamefont {Chou}, \citenamefont {Hasan}, \citenamefont
  {Durakiewicz},\ and\ \citenamefont {Neupane}}]{Hosen2017}%
  \BibitemOpen
  \bibfield  {author} {\bibinfo {author} {\bibfnamefont {M.~M.}\ \bibnamefont
  {Hosen}}, \bibinfo {author} {\bibfnamefont {K.}~\bibnamefont {Dimitri}},
  \bibinfo {author} {\bibfnamefont {I.}~\bibnamefont {Belopolski}}, \bibinfo
  {author} {\bibfnamefont {P.}~\bibnamefont {Maldonado}}, \bibinfo {author}
  {\bibfnamefont {R.}~\bibnamefont {Sankar}}, \bibinfo {author} {\bibfnamefont
  {N.}~\bibnamefont {Dhakal}}, \bibinfo {author} {\bibfnamefont
  {G.}~\bibnamefont {Dhakal}}, \bibinfo {author} {\bibfnamefont
  {T.}~\bibnamefont {Cole}}, \bibinfo {author} {\bibfnamefont {P.~M.}\
  \bibnamefont {Oppeneer}}, \bibinfo {author} {\bibfnamefont {D.}~\bibnamefont
  {Kaczorowski}}, \bibinfo {author} {\bibfnamefont {F.}~\bibnamefont {Chou}},
  \bibinfo {author} {\bibfnamefont {M.~Z.}\ \bibnamefont {Hasan}}, \bibinfo
  {author} {\bibfnamefont {T.}~\bibnamefont {Durakiewicz}},\ and\ \bibinfo
  {author} {\bibfnamefont {M.}~\bibnamefont {Neupane}},\ }\bibfield  {title}
  {\bibinfo {title} {Tunability of the topological nodal-line semimetal phase
  in {$\mathrm{ZrSi}X$}-type materials ({$X=\mathrm{S}, \mathrm{Se},
  \mathrm{Te}$})},\ }\href {https://doi.org/10.1103/PhysRevB.95.161101}
  {\bibfield  {journal} {\bibinfo  {journal} {Phys. Rev. B}\ }\textbf {\bibinfo
  {volume} {95}},\ \bibinfo {pages} {161101} (\bibinfo {year}
  {2017})}\BibitemShut {NoStop}%
\bibitem [{\citenamefont {Takane}\ \emph {et~al.}(2018)\citenamefont {Takane},
  \citenamefont {Nakayama}, \citenamefont {Souma}, \citenamefont {Wada},
  \citenamefont {Okamoto}, \citenamefont {Takenaka}, \citenamefont {Yamakawa},
  \citenamefont {Yamakage}, \citenamefont {Mitsuhashi}, \citenamefont {Horiba},
  \citenamefont {Kumigashira}, \citenamefont {Takahashi},\ and\ \citenamefont
  {Sato}}]{Takane2018}%
  \BibitemOpen
  \bibfield  {author} {\bibinfo {author} {\bibfnamefont {D.}~\bibnamefont
  {Takane}}, \bibinfo {author} {\bibfnamefont {K.}~\bibnamefont {Nakayama}},
  \bibinfo {author} {\bibfnamefont {S.}~\bibnamefont {Souma}}, \bibinfo
  {author} {\bibfnamefont {T.}~\bibnamefont {Wada}}, \bibinfo {author}
  {\bibfnamefont {Y.}~\bibnamefont {Okamoto}}, \bibinfo {author} {\bibfnamefont
  {K.}~\bibnamefont {Takenaka}}, \bibinfo {author} {\bibfnamefont
  {Y.}~\bibnamefont {Yamakawa}}, \bibinfo {author} {\bibfnamefont
  {A.}~\bibnamefont {Yamakage}}, \bibinfo {author} {\bibfnamefont
  {T.}~\bibnamefont {Mitsuhashi}}, \bibinfo {author} {\bibfnamefont
  {K.}~\bibnamefont {Horiba}}, \bibinfo {author} {\bibfnamefont
  {H.}~\bibnamefont {Kumigashira}}, \bibinfo {author} {\bibfnamefont
  {T.}~\bibnamefont {Takahashi}},\ and\ \bibinfo {author} {\bibfnamefont
  {T.}~\bibnamefont {Sato}},\ }\bibfield  {title} {\bibinfo {title}
  {Observation of {Dirac}-like energy band and ring-torus {Fermi} surface
  associated with the nodal line in topological insulator {CaAgAs}},\ }\href
  {https://doi.org/10.1038/s41535-017-0074-z} {\bibfield  {journal} {\bibinfo
  {journal} {npj Quantum Materials}\ }\textbf {\bibinfo {volume} {3}},\
  \bibinfo {pages} {1} (\bibinfo {year} {2018})}\BibitemShut {NoStop}%
\bibitem [{\citenamefont {Belopolski}\ \emph {et~al.}(2019)\citenamefont
  {Belopolski}, \citenamefont {Manna}, \citenamefont {Sanchez}, \citenamefont
  {Chang}, \citenamefont {Ernst}, \citenamefont {Yin}, \citenamefont {Zhang},
  \citenamefont {Cochran}, \citenamefont {Shumiya}, \citenamefont {Zheng},
  \citenamefont {Singh}, \citenamefont {Bian}, \citenamefont {Multer},
  \citenamefont {Litskevich}, \citenamefont {Zhou}, \citenamefont {Huang},
  \citenamefont {Wang}, \citenamefont {Chang}, \citenamefont {Xu},
  \citenamefont {Bansil}, \citenamefont {Felser}, \citenamefont {Lin},\ and\
  \citenamefont {Hasan}}]{Belopolski2019}%
  \BibitemOpen
  \bibfield  {author} {\bibinfo {author} {\bibfnamefont {I.}~\bibnamefont
  {Belopolski}}, \bibinfo {author} {\bibfnamefont {K.}~\bibnamefont {Manna}},
  \bibinfo {author} {\bibfnamefont {D.~S.}\ \bibnamefont {Sanchez}}, \bibinfo
  {author} {\bibfnamefont {G.}~\bibnamefont {Chang}}, \bibinfo {author}
  {\bibfnamefont {B.}~\bibnamefont {Ernst}}, \bibinfo {author} {\bibfnamefont
  {J.}~\bibnamefont {Yin}}, \bibinfo {author} {\bibfnamefont {S.~S.}\
  \bibnamefont {Zhang}}, \bibinfo {author} {\bibfnamefont {T.}~\bibnamefont
  {Cochran}}, \bibinfo {author} {\bibfnamefont {N.}~\bibnamefont {Shumiya}},
  \bibinfo {author} {\bibfnamefont {H.}~\bibnamefont {Zheng}}, \bibinfo
  {author} {\bibfnamefont {B.}~\bibnamefont {Singh}}, \bibinfo {author}
  {\bibfnamefont {G.}~\bibnamefont {Bian}}, \bibinfo {author} {\bibfnamefont
  {D.}~\bibnamefont {Multer}}, \bibinfo {author} {\bibfnamefont
  {M.}~\bibnamefont {Litskevich}}, \bibinfo {author} {\bibfnamefont
  {X.}~\bibnamefont {Zhou}}, \bibinfo {author} {\bibfnamefont {S.-M.}\
  \bibnamefont {Huang}}, \bibinfo {author} {\bibfnamefont {B.}~\bibnamefont
  {Wang}}, \bibinfo {author} {\bibfnamefont {T.-R.}\ \bibnamefont {Chang}},
  \bibinfo {author} {\bibfnamefont {S.-Y.}\ \bibnamefont {Xu}}, \bibinfo
  {author} {\bibfnamefont {A.}~\bibnamefont {Bansil}}, \bibinfo {author}
  {\bibfnamefont {C.}~\bibnamefont {Felser}}, \bibinfo {author} {\bibfnamefont
  {H.}~\bibnamefont {Lin}},\ and\ \bibinfo {author} {\bibfnamefont {M.~Z.}\
  \bibnamefont {Hasan}},\ }\bibfield  {title} {\bibinfo {title} {Discovery of
  topological {Weyl} fermion lines and drumhead surface states in a room
  temperature magnet},\ }\href {https://doi.org/10.1126/science.aav2327}
  {\bibfield  {journal} {\bibinfo  {journal} {Science}\ }\textbf {\bibinfo
  {volume} {365}},\ \bibinfo {pages} {1278} (\bibinfo {year}
  {2019})}\BibitemShut {NoStop}%
\bibitem [{\citenamefont {Liu}\ \emph {et~al.}(2018)\citenamefont {Liu},
  \citenamefont {Lou}, \citenamefont {Guo}, \citenamefont {Wang}, \citenamefont
  {Sun}, \citenamefont {Li}, \citenamefont {Thirupathaiah}, \citenamefont
  {Fedorov}, \citenamefont {Shen}, \citenamefont {Liu}, \citenamefont {Lei},\
  and\ \citenamefont {Wang}}]{Liu2018TiB2}%
  \BibitemOpen
  \bibfield  {author} {\bibinfo {author} {\bibfnamefont {Z.}~\bibnamefont
  {Liu}}, \bibinfo {author} {\bibfnamefont {R.}~\bibnamefont {Lou}}, \bibinfo
  {author} {\bibfnamefont {P.}~\bibnamefont {Guo}}, \bibinfo {author}
  {\bibfnamefont {Q.}~\bibnamefont {Wang}}, \bibinfo {author} {\bibfnamefont
  {S.}~\bibnamefont {Sun}}, \bibinfo {author} {\bibfnamefont {C.}~\bibnamefont
  {Li}}, \bibinfo {author} {\bibfnamefont {S.}~\bibnamefont {Thirupathaiah}},
  \bibinfo {author} {\bibfnamefont {A.}~\bibnamefont {Fedorov}}, \bibinfo
  {author} {\bibfnamefont {D.}~\bibnamefont {Shen}}, \bibinfo {author}
  {\bibfnamefont {K.}~\bibnamefont {Liu}}, \bibinfo {author} {\bibfnamefont
  {H.}~\bibnamefont {Lei}},\ and\ \bibinfo {author} {\bibfnamefont
  {S.}~\bibnamefont {Wang}},\ }\bibfield  {title} {\bibinfo {title}
  {Experimental observation of {Dirac} nodal links in centrosymmetric semimetal
  {TiB$_2$}},\ }\href {https://doi.org/10.1103/PhysRevX.8.031044} {\bibfield
  {journal} {\bibinfo  {journal} {Phys. Rev. X}\ }\textbf {\bibinfo {volume}
  {8}},\ \bibinfo {pages} {031044} (\bibinfo {year} {2018})}\BibitemShut
  {NoStop}%
\bibitem [{\citenamefont {Yi}\ \emph {et~al.}(2018)\citenamefont {Yi},
  \citenamefont {Lv}, \citenamefont {Wu}, \citenamefont {Fu}, \citenamefont
  {Gao}, \citenamefont {Yang}, \citenamefont {Peng}, \citenamefont {Li},
  \citenamefont {Huang}, \citenamefont {Richard}, \citenamefont {Shi},
  \citenamefont {Li}, \citenamefont {Yazyev}, \citenamefont {Shi},
  \citenamefont {Qian},\ and\ \citenamefont {Ding}}]{Yi2019TiB2}%
  \BibitemOpen
  \bibfield  {author} {\bibinfo {author} {\bibfnamefont {C.-J.}\ \bibnamefont
  {Yi}}, \bibinfo {author} {\bibfnamefont {B.~Q.}\ \bibnamefont {Lv}}, \bibinfo
  {author} {\bibfnamefont {Q.~S.}\ \bibnamefont {Wu}}, \bibinfo {author}
  {\bibfnamefont {B.-B.}\ \bibnamefont {Fu}}, \bibinfo {author} {\bibfnamefont
  {X.}~\bibnamefont {Gao}}, \bibinfo {author} {\bibfnamefont {M.}~\bibnamefont
  {Yang}}, \bibinfo {author} {\bibfnamefont {X.-L.}\ \bibnamefont {Peng}},
  \bibinfo {author} {\bibfnamefont {M.}~\bibnamefont {Li}}, \bibinfo {author}
  {\bibfnamefont {Y.-B.}\ \bibnamefont {Huang}}, \bibinfo {author}
  {\bibfnamefont {P.}~\bibnamefont {Richard}}, \bibinfo {author} {\bibfnamefont
  {M.}~\bibnamefont {Shi}}, \bibinfo {author} {\bibfnamefont {G.}~\bibnamefont
  {Li}}, \bibinfo {author} {\bibfnamefont {O.~V.}\ \bibnamefont {Yazyev}},
  \bibinfo {author} {\bibfnamefont {Y.-G.}\ \bibnamefont {Shi}}, \bibinfo
  {author} {\bibfnamefont {T.}~\bibnamefont {Qian}},\ and\ \bibinfo {author}
  {\bibfnamefont {H.}~\bibnamefont {Ding}},\ }\bibfield  {title} {\bibinfo
  {title} {Observation of a nodal chain with {Dirac} surface states in
  {TiB$_2$}},\ }\href {https://doi.org/10.1103/PhysRevB.97.201107} {\bibfield
  {journal} {\bibinfo  {journal} {Phys. Rev. B}\ }\textbf {\bibinfo {volume}
  {97}},\ \bibinfo {pages} {201107(R)} (\bibinfo {year} {2018})}\BibitemShut
  {NoStop}%
\bibitem [{\citenamefont {Lou}\ \emph {et~al.}(2018)\citenamefont {Lou},
  \citenamefont {Guo}, \citenamefont {Li}, \citenamefont {Wang}, \citenamefont
  {Liu}, \citenamefont {Sun}, \citenamefont {Li}, \citenamefont {Wu},
  \citenamefont {Wang}, \citenamefont {Sun}, \citenamefont {Shen},
  \citenamefont {Huang}, \citenamefont {Liu}, \citenamefont {Lu}, \citenamefont
  {Lei}, \citenamefont {Ding},\ and\ \citenamefont {Wang}}]{Lou2018}%
  \BibitemOpen
  \bibfield  {author} {\bibinfo {author} {\bibfnamefont {R.}~\bibnamefont
  {Lou}}, \bibinfo {author} {\bibfnamefont {P.}~\bibnamefont {Guo}}, \bibinfo
  {author} {\bibfnamefont {M.}~\bibnamefont {Li}}, \bibinfo {author}
  {\bibfnamefont {Q.}~\bibnamefont {Wang}}, \bibinfo {author} {\bibfnamefont
  {Z.}~\bibnamefont {Liu}}, \bibinfo {author} {\bibfnamefont {S.}~\bibnamefont
  {Sun}}, \bibinfo {author} {\bibfnamefont {C.}~\bibnamefont {Li}}, \bibinfo
  {author} {\bibfnamefont {X.}~\bibnamefont {Wu}}, \bibinfo {author}
  {\bibfnamefont {Z.}~\bibnamefont {Wang}}, \bibinfo {author} {\bibfnamefont
  {Z.}~\bibnamefont {Sun}}, \bibinfo {author} {\bibfnamefont {D.}~\bibnamefont
  {Shen}}, \bibinfo {author} {\bibfnamefont {Y.}~\bibnamefont {Huang}},
  \bibinfo {author} {\bibfnamefont {K.}~\bibnamefont {Liu}}, \bibinfo {author}
  {\bibfnamefont {Z.-Y.}\ \bibnamefont {Lu}}, \bibinfo {author} {\bibfnamefont
  {H.}~\bibnamefont {Lei}}, \bibinfo {author} {\bibfnamefont {H.}~\bibnamefont
  {Ding}},\ and\ \bibinfo {author} {\bibfnamefont {S.}~\bibnamefont {Wang}},\
  }\bibfield  {title} {\bibinfo {title} {Experimental observation of bulk nodal
  lines and electronic surface states in {ZrB}$_2$},\ }\href@noop {} {\bibfield
   {journal} {\bibinfo  {journal} {npj Quantum Materials}\ }\textbf {\bibinfo
  {volume} {3}},\ \bibinfo {pages} {43} (\bibinfo {year} {2018})}\BibitemShut
  {NoStop}%
\bibitem [{\citenamefont {Hosen}\ \emph {et~al.}(2020)\citenamefont {Hosen},
  \citenamefont {Dhakal}, \citenamefont {Dimitri}, \citenamefont {Amft},
  \citenamefont {Maldonado}, \citenamefont {Dey}, \citenamefont {Ahamed},
  \citenamefont {Sapkota}, \citenamefont {Weerasinghe}, \citenamefont {Kabir},
  \citenamefont {Simayi},\ and\ \citenamefont {Neupane}}]{Hosen2020}%
  \BibitemOpen
  \bibfield  {author} {\bibinfo {author} {\bibfnamefont {M.~M.}\ \bibnamefont
  {Hosen}}, \bibinfo {author} {\bibfnamefont {G.}~\bibnamefont {Dhakal}},
  \bibinfo {author} {\bibfnamefont {K.}~\bibnamefont {Dimitri}}, \bibinfo
  {author} {\bibfnamefont {M.}~\bibnamefont {Amft}}, \bibinfo {author}
  {\bibfnamefont {F.}~\bibnamefont {Maldonado}}, \bibinfo {author}
  {\bibfnamefont {P.}~\bibnamefont {Dey}}, \bibinfo {author} {\bibfnamefont
  {S.}~\bibnamefont {Ahamed}}, \bibinfo {author} {\bibfnamefont
  {B.}~\bibnamefont {Sapkota}}, \bibinfo {author} {\bibfnamefont
  {D.}~\bibnamefont {Weerasinghe}}, \bibinfo {author} {\bibfnamefont
  {F.}~\bibnamefont {Kabir}}, \bibinfo {author} {\bibfnamefont
  {N.}~\bibnamefont {Simayi}},\ and\ \bibinfo {author} {\bibfnamefont
  {M.}~\bibnamefont {Neupane}},\ }\bibfield  {title} {\bibinfo {title}
  {Experimental observation of drumhead surface states in {SrAs$_3$}},\ }\href
  {https://doi.org/10.1038/s41598-020-59200-2} {\bibfield  {journal} {\bibinfo
  {journal} {Sci. Rep.}\ }\textbf {\bibinfo {volume} {10}},\ \bibinfo {pages}
  {2776} (\bibinfo {year} {2020})}\BibitemShut {NoStop}%
\bibitem [{\citenamefont {Mori}\ \emph {et~al.}(2022)\citenamefont {Mori},
  \citenamefont {Wang}, \citenamefont {Morimoto}, \citenamefont {Ciocys},
  \citenamefont {Denlinger}, \citenamefont {Paglione},\ and\ \citenamefont
  {Lanzara}}]{Mori2022}%
  \BibitemOpen
  \bibfield  {author} {\bibinfo {author} {\bibfnamefont {R.}~\bibnamefont
  {Mori}}, \bibinfo {author} {\bibfnamefont {K.}~\bibnamefont {Wang}}, \bibinfo
  {author} {\bibfnamefont {T.}~\bibnamefont {Morimoto}}, \bibinfo {author}
  {\bibfnamefont {S.}~\bibnamefont {Ciocys}}, \bibinfo {author} {\bibfnamefont
  {J.~D.}\ \bibnamefont {Denlinger}}, \bibinfo {author} {\bibfnamefont
  {J.}~\bibnamefont {Paglione}},\ and\ \bibinfo {author} {\bibfnamefont
  {A.}~\bibnamefont {Lanzara}},\ }\bibfield  {title} {\bibinfo {title}
  {Observation of a flat and extended surface state in a topological
  semimetal},\ }\href {https://doi.org/10.3390/ma15082744} {\bibfield
  {journal} {\bibinfo  {journal} {Materials}\ }\textbf {\bibinfo {volume}
  {15}},\ \bibinfo {pages} {2744} (\bibinfo {year} {2022})}\BibitemShut
  {NoStop}%
\bibitem [{\citenamefont {Song}\ \emph {et~al.}(2023)\citenamefont {Song},
  \citenamefont {Liu}, \citenamefont {Cui}, \citenamefont {Gao}, \citenamefont
  {Song}, \citenamefont {Jin}, \citenamefont {Zhao}, \citenamefont {Sun},
  \citenamefont {Zhang}, \citenamefont {Zhao}, \citenamefont {Luo},
  \citenamefont {Sun}, \citenamefont {Shi}, \citenamefont {Zhang},
  \citenamefont {Liu},\ and\ \citenamefont {Zhou}}]{Song2023}%
  \BibitemOpen
  \bibfield  {author} {\bibinfo {author} {\bibfnamefont {C.}~\bibnamefont
  {Song}}, \bibinfo {author} {\bibfnamefont {L.}~\bibnamefont {Liu}}, \bibinfo
  {author} {\bibfnamefont {S.}~\bibnamefont {Cui}}, \bibinfo {author}
  {\bibfnamefont {J.}~\bibnamefont {Gao}}, \bibinfo {author} {\bibfnamefont
  {P.}~\bibnamefont {Song}}, \bibinfo {author} {\bibfnamefont {L.}~\bibnamefont
  {Jin}}, \bibinfo {author} {\bibfnamefont {W.}~\bibnamefont {Zhao}}, \bibinfo
  {author} {\bibfnamefont {Z.}~\bibnamefont {Sun}}, \bibinfo {author}
  {\bibfnamefont {X.}~\bibnamefont {Zhang}}, \bibinfo {author} {\bibfnamefont
  {L.}~\bibnamefont {Zhao}}, \bibinfo {author} {\bibfnamefont {X.}~\bibnamefont
  {Luo}}, \bibinfo {author} {\bibfnamefont {Y.}~\bibnamefont {Sun}}, \bibinfo
  {author} {\bibfnamefont {Y.}~\bibnamefont {Shi}}, \bibinfo {author}
  {\bibfnamefont {H.}~\bibnamefont {Zhang}}, \bibinfo {author} {\bibfnamefont
  {G.}~\bibnamefont {Liu}},\ and\ \bibinfo {author} {\bibfnamefont {X.~J.}\
  \bibnamefont {Zhou}},\ }\bibfield  {title} {\bibinfo {title} {Observation of
  spin-polarized surface states in the nodal-line semimetal {SnTaS$_2$}},\
  }\href {https://doi.org/10.1103/PhysRevB.107.045142} {\bibfield  {journal}
  {\bibinfo  {journal} {Phys. Rev. B}\ }\textbf {\bibinfo {volume} {107}},\
  \bibinfo {pages} {045142} (\bibinfo {year} {2023})}\BibitemShut {NoStop}%
\bibitem [{\citenamefont {Yang}\ \emph {et~al.}(2024)\citenamefont {Yang},
  \citenamefont {Yao}, \citenamefont {Zheng}, \citenamefont {Guan},
  \citenamefont {Zhou}, \citenamefont {Cochran}, \citenamefont {Lin},
  \citenamefont {Yin}, \citenamefont {Zhou}, \citenamefont {Cheng},
  \citenamefont {Li}, \citenamefont {Shi}, \citenamefont {Hossain},
  \citenamefont {Chi}, \citenamefont {Belopolski}, \citenamefont {Jiang},
  \citenamefont {Litskevich}, \citenamefont {Xu}, \citenamefont {Tian},
  \citenamefont {Bansil}, \citenamefont {Yin}, \citenamefont {Jia},
  \citenamefont {Chang},\ and\ \citenamefont {Hasan}}]{Yang2024}%
  \BibitemOpen
  \bibfield  {author} {\bibinfo {author} {\bibfnamefont {X.~P.}\ \bibnamefont
  {Yang}}, \bibinfo {author} {\bibfnamefont {Y.-T.}\ \bibnamefont {Yao}},
  \bibinfo {author} {\bibfnamefont {P.}~\bibnamefont {Zheng}}, \bibinfo
  {author} {\bibfnamefont {S.}~\bibnamefont {Guan}}, \bibinfo {author}
  {\bibfnamefont {H.}~\bibnamefont {Zhou}}, \bibinfo {author} {\bibfnamefont
  {T.~A.}\ \bibnamefont {Cochran}}, \bibinfo {author} {\bibfnamefont {C.-M.}\
  \bibnamefont {Lin}}, \bibinfo {author} {\bibfnamefont {J.-X.}\ \bibnamefont
  {Yin}}, \bibinfo {author} {\bibfnamefont {X.}~\bibnamefont {Zhou}}, \bibinfo
  {author} {\bibfnamefont {Z.-J.}\ \bibnamefont {Cheng}}, \bibinfo {author}
  {\bibfnamefont {Z.}~\bibnamefont {Li}}, \bibinfo {author} {\bibfnamefont
  {T.}~\bibnamefont {Shi}}, \bibinfo {author} {\bibfnamefont {M.~S.}\
  \bibnamefont {Hossain}}, \bibinfo {author} {\bibfnamefont {S.}~\bibnamefont
  {Chi}}, \bibinfo {author} {\bibfnamefont {I.}~\bibnamefont {Belopolski}},
  \bibinfo {author} {\bibfnamefont {Y.-X.}\ \bibnamefont {Jiang}}, \bibinfo
  {author} {\bibfnamefont {M.}~\bibnamefont {Litskevich}}, \bibinfo {author}
  {\bibfnamefont {G.}~\bibnamefont {Xu}}, \bibinfo {author} {\bibfnamefont
  {Z.}~\bibnamefont {Tian}}, \bibinfo {author} {\bibfnamefont {A.}~\bibnamefont
  {Bansil}}, \bibinfo {author} {\bibfnamefont {Z.}~\bibnamefont {Yin}},
  \bibinfo {author} {\bibfnamefont {S.}~\bibnamefont {Jia}}, \bibinfo {author}
  {\bibfnamefont {T.-R.}\ \bibnamefont {Chang}},\ and\ \bibinfo {author}
  {\bibfnamefont {M.~Z.}\ \bibnamefont {Hasan}},\ }\bibfield  {title} {\bibinfo
  {title} {A topological {Hund} nodal line antiferromagnet},\ }\href
  {https://doi.org/10.1038/s41467-024-51255-3} {\bibfield  {journal} {\bibinfo
  {journal} {Nat. Commun.}\ }\textbf {\bibinfo {volume} {15}},\ \bibinfo
  {pages} {7052} (\bibinfo {year} {2024})}\BibitemShut {NoStop}%
\bibitem [{\citenamefont {Akiba}\ \emph {et~al.}(2024)\citenamefont {Akiba},
  \citenamefont {Akahama}, \citenamefont {Tokunaga},\ and\ \citenamefont
  {Kobayashi}}]{Akiba2024}%
  \BibitemOpen
  \bibfield  {author} {\bibinfo {author} {\bibfnamefont {K.}~\bibnamefont
  {Akiba}}, \bibinfo {author} {\bibfnamefont {Y.}~\bibnamefont {Akahama}},
  \bibinfo {author} {\bibfnamefont {M.}~\bibnamefont {Tokunaga}},\ and\
  \bibinfo {author} {\bibfnamefont {T.~C.}\ \bibnamefont {Kobayashi}},\
  }\bibfield  {title} {\bibinfo {title} {Realization of nodal-ring semimetal in
  pressurized black phosphorus},\ }\href
  {https://doi.org/10.1103/PhysRevB.109.L201103} {\bibfield  {journal}
  {\bibinfo  {journal} {Phys. Rev. B}\ }\textbf {\bibinfo {volume} {109}},\
  \bibinfo {pages} {L201103} (\bibinfo {year} {2024})}\BibitemShut {NoStop}%
\bibitem [{\citenamefont {Song}\ \emph {et~al.}(2019)\citenamefont {Song},
  \citenamefont {He}, \citenamefont {Niu}, \citenamefont {Zhang}, \citenamefont
  {Ren}, \citenamefont {Liu},\ and\ \citenamefont {Jo}}]{Song2019}%
  \BibitemOpen
  \bibfield  {author} {\bibinfo {author} {\bibfnamefont {B.}~\bibnamefont
  {Song}}, \bibinfo {author} {\bibfnamefont {C.}~\bibnamefont {He}}, \bibinfo
  {author} {\bibfnamefont {S.}~\bibnamefont {Niu}}, \bibinfo {author}
  {\bibfnamefont {L.}~\bibnamefont {Zhang}}, \bibinfo {author} {\bibfnamefont
  {Z.}~\bibnamefont {Ren}}, \bibinfo {author} {\bibfnamefont {X.-J.}\
  \bibnamefont {Liu}},\ and\ \bibinfo {author} {\bibfnamefont {G.-B.}\
  \bibnamefont {Jo}},\ }\bibfield  {title} {\bibinfo {title} {Observation of
  nodal-line semimetal with ultracold fermions in an optical lattice},\ }\href
  {https://doi.org/10.1038/s41567-019-0564-y} {\bibfield  {journal} {\bibinfo
  {journal} {Nat. Phys.}\ }\textbf {\bibinfo {volume} {15}},\ \bibinfo {pages}
  {911} (\bibinfo {year} {2019})}\BibitemShut {NoStop}%
\bibitem [{\citenamefont {Deng}\ \emph {et~al.}(2019)\citenamefont {Deng},
  \citenamefont {Lu}, \citenamefont {Li}, \citenamefont {Huang}, \citenamefont
  {Yan}, \citenamefont {Ma},\ and\ \citenamefont {Liu}}]{Deng2019}%
  \BibitemOpen
  \bibfield  {author} {\bibinfo {author} {\bibfnamefont {W.}~\bibnamefont
  {Deng}}, \bibinfo {author} {\bibfnamefont {J.}~\bibnamefont {Lu}}, \bibinfo
  {author} {\bibfnamefont {F.}~\bibnamefont {Li}}, \bibinfo {author}
  {\bibfnamefont {X.}~\bibnamefont {Huang}}, \bibinfo {author} {\bibfnamefont
  {M.}~\bibnamefont {Yan}}, \bibinfo {author} {\bibfnamefont {J.}~\bibnamefont
  {Ma}},\ and\ \bibinfo {author} {\bibfnamefont {Z.}~\bibnamefont {Liu}},\
  }\bibfield  {title} {\bibinfo {title} {Nodal rings and drumhead surface
  states in phononic crystals},\ }\href
  {https://doi.org/10.1038/s41467-019-09820-8} {\bibfield  {journal} {\bibinfo
  {journal} {Nat. Commun.}\ }\textbf {\bibinfo {volume} {10}},\ \bibinfo
  {pages} {1769} (\bibinfo {year} {2019})}\BibitemShut {NoStop}%
\bibitem [{\citenamefont {Park}\ \emph {et~al.}(2022)\citenamefont {Park},
  \citenamefont {Gao}, \citenamefont {Zhang},\ and\ \citenamefont
  {Oh}}]{Park2022}%
  \BibitemOpen
  \bibfield  {author} {\bibinfo {author} {\bibfnamefont {H.}~\bibnamefont
  {Park}}, \bibinfo {author} {\bibfnamefont {W.}~\bibnamefont {Gao}}, \bibinfo
  {author} {\bibfnamefont {X.}~\bibnamefont {Zhang}},\ and\ \bibinfo {author}
  {\bibfnamefont {S.~S.}\ \bibnamefont {Oh}},\ }\bibfield  {title} {\bibinfo
  {title} {Nodal lines in momentum space: topological invariants and recent
  realizations in photonic and other systems},\ }\href
  {https://doi.org/https://doi.org/10.1515/nanoph-2021-0692} {\bibfield
  {journal} {\bibinfo  {journal} {Nanophotonics}\ }\textbf {\bibinfo {volume}
  {11}},\ \bibinfo {pages} {2779} (\bibinfo {year} {2022})}\BibitemShut
  {NoStop}%
\bibitem [{\citenamefont {Silkin}\ \emph {et~al.}(2004)\citenamefont {Silkin},
  \citenamefont {García-Lekue}, \citenamefont {Pitarke}, \citenamefont
  {Chulkov}, \citenamefont {Zaremba},\ and\ \citenamefont
  {Echenique}}]{Silkin2004}%
  \BibitemOpen
  \bibfield  {author} {\bibinfo {author} {\bibfnamefont {V.~M.}\ \bibnamefont
  {Silkin}}, \bibinfo {author} {\bibfnamefont {A.}~\bibnamefont
  {García-Lekue}}, \bibinfo {author} {\bibfnamefont {J.~M.}\ \bibnamefont
  {Pitarke}}, \bibinfo {author} {\bibfnamefont {E.~V.}\ \bibnamefont
  {Chulkov}}, \bibinfo {author} {\bibfnamefont {E.}~\bibnamefont {Zaremba}},\
  and\ \bibinfo {author} {\bibfnamefont {P.~M.}\ \bibnamefont {Echenique}},\
  }\bibfield  {title} {\bibinfo {title} {Novel low-energy collective excitation
  at metal surfaces},\ }\href {https://doi.org/10.1209/epl/i2003-10184-1}
  {\bibfield  {journal} {\bibinfo  {journal} {Europhys. Lett.}\ }\textbf
  {\bibinfo {volume} {66}},\ \bibinfo {pages} {260} (\bibinfo {year}
  {2004})}\BibitemShut {NoStop}%
\bibitem [{\citenamefont {Pitarke}\ \emph {et~al.}(2004)\citenamefont
  {Pitarke}, \citenamefont {Nazarov}, \citenamefont {Silkin}, \citenamefont
  {Chulkov}, \citenamefont {Zaremba},\ and\ \citenamefont
  {Echenique}}]{Pitarke2004}%
  \BibitemOpen
  \bibfield  {author} {\bibinfo {author} {\bibfnamefont {J.~M.}\ \bibnamefont
  {Pitarke}}, \bibinfo {author} {\bibfnamefont {V.~U.}\ \bibnamefont
  {Nazarov}}, \bibinfo {author} {\bibfnamefont {V.~M.}\ \bibnamefont {Silkin}},
  \bibinfo {author} {\bibfnamefont {E.~V.}\ \bibnamefont {Chulkov}}, \bibinfo
  {author} {\bibfnamefont {E.}~\bibnamefont {Zaremba}},\ and\ \bibinfo {author}
  {\bibfnamefont {P.~M.}\ \bibnamefont {Echenique}},\ }\bibfield  {title}
  {\bibinfo {title} {Theory of acoustic surface plasmons},\ }\href
  {https://doi.org/10.1103/PhysRevB.70.205403} {\bibfield  {journal} {\bibinfo
  {journal} {Phys. Rev. B}\ }\textbf {\bibinfo {volume} {70}},\ \bibinfo
  {pages} {205403} (\bibinfo {year} {2004})}\BibitemShut {NoStop}%
\bibitem [{\citenamefont {Pohl}\ \emph {et~al.}(2010)\citenamefont {Pohl},
  \citenamefont {Diaconescu}, \citenamefont {Vercelli}, \citenamefont
  {Vattuone}, \citenamefont {Silkin}, \citenamefont {Chulkov}, \citenamefont
  {Echenique},\ and\ \citenamefont {Rocca}}]{Pohl2010}%
  \BibitemOpen
  \bibfield  {author} {\bibinfo {author} {\bibfnamefont {K.}~\bibnamefont
  {Pohl}}, \bibinfo {author} {\bibfnamefont {B.}~\bibnamefont {Diaconescu}},
  \bibinfo {author} {\bibfnamefont {G.}~\bibnamefont {Vercelli}}, \bibinfo
  {author} {\bibfnamefont {L.}~\bibnamefont {Vattuone}}, \bibinfo {author}
  {\bibfnamefont {V.~M.}\ \bibnamefont {Silkin}}, \bibinfo {author}
  {\bibfnamefont {E.~V.}\ \bibnamefont {Chulkov}}, \bibinfo {author}
  {\bibfnamefont {P.~M.}\ \bibnamefont {Echenique}},\ and\ \bibinfo {author}
  {\bibfnamefont {M.}~\bibnamefont {Rocca}},\ }\bibfield  {title} {\bibinfo
  {title} {Acoustic surface plasmon on {Cu(111)}},\ }\href
  {https://doi.org/10.1209/0295-5075/90/57006} {\bibfield  {journal} {\bibinfo
  {journal} {Europhysics Letters}\ }\textbf {\bibinfo {volume} {90}},\ \bibinfo
  {pages} {57006} (\bibinfo {year} {2010})}\BibitemShut {NoStop}%
\bibitem [{\citenamefont {Xue}\ \emph {et~al.}(2026)\citenamefont {Xue},
  \citenamefont {Li}, \citenamefont {Wang}, \citenamefont {Deng}, \citenamefont
  {Li}, \citenamefont {Tao}, \citenamefont {Lin}, \citenamefont {Su},
  \citenamefont {Shi}, \citenamefont {Zhou}, \citenamefont {Shi}, \citenamefont
  {Guo},\ and\ \citenamefont {Zhu}}]{Xue2026}%
  \BibitemOpen
  \bibfield  {author} {\bibinfo {author} {\bibfnamefont {S.}~\bibnamefont
  {Xue}}, \bibinfo {author} {\bibfnamefont {Y.}~\bibnamefont {Li}}, \bibinfo
  {author} {\bibfnamefont {M.}~\bibnamefont {Wang}}, \bibinfo {author}
  {\bibfnamefont {X.}~\bibnamefont {Deng}}, \bibinfo {author} {\bibfnamefont
  {J.}~\bibnamefont {Li}}, \bibinfo {author} {\bibfnamefont {Z.}~\bibnamefont
  {Tao}}, \bibinfo {author} {\bibfnamefont {Z.}~\bibnamefont {Lin}}, \bibinfo
  {author} {\bibfnamefont {Z.}~\bibnamefont {Su}}, \bibinfo {author}
  {\bibfnamefont {X.}~\bibnamefont {Shi}}, \bibinfo {author} {\bibfnamefont
  {J.}~\bibnamefont {Zhou}}, \bibinfo {author} {\bibfnamefont {Y.}~\bibnamefont
  {Shi}}, \bibinfo {author} {\bibfnamefont {J.}~\bibnamefont {Guo}},\ and\
  \bibinfo {author} {\bibfnamefont {X.}~\bibnamefont {Zhu}},\ }\bibfield
  {title} {\bibinfo {title} {Observation of acoustic surface plasmon in a
  nodal-line semimetal},\ }\href {https://doi.org/10.1103/3v78-6ltn} {\bibfield
   {journal} {\bibinfo  {journal} {Phys. Rev. B}\ }\textbf {\bibinfo {volume}
  {114}},\ \bibinfo {pages} {L111408} (\bibinfo {year} {2026})}\BibitemShut
  {NoStop}%
\bibitem [{\citenamefont {Pitarke}\ \emph {et~al.}(2007)\citenamefont
  {Pitarke}, \citenamefont {Silkin}, \citenamefont {Chulkov},\ and\
  \citenamefont {Echenique}}]{Pitarke2007}%
  \BibitemOpen
  \bibfield  {author} {\bibinfo {author} {\bibfnamefont {J.~M.}\ \bibnamefont
  {Pitarke}}, \bibinfo {author} {\bibfnamefont {V.~M.}\ \bibnamefont {Silkin}},
  \bibinfo {author} {\bibfnamefont {E.~V.}\ \bibnamefont {Chulkov}},\ and\
  \bibinfo {author} {\bibfnamefont {P.~M.}\ \bibnamefont {Echenique}},\
  }\bibfield  {title} {\bibinfo {title} {Theory of surface plasmons and
  surface-plasmon polaritons},\ }\href
  {https://doi.org/10.1088/0034-4885/70/1/R01} {\bibfield  {journal} {\bibinfo
  {journal} {Reports on Progress in Physics}\ }\textbf {\bibinfo {volume}
  {70}},\ \bibinfo {pages} {1} (\bibinfo {year} {2007})}\BibitemShut {NoStop}%
\bibitem [{\citenamefont {Rahimpoor}\ and\ \citenamefont
  {Abedinpour}(2024)}]{Rahimpoor2024}%
  \BibitemOpen
  \bibfield  {author} {\bibinfo {author} {\bibfnamefont {H.}~\bibnamefont
  {Rahimpoor}}\ and\ \bibinfo {author} {\bibfnamefont {S.~H.}\ \bibnamefont
  {Abedinpour}},\ }\bibfield  {title} {\bibinfo {title} {Collective excitations
  and screening in two-dimensional tilted nodal-line semimetals},\ }\href
  {https://doi.org/10.1103/PhysRevB.109.045120} {\bibfield  {journal} {\bibinfo
   {journal} {Phys. Rev. B}\ }\textbf {\bibinfo {volume} {109}},\ \bibinfo
  {pages} {045120} (\bibinfo {year} {2024})}\BibitemShut {NoStop}%
\bibitem [{\citenamefont {Rudi}\ \emph {et~al.}(2024)\citenamefont {Rudi},
  \citenamefont {De~Martino}, \citenamefont {Moors}, \citenamefont {Giuliano},\
  and\ \citenamefont {Buccheri}}]{Rudi2024}%
  \BibitemOpen
  \bibfield  {author} {\bibinfo {author} {\bibfnamefont {M.}~\bibnamefont
  {Rudi}}, \bibinfo {author} {\bibfnamefont {A.}~\bibnamefont {De~Martino}},
  \bibinfo {author} {\bibfnamefont {K.}~\bibnamefont {Moors}}, \bibinfo
  {author} {\bibfnamefont {D.}~\bibnamefont {Giuliano}},\ and\ \bibinfo
  {author} {\bibfnamefont {F.}~\bibnamefont {Buccheri}},\ }\bibfield  {title}
  {\bibinfo {title} {Interfaces of nodal-line semimetals: Drum states,
  transport, and refraction},\ }\href
  {https://doi.org/10.1103/PhysRevB.109.195144} {\bibfield  {journal} {\bibinfo
   {journal} {Phys. Rev. B}\ }\textbf {\bibinfo {volume} {109}},\ \bibinfo
  {pages} {195144} (\bibinfo {year} {2024})}\BibitemShut {NoStop}%
\bibitem [{\citenamefont {Chiu}\ \emph {et~al.}(2016)\citenamefont {Chiu},
  \citenamefont {Teo}, \citenamefont {Schnyder},\ and\ \citenamefont
  {Ryu}}]{Chiu2016}%
  \BibitemOpen
  \bibfield  {author} {\bibinfo {author} {\bibfnamefont {C.-K.}\ \bibnamefont
  {Chiu}}, \bibinfo {author} {\bibfnamefont {J.~C.~Y.}\ \bibnamefont {Teo}},
  \bibinfo {author} {\bibfnamefont {A.~P.}\ \bibnamefont {Schnyder}},\ and\
  \bibinfo {author} {\bibfnamefont {S.}~\bibnamefont {Ryu}},\ }\bibfield
  {title} {\bibinfo {title} {Classification of topological quantum matter with
  symmetries},\ }\href {https://doi.org/10.1103/RevModPhys.88.035005}
  {\bibfield  {journal} {\bibinfo  {journal} {Rev. Mod. Phys.}\ }\textbf
  {\bibinfo {volume} {88}},\ \bibinfo {pages} {035005} (\bibinfo {year}
  {2016})}\BibitemShut {NoStop}%
\bibitem [{\citenamefont {Schnyder}\ \emph {et~al.}(2008)\citenamefont
  {Schnyder}, \citenamefont {Ryu}, \citenamefont {Furusaki},\ and\
  \citenamefont {Ludwig}}]{Schnyder2008}%
  \BibitemOpen
  \bibfield  {author} {\bibinfo {author} {\bibfnamefont {A.~P.}\ \bibnamefont
  {Schnyder}}, \bibinfo {author} {\bibfnamefont {S.}~\bibnamefont {Ryu}},
  \bibinfo {author} {\bibfnamefont {A.}~\bibnamefont {Furusaki}},\ and\
  \bibinfo {author} {\bibfnamefont {A.~W.~W.}\ \bibnamefont {Ludwig}},\
  }\bibfield  {title} {\bibinfo {title} {Classification of topological
  insulators and superconductors in three spatial dimensions},\ }\href
  {https://doi.org/10.1103/PhysRevB.78.195125} {\bibfield  {journal} {\bibinfo
  {journal} {Phys. Rev. B}\ }\textbf {\bibinfo {volume} {78}},\ \bibinfo
  {pages} {195125} (\bibinfo {year} {2008})}\BibitemShut {NoStop}%
\bibitem [{\citenamefont {Giuliani}\ and\ \citenamefont
  {Vignale}(2005)}]{Giuliani}%
  \BibitemOpen
  \bibfield  {author} {\bibinfo {author} {\bibfnamefont {G.}~\bibnamefont
  {Giuliani}}\ and\ \bibinfo {author} {\bibfnamefont {G.}~\bibnamefont
  {Vignale}},\ }\href@noop {} {\emph {\bibinfo {title} {Quantum Theory of the
  Electron Liquid}}}\ (\bibinfo  {publisher} {Cambridge University Press},\
  \bibinfo {year} {2005})\BibitemShut {NoStop}%
\bibitem [{\citenamefont {Newns}(1970)}]{Newns1970}%
  \BibitemOpen
  \bibfield  {author} {\bibinfo {author} {\bibfnamefont {D.~M.}\ \bibnamefont
  {Newns}},\ }\bibfield  {title} {\bibinfo {title} {Dielectric response of a
  semi-infinite degenerate electron gas},\ }\href
  {https://doi.org/10.1103/PhysRevB.1.3304} {\bibfield  {journal} {\bibinfo
  {journal} {Phys. Rev. B}\ }\textbf {\bibinfo {volume} {1}},\ \bibinfo {pages}
  {3304} (\bibinfo {year} {1970})}\BibitemShut {NoStop}%
\bibitem [{\citenamefont {Aryasetiawan}\ \emph {et~al.}(2004)\citenamefont
  {Aryasetiawan}, \citenamefont {Imada}, \citenamefont {Georges}, \citenamefont
  {Kotliar}, \citenamefont {Biermann},\ and\ \citenamefont
  {Lichtenstein}}]{Aryasetiawan2004}%
  \BibitemOpen
  \bibfield  {author} {\bibinfo {author} {\bibfnamefont {F.}~\bibnamefont
  {Aryasetiawan}}, \bibinfo {author} {\bibfnamefont {M.}~\bibnamefont {Imada}},
  \bibinfo {author} {\bibfnamefont {A.}~\bibnamefont {Georges}}, \bibinfo
  {author} {\bibfnamefont {G.}~\bibnamefont {Kotliar}}, \bibinfo {author}
  {\bibfnamefont {S.}~\bibnamefont {Biermann}},\ and\ \bibinfo {author}
  {\bibfnamefont {A.~I.}\ \bibnamefont {Lichtenstein}},\ }\bibfield  {title}
  {\bibinfo {title} {Frequency-dependent local interactions and low-energy
  effective models from electronic structure calculations},\ }\href
  {https://doi.org/10.1103/PhysRevB.70.195104} {\bibfield  {journal} {\bibinfo
  {journal} {Phys. Rev. B}\ }\textbf {\bibinfo {volume} {70}},\ \bibinfo
  {pages} {195104} (\bibinfo {year} {2004})}\BibitemShut {NoStop}%
\bibitem [{\citenamefont {Luryi}(1988)}]{Luryi1988}%
  \BibitemOpen
  \bibfield  {author} {\bibinfo {author} {\bibfnamefont {S.}~\bibnamefont
  {Luryi}},\ }\bibfield  {title} {\bibinfo {title} {Quantum capacitance
  devices},\ }\href {https://doi.org/10.1063/1.99649} {\bibfield  {journal}
  {\bibinfo  {journal} {Applied Physics Letters}\ }\textbf {\bibinfo {volume}
  {52}},\ \bibinfo {pages} {501} (\bibinfo {year} {1988})}\BibitemShut
  {NoStop}%
\bibitem [{\citenamefont {Ando}\ \emph {et~al.}(1982)\citenamefont {Ando},
  \citenamefont {Fowler},\ and\ \citenamefont {Stern}}]{Ando1982}%
  \BibitemOpen
  \bibfield  {author} {\bibinfo {author} {\bibfnamefont {T.}~\bibnamefont
  {Ando}}, \bibinfo {author} {\bibfnamefont {A.~B.}\ \bibnamefont {Fowler}},\
  and\ \bibinfo {author} {\bibfnamefont {F.}~\bibnamefont {Stern}},\ }\bibfield
   {title} {\bibinfo {title} {Electronic properties of two-dimensional
  systems},\ }\href {https://doi.org/10.1103/RevModPhys.54.437} {\bibfield
  {journal} {\bibinfo  {journal} {Rev. Mod. Phys.}\ }\textbf {\bibinfo {volume}
  {54}},\ \bibinfo {pages} {437} (\bibinfo {year} {1982})}\BibitemShut
  {NoStop}%
\bibitem [{\citenamefont {Mukherjee}\ \emph {et~al.}(2011)\citenamefont
  {Mukherjee}, \citenamefont {Manninen},\ and\ \citenamefont {{Singha
  Deo}}}]{Mukherjee2011}%
  \BibitemOpen
  \bibfield  {author} {\bibinfo {author} {\bibfnamefont {S.}~\bibnamefont
  {Mukherjee}}, \bibinfo {author} {\bibfnamefont {M.}~\bibnamefont
  {Manninen}},\ and\ \bibinfo {author} {\bibfnamefont {P.}~\bibnamefont
  {{Singha Deo}}},\ }\bibfield  {title} {\bibinfo {title} {Quantum capacitance:
  A microscopic derivation},\ }\href
  {https://doi.org/https://doi.org/10.1016/j.physe.2011.07.005} {\bibfield
  {journal} {\bibinfo  {journal} {Physica E: Low-dimensional Systems and
  Nanostructures}\ }\textbf {\bibinfo {volume} {44}},\ \bibinfo {pages} {62}
  (\bibinfo {year} {2011})}\BibitemShut {NoStop}%
\bibitem [{\citenamefont {Christen}\ and\ \citenamefont
  {B{\"u}ttiker}(1996)}]{Christen1996}%
  \BibitemOpen
  \bibfield  {author} {\bibinfo {author} {\bibfnamefont {T.}~\bibnamefont
  {Christen}}\ and\ \bibinfo {author} {\bibfnamefont {M.}~\bibnamefont
  {B{\"u}ttiker}},\ }\bibfield  {title} {\bibinfo {title} {Low frequency
  admittance of a quantum point contact},\ }\href
  {https://doi.org/10.1103/PhysRevLett.77.143} {\bibfield  {journal} {\bibinfo
  {journal} {Phys. Rev. Lett.}\ }\textbf {\bibinfo {volume} {77}},\ \bibinfo
  {pages} {143} (\bibinfo {year} {1996})}\BibitemShut {NoStop}%
\bibitem [{\citenamefont {Ibach}\ and\ \citenamefont {Mills}(1982)}]{Ibach}%
  \BibitemOpen
  \bibfield  {author} {\bibinfo {author} {\bibfnamefont {H.}~\bibnamefont
  {Ibach}}\ and\ \bibinfo {author} {\bibfnamefont {D.~L.}\ \bibnamefont
  {Mills}},\ }\bibfield  {title} {\bibinfo {title} {Basic theory of electron
  energy loss spectroscopy},\ }in\ \href
  {https://doi.org/https://doi.org/10.1016/B978-0-12-369350-1.50007-5} {\emph
  {\bibinfo {booktitle} {Electron Energy Loss Spectroscopy and Surface
  Vibrations}}}\ (\bibinfo  {publisher} {Academic Press},\ \bibinfo {year}
  {1982})\ pp.\ \bibinfo {pages} {63--126}\BibitemShut {NoStop}%
\bibitem [{\citenamefont {Brenner}\ and\ \citenamefont
  {Scott}(2008)}]{BrennerScott2008}%
  \BibitemOpen
  \bibfield  {author} {\bibinfo {author} {\bibfnamefont {S.~C.}\ \bibnamefont
  {Brenner}}\ and\ \bibinfo {author} {\bibfnamefont {L.~R.}\ \bibnamefont
  {Scott}},\ }\href {https://doi.org/10.1007/978-0-387-75934-0} {\emph
  {\bibinfo {title} {The Mathematical Theory of Finite Element Methods}}},\
  \bibinfo {edition} {3rd}\ ed.,\ \bibinfo {series} {Texts in Applied
  Mathematics}, Vol.~\bibinfo {volume} {15}\ (\bibinfo  {publisher}
  {Springer},\ \bibinfo {address} {New York},\ \bibinfo {year}
  {2008})\BibitemShut {NoStop}%
\bibitem [{\citenamefont {Boyd}(2001)}]{Boyd2001}%
  \BibitemOpen
  \bibfield  {author} {\bibinfo {author} {\bibfnamefont {J.~P.}\ \bibnamefont
  {Boyd}},\ }\href {https://books.google.de/books?id=i9UoAwAAQBAJ} {\emph
  {\bibinfo {title} {Chebyshev and Fourier Spectral Methods: Second Revised
  Edition}}},\ Dover Books on Mathematics\ (\bibinfo  {publisher} {Dover
  Publications},\ \bibinfo {year} {2001})\BibitemShut {NoStop}%
\bibitem [{\citenamefont {Atkinson}(1997)}]{Atkinson1997}%
  \BibitemOpen
  \bibfield  {author} {\bibinfo {author} {\bibfnamefont {K.~E.}\ \bibnamefont
  {Atkinson}},\ }\href@noop {} {\emph {\bibinfo {title} {The Numerical Solution
  of Integral Equations of the Second Kind}}},\ Cambridge Monographs on Applied
  and Computational Mathematics\ (\bibinfo  {publisher} {Cambridge University
  Press},\ \bibinfo {year} {1997})\BibitemShut {NoStop}%
\bibitem [{\citenamefont {Yan}\ \emph {et~al.}(2012)\citenamefont {Yan},
  \citenamefont {Jacobsen},\ and\ \citenamefont {Thygesen}}]{Yan2012}%
  \BibitemOpen
  \bibfield  {author} {\bibinfo {author} {\bibfnamefont {J.}~\bibnamefont
  {Yan}}, \bibinfo {author} {\bibfnamefont {K.~W.}\ \bibnamefont {Jacobsen}},\
  and\ \bibinfo {author} {\bibfnamefont {K.~S.}\ \bibnamefont {Thygesen}},\
  }\bibfield  {title} {\bibinfo {title} {Conventional and acoustic surface
  plasmons on noble metal surfaces: A time-dependent density functional theory
  study},\ }\href {https://doi.org/10.1103/PhysRevB.86.241404} {\bibfield
  {journal} {\bibinfo  {journal} {Phys. Rev. B}\ }\textbf {\bibinfo {volume}
  {86}},\ \bibinfo {pages} {241404(R)} (\bibinfo {year} {2012})}\BibitemShut
  {NoStop}%
\bibitem [{\citenamefont {Witten}(2016)}]{Witten2016}%
  \BibitemOpen
  \bibfield  {author} {\bibinfo {author} {\bibfnamefont {E.}~\bibnamefont
  {Witten}},\ }\bibfield  {title} {\bibinfo {title} {Three lectures on
  topological phases of matter},\ }\href
  {https://doi.org/10.1393/ncr/i2016-10125-3} {\bibfield  {journal} {\bibinfo
  {journal} {La Rivista del Nuovo Cimento}\ }\textbf {\bibinfo {volume} {39}},\
  \bibinfo {pages} {313} (\bibinfo {year} {2016})}\BibitemShut {NoStop}%
\bibitem [{\citenamefont {Jishi}(2013)}]{Jishi}%
  \BibitemOpen
  \bibfield  {author} {\bibinfo {author} {\bibfnamefont {R.~A.}\ \bibnamefont
  {Jishi}},\ }\href@noop {} {\emph {\bibinfo {title} {Feynman Diagram
  Techniques in Condensed Matter Physics}}}\ (\bibinfo  {publisher} {Cambridge
  University Press},\ \bibinfo {address} {Cambridge},\ \bibinfo {year}
  {2013})\BibitemShut {NoStop}%
\bibitem [{\citenamefont {Mahan}(2000)}]{Mahan}%
  \BibitemOpen
  \bibfield  {author} {\bibinfo {author} {\bibfnamefont {G.~D.}\ \bibnamefont
  {Mahan}},\ }\href@noop {} {\emph {\bibinfo {title} {Many Particle
  Physics}}},\ \bibinfo {edition} {3rd}\ ed.\ (\bibinfo  {publisher} {Plenum},\
  \bibinfo {address} {New York},\ \bibinfo {year} {2000})\BibitemShut {NoStop}%
\bibitem [{\citenamefont {Choi}\ and\ \citenamefont {Kim}(2014)}]{ChoiKim2014}%
  \BibitemOpen
  \bibfield  {author} {\bibinfo {author} {\bibfnamefont {J.}~\bibnamefont
  {Choi}}\ and\ \bibinfo {author} {\bibfnamefont {S.}~\bibnamefont {Kim}},\
  }\bibfield  {title} {\bibinfo {title} {Green's functions for elliptic and
  parabolic systems with {Robin}-type boundary conditions},\ }\href
  {https://doi.org/10.1016/j.jfa.2014.08.011} {\bibfield  {journal} {\bibinfo
  {journal} {Journal of Functional Analysis}\ }\textbf {\bibinfo {volume}
  {267}},\ \bibinfo {pages} {3205} (\bibinfo {year} {2014})},\ \Eprint
  {https://arxiv.org/abs/1404.2418} {arXiv:1404.2418 [math.AP]} \BibitemShut
  {NoStop}%
\bibitem [{\citenamefont {Scalfi}\ \emph {et~al.}(2020)\citenamefont {Scalfi},
  \citenamefont {Dufils}, \citenamefont {Reeves}, \citenamefont {Rotenberg},\
  and\ \citenamefont {Salanne}}]{Scalfi2020}%
  \BibitemOpen
  \bibfield  {author} {\bibinfo {author} {\bibfnamefont {L.}~\bibnamefont
  {Scalfi}}, \bibinfo {author} {\bibfnamefont {T.}~\bibnamefont {Dufils}},
  \bibinfo {author} {\bibfnamefont {K.~G.}\ \bibnamefont {Reeves}}, \bibinfo
  {author} {\bibfnamefont {B.}~\bibnamefont {Rotenberg}},\ and\ \bibinfo
  {author} {\bibfnamefont {M.}~\bibnamefont {Salanne}},\ }\bibfield  {title}
  {\bibinfo {title} {A semiclassical {Thomas–Fermi} model to tune the
  metallicity of electrodes in molecular simulations},\ }\href
  {https://doi.org/10.1063/5.0028232} {\bibfield  {journal} {\bibinfo
  {journal} {The Journal of Chemical Physics}\ }\textbf {\bibinfo {volume}
  {153}},\ \bibinfo {pages} {174704} (\bibinfo {year} {2020})}\BibitemShut
  {NoStop}%
\bibitem [{\citenamefont {Silkin}\ \emph {et~al.}(2005)\citenamefont {Silkin},
  \citenamefont {Pitarke}, \citenamefont {Chulkov},\ and\ \citenamefont
  {Echenique}}]{Silkin2005}%
  \BibitemOpen
  \bibfield  {author} {\bibinfo {author} {\bibfnamefont {V.~M.}\ \bibnamefont
  {Silkin}}, \bibinfo {author} {\bibfnamefont {J.~M.}\ \bibnamefont {Pitarke}},
  \bibinfo {author} {\bibfnamefont {E.~V.}\ \bibnamefont {Chulkov}},\ and\
  \bibinfo {author} {\bibfnamefont {P.~M.}\ \bibnamefont {Echenique}},\
  }\bibfield  {title} {\bibinfo {title} {Acoustic surface plasmons in the noble
  metals {Cu}, {Ag}, and {Au}},\ }\href
  {https://doi.org/10.1103/PhysRevB.72.115435} {\bibfield  {journal} {\bibinfo
  {journal} {Phys. Rev. B}\ }\textbf {\bibinfo {volume} {72}},\ \bibinfo
  {pages} {115435} (\bibinfo {year} {2005})}\BibitemShut {NoStop}%
\bibitem [{\citenamefont {Alassaf}\ \emph {et~al.}(2024)\citenamefont
  {Alassaf}, \citenamefont {Koltai}, \citenamefont {Garc{\'\i}a-Fuente},\ and\
  \citenamefont {Oroszl{\'a}ny}}]{Alassaf2024}%
  \BibitemOpen
  \bibfield  {author} {\bibinfo {author} {\bibfnamefont {A.}~\bibnamefont
  {Alassaf}}, \bibinfo {author} {\bibfnamefont {J.}~\bibnamefont {Koltai}},
  \bibinfo {author} {\bibfnamefont {A.}~\bibnamefont {Garc{\'\i}a-Fuente}},\
  and\ \bibinfo {author} {\bibfnamefont {L.}~\bibnamefont {Oroszl{\'a}ny}},\
  }\bibfield  {title} {\bibinfo {title} {Surface reconstruction limited
  magnetism of the nodal loop semimetal {Ca}$_3${P}$_2$},\ }\href@noop {}
  {\bibfield  {journal} {\bibinfo  {journal} {Sci. Rep.}\ }\textbf {\bibinfo
  {volume} {14}},\ \bibinfo {pages} {31845} (\bibinfo {year}
  {2024})}\BibitemShut {NoStop}%
\bibitem [{\citenamefont {Fumega}\ \emph {et~al.}(2021)\citenamefont {Fumega},
  \citenamefont {Pardo},\ and\ \citenamefont {Cortijo}}]{Fumega2021}%
  \BibitemOpen
  \bibfield  {author} {\bibinfo {author} {\bibfnamefont {A.~O.}\ \bibnamefont
  {Fumega}}, \bibinfo {author} {\bibfnamefont {V.}~\bibnamefont {Pardo}},\ and\
  \bibinfo {author} {\bibfnamefont {A.}~\bibnamefont {Cortijo}},\ }\bibfield
  {title} {\bibinfo {title} {Increasing the number of topological nodal lines
  in semimetals via uniaxial pressure},\ }\href
  {https://doi.org/10.1038/s41598-021-90165-y} {\bibfield  {journal} {\bibinfo
  {journal} {Sci. Rep.}\ }\textbf {\bibinfo {volume} {11}},\ \bibinfo {pages}
  {10574} (\bibinfo {year} {2021})}\BibitemShut {NoStop}%
\bibitem [{\citenamefont {Kim}\ and\ \citenamefont
  {F{\"a}ssler}(2008)}]{Kim2008}%
  \BibitemOpen
  \bibfield  {author} {\bibinfo {author} {\bibfnamefont {S.-J.}\ \bibnamefont
  {Kim}}\ and\ \bibinfo {author} {\bibfnamefont {T.~F.}\ \bibnamefont
  {F{\"a}ssler}},\ }\bibfield  {title} {\bibinfo {title} {Crystal structure of
  barium distannide, {BaSn2}},\ }\href {https://doi.org/10.1524/ncrs.2008.0140}
  {\bibfield  {journal} {\bibinfo  {journal} {Z. Kristallogr. New Cryst.
  Struct.}\ }\textbf {\bibinfo {volume} {223}},\ \bibinfo {pages} {325}
  (\bibinfo {year} {2008})}\BibitemShut {NoStop}%
\bibitem [{\citenamefont {Zhao}\ \emph {et~al.}(2023)\citenamefont {Zhao},
  \citenamefont {Li}, \citenamefont {Rui}, \citenamefont {Wang}, \citenamefont
  {Lu},\ and\ \citenamefont {Xie}}]{Zhao2023}%
  \BibitemOpen
  \bibfield  {author} {\bibinfo {author} {\bibfnamefont {G.-Q.}\ \bibnamefont
  {Zhao}}, \bibinfo {author} {\bibfnamefont {S.}~\bibnamefont {Li}}, \bibinfo
  {author} {\bibfnamefont {W.~B.}\ \bibnamefont {Rui}}, \bibinfo {author}
  {\bibfnamefont {C.~M.}\ \bibnamefont {Wang}}, \bibinfo {author}
  {\bibfnamefont {H.-Z.}\ \bibnamefont {Lu}},\ and\ \bibinfo {author}
  {\bibfnamefont {X.~C.}\ \bibnamefont {Xie}},\ }\bibfield  {title} {\bibinfo
  {title} {{3D} quantum {Hall} effect in a topological nodal-ring semimetal},\
  }\href {https://doi.org/10.1007/s44214-023-00046-w} {\bibfield  {journal}
  {\bibinfo  {journal} {Quantum Frontiers}\ }\textbf {\bibinfo {volume} {2}},\
  \bibinfo {pages} {22} (\bibinfo {year} {2023})}\BibitemShut {NoStop}%
\bibitem [{\citenamefont {Wang}\ \emph {et~al.}(2017)\citenamefont {Wang},
  \citenamefont {Ma}, \citenamefont {Emmanouilidou}, \citenamefont {Shen},
  \citenamefont {Hsu}, \citenamefont {Zhou}, \citenamefont {Zuo}, \citenamefont
  {Song}, \citenamefont {Xu}, \citenamefont {Wang}, \citenamefont {Huang},
  \citenamefont {Ni},\ and\ \citenamefont {Liu}}]{Wang2017}%
  \BibitemOpen
  \bibfield  {author} {\bibinfo {author} {\bibfnamefont {X.-B.}\ \bibnamefont
  {Wang}}, \bibinfo {author} {\bibfnamefont {X.-M.}\ \bibnamefont {Ma}},
  \bibinfo {author} {\bibfnamefont {E.}~\bibnamefont {Emmanouilidou}}, \bibinfo
  {author} {\bibfnamefont {B.}~\bibnamefont {Shen}}, \bibinfo {author}
  {\bibfnamefont {C.-H.}\ \bibnamefont {Hsu}}, \bibinfo {author} {\bibfnamefont
  {C.-S.}\ \bibnamefont {Zhou}}, \bibinfo {author} {\bibfnamefont
  {Y.}~\bibnamefont {Zuo}}, \bibinfo {author} {\bibfnamefont {R.-R.}\
  \bibnamefont {Song}}, \bibinfo {author} {\bibfnamefont {S.-Y.}\ \bibnamefont
  {Xu}}, \bibinfo {author} {\bibfnamefont {G.}~\bibnamefont {Wang}}, \bibinfo
  {author} {\bibfnamefont {L.}~\bibnamefont {Huang}}, \bibinfo {author}
  {\bibfnamefont {N.}~\bibnamefont {Ni}},\ and\ \bibinfo {author}
  {\bibfnamefont {C.}~\bibnamefont {Liu}},\ }\bibfield  {title} {\bibinfo
  {title} {Topological surface electronic states in candidate nodal-line
  semimetal {CaAgAs}},\ }\href {https://doi.org/10.1103/PhysRevB.96.161112}
  {\bibfield  {journal} {\bibinfo  {journal} {Phys. Rev. B}\ }\textbf {\bibinfo
  {volume} {96}},\ \bibinfo {pages} {161112(R)} (\bibinfo {year}
  {2017})}\BibitemShut {NoStop}%
\bibitem [{\citenamefont {Topp}\ \emph {et~al.}(2017)\citenamefont {Topp},
  \citenamefont {Queiroz}, \citenamefont {Gr\"uneis}, \citenamefont
  {M\"uchler}, \citenamefont {Rost}, \citenamefont {Varykhalov}, \citenamefont
  {Marchenko}, \citenamefont {Krivenkov}, \citenamefont {Rodolakis},
  \citenamefont {McChesney}, \citenamefont {Lotsch}, \citenamefont {Schoop},\
  and\ \citenamefont {Ast}}]{Topp2017}%
  \BibitemOpen
  \bibfield  {author} {\bibinfo {author} {\bibfnamefont {A.}~\bibnamefont
  {Topp}}, \bibinfo {author} {\bibfnamefont {R.}~\bibnamefont {Queiroz}},
  \bibinfo {author} {\bibfnamefont {A.}~\bibnamefont {Gr\"uneis}}, \bibinfo
  {author} {\bibfnamefont {L.}~\bibnamefont {M\"uchler}}, \bibinfo {author}
  {\bibfnamefont {A.~W.}\ \bibnamefont {Rost}}, \bibinfo {author}
  {\bibfnamefont {A.}~\bibnamefont {Varykhalov}}, \bibinfo {author}
  {\bibfnamefont {D.}~\bibnamefont {Marchenko}}, \bibinfo {author}
  {\bibfnamefont {M.}~\bibnamefont {Krivenkov}}, \bibinfo {author}
  {\bibfnamefont {F.}~\bibnamefont {Rodolakis}}, \bibinfo {author}
  {\bibfnamefont {J.~L.}\ \bibnamefont {McChesney}}, \bibinfo {author}
  {\bibfnamefont {B.~V.}\ \bibnamefont {Lotsch}}, \bibinfo {author}
  {\bibfnamefont {L.~M.}\ \bibnamefont {Schoop}},\ and\ \bibinfo {author}
  {\bibfnamefont {C.~R.}\ \bibnamefont {Ast}},\ }\bibfield  {title} {\bibinfo
  {title} {Surface floating {2D} bands in layered nonsymmorphic semimetals:
  {ZrSiS} and related compounds},\ }\href
  {https://doi.org/10.1103/PhysRevX.7.041073} {\bibfield  {journal} {\bibinfo
  {journal} {Phys. Rev. X}\ }\textbf {\bibinfo {volume} {7}},\ \bibinfo {pages}
  {041073} (\bibinfo {year} {2017})}\BibitemShut {NoStop}%
\bibitem [{\citenamefont {Zhu}\ \emph {et~al.}(2018)\citenamefont {Zhu},
  \citenamefont {Chang}, \citenamefont {Huang}, \citenamefont {Pan},
  \citenamefont {Nie}, \citenamefont {Wang}, \citenamefont {Jin}, \citenamefont
  {Xu}, \citenamefont {Huang}, \citenamefont {Guan}, \citenamefont {Wang},
  \citenamefont {Li}, \citenamefont {Liu}, \citenamefont {Qian}, \citenamefont
  {Ku}, \citenamefont {Song}, \citenamefont {Lin}, \citenamefont {Zheng},\ and\
  \citenamefont {Jia}}]{Zhu2018}%
  \BibitemOpen
  \bibfield  {author} {\bibinfo {author} {\bibfnamefont {Z.}~\bibnamefont
  {Zhu}}, \bibinfo {author} {\bibfnamefont {T.-R.}\ \bibnamefont {Chang}},
  \bibinfo {author} {\bibfnamefont {C.-Y.}\ \bibnamefont {Huang}}, \bibinfo
  {author} {\bibfnamefont {H.}~\bibnamefont {Pan}}, \bibinfo {author}
  {\bibfnamefont {X.-A.}\ \bibnamefont {Nie}}, \bibinfo {author} {\bibfnamefont
  {X.-Z.}\ \bibnamefont {Wang}}, \bibinfo {author} {\bibfnamefont {Z.-T.}\
  \bibnamefont {Jin}}, \bibinfo {author} {\bibfnamefont {S.-Y.}\ \bibnamefont
  {Xu}}, \bibinfo {author} {\bibfnamefont {S.-M.}\ \bibnamefont {Huang}},
  \bibinfo {author} {\bibfnamefont {D.-D.}\ \bibnamefont {Guan}}, \bibinfo
  {author} {\bibfnamefont {S.}~\bibnamefont {Wang}}, \bibinfo {author}
  {\bibfnamefont {Y.-Y.}\ \bibnamefont {Li}}, \bibinfo {author} {\bibfnamefont
  {C.}~\bibnamefont {Liu}}, \bibinfo {author} {\bibfnamefont {D.}~\bibnamefont
  {Qian}}, \bibinfo {author} {\bibfnamefont {W.}~\bibnamefont {Ku}}, \bibinfo
  {author} {\bibfnamefont {F.}~\bibnamefont {Song}}, \bibinfo {author}
  {\bibfnamefont {H.}~\bibnamefont {Lin}}, \bibinfo {author} {\bibfnamefont
  {H.}~\bibnamefont {Zheng}},\ and\ \bibinfo {author} {\bibfnamefont {J.-F.}\
  \bibnamefont {Jia}},\ }\bibfield  {title} {\bibinfo {title} {Quasiparticle
  interference and nonsymmorphic effect on a floating band surface state of
  {ZrSiSe}},\ }\href {https://doi.org/10.1038/s41467-018-06661-9} {\bibfield
  {journal} {\bibinfo  {journal} {Nat. Commun.}\ }\textbf {\bibinfo {volume}
  {9}},\ \bibinfo {pages} {4153} (\bibinfo {year} {2018})}\BibitemShut
  {NoStop}%
\bibitem [{\citenamefont {Onsager}(1952)}]{Onsager1952}%
  \BibitemOpen
  \bibfield  {author} {\bibinfo {author} {\bibfnamefont {L.}~\bibnamefont
  {Onsager}},\ }\bibfield  {title} {\bibinfo {title} {Interpretation of the {de
  Haas}-{van Alphen} effect},\ }\href
  {https://doi.org/10.1080/14786440908521019} {\bibfield  {journal} {\bibinfo
  {journal} {The London, Edinburgh, and Dublin Philosophical Magazine and
  Journal of Science}\ }\textbf {\bibinfo {volume} {43}},\ \bibinfo {pages}
  {1006} (\bibinfo {year} {1952})}\BibitemShut {NoStop}%
\bibitem [{\citenamefont {Girvin}\ and\ \citenamefont
  {Yang}(2019)}]{GirvinYang2019}%
  \BibitemOpen
  \bibfield  {author} {\bibinfo {author} {\bibfnamefont {S.~M.}\ \bibnamefont
  {Girvin}}\ and\ \bibinfo {author} {\bibfnamefont {K.}~\bibnamefont {Yang}},\
  }\href {https://doi.org/10.1017/9781316480649} {\emph {\bibinfo {title}
  {Modern Condensed Matter Physics}}}\ (\bibinfo  {publisher} {Cambridge
  University Press},\ \bibinfo {address} {Cambridge},\ \bibinfo {year}
  {2019})\BibitemShut {NoStop}%
\bibitem [{\citenamefont {Lifshitz}\ and\ \citenamefont
  {Kosevich}(1956)}]{LifshitzKosevich1956}%
  \BibitemOpen
  \bibfield  {author} {\bibinfo {author} {\bibfnamefont {I.~M.}\ \bibnamefont
  {Lifshitz}}\ and\ \bibinfo {author} {\bibfnamefont {A.~M.}\ \bibnamefont
  {Kosevich}},\ }\bibfield  {title} {\bibinfo {title} {Theory of magnetic
  susceptibility in metals at low temperatures},\ }\href@noop {} {\bibfield
  {journal} {\bibinfo  {journal} {Sov. Phys. JETP}\ }\textbf {\bibinfo {volume}
  {2}},\ \bibinfo {pages} {636} (\bibinfo {year} {1956})}\BibitemShut {NoStop}%
\bibitem [{\citenamefont {Shoenberg}(1984)}]{Shoenberg1984}%
  \BibitemOpen
  \bibfield  {author} {\bibinfo {author} {\bibfnamefont {D.}~\bibnamefont
  {Shoenberg}},\ }\href {https://doi.org/10.1017/CBO9780511897870} {\emph
  {\bibinfo {title} {Magnetic Oscillations in Metals}}}\ (\bibinfo  {publisher}
  {Cambridge University Press},\ \bibinfo {address} {Cambridge},\ \bibinfo
  {year} {1984})\BibitemShut {NoStop}%
\bibitem [{\citenamefont {Liu}\ \emph {et~al.}(2021)\citenamefont {Liu},
  \citenamefont {Yue}, \citenamefont {Erohin}, \citenamefont {Zhu},
  \citenamefont {Joshy}, \citenamefont {Liu}, \citenamefont {S{\'{a}}nchez},
  \citenamefont {Graf}, \citenamefont {Sorokin}, \citenamefont {Mao},
  \citenamefont {Hu},\ and\ \citenamefont {Wei}}]{Liu2021}%
  \BibitemOpen
  \bibfield  {author} {\bibinfo {author} {\bibfnamefont {X.}~\bibnamefont
  {Liu}}, \bibinfo {author} {\bibfnamefont {C.}~\bibnamefont {Yue}}, \bibinfo
  {author} {\bibfnamefont {S.~V.}\ \bibnamefont {Erohin}}, \bibinfo {author}
  {\bibfnamefont {Y.}~\bibnamefont {Zhu}}, \bibinfo {author} {\bibfnamefont
  {A.}~\bibnamefont {Joshy}}, \bibinfo {author} {\bibfnamefont
  {J.}~\bibnamefont {Liu}}, \bibinfo {author} {\bibfnamefont {A.~M.}\
  \bibnamefont {S{\'{a}}nchez}}, \bibinfo {author} {\bibfnamefont
  {D.}~\bibnamefont {Graf}}, \bibinfo {author} {\bibfnamefont {P.~B.}\
  \bibnamefont {Sorokin}}, \bibinfo {author} {\bibfnamefont {Z.}~\bibnamefont
  {Mao}}, \bibinfo {author} {\bibfnamefont {J.}~\bibnamefont {Hu}},\ and\
  \bibinfo {author} {\bibfnamefont {J.}~\bibnamefont {Wei}},\ }\bibfield
  {title} {\bibinfo {title} {Quantum transport of the {2D} surface state in a
  nonsymmorphic semimetal},\ }\href
  {https://doi.org/10.1021/acs.nanolett.0c04946} {\bibfield  {journal}
  {\bibinfo  {journal} {Nano Lett.}\ }\textbf {\bibinfo {volume} {21}},\
  \bibinfo {pages} {4887} (\bibinfo {year} {2021})}\BibitemShut {NoStop}%
\bibitem [{\citenamefont {Yen}\ and\ \citenamefont {Guo}(2020)}]{Yen2020}%
  \BibitemOpen
  \bibfield  {author} {\bibinfo {author} {\bibfnamefont {Y.}~\bibnamefont
  {Yen}}\ and\ \bibinfo {author} {\bibfnamefont {G.-Y.}\ \bibnamefont {Guo}},\
  }\bibfield  {title} {\bibinfo {title} {Tunable large spin {Hall} and spin
  {Nernst} effects in the {Dirac} semimetals {ZrXY} ({X} = {Si}, {Ge}; {Y} =
  {S}, {Se}, {Te})},\ }\href {https://doi.org/10.1103/PhysRevB.101.064430}
  {\bibfield  {journal} {\bibinfo  {journal} {Phys. Rev. B}\ }\textbf {\bibinfo
  {volume} {101}},\ \bibinfo {pages} {064430} (\bibinfo {year}
  {2020})}\BibitemShut {NoStop}%
\bibitem [{\citenamefont {Ebad-Allah}\ \emph {et~al.}(2022)\citenamefont
  {Ebad-Allah}, \citenamefont {Rojewski}, \citenamefont {Zhu}, \citenamefont
  {Mao},\ and\ \citenamefont {Kuntscher}}]{EbadAllah2022}%
  \BibitemOpen
  \bibfield  {author} {\bibinfo {author} {\bibfnamefont {J.}~\bibnamefont
  {Ebad-Allah}}, \bibinfo {author} {\bibfnamefont {S.}~\bibnamefont
  {Rojewski}}, \bibinfo {author} {\bibfnamefont {Y.~L.}\ \bibnamefont {Zhu}},
  \bibinfo {author} {\bibfnamefont {Z.~Q.}\ \bibnamefont {Mao}},\ and\ \bibinfo
  {author} {\bibfnamefont {C.~A.}\ \bibnamefont {Kuntscher}},\ }\bibfield
  {title} {\bibinfo {title} {In-plane and out-of-plane optical response of the
  nodal-line semimetals {ZrGeS} and {ZrGeSe}},\ }\href
  {https://doi.org/10.1103/PhysRevB.106.075143} {\bibfield  {journal} {\bibinfo
   {journal} {Phys. Rev. B}\ }\textbf {\bibinfo {volume} {106}},\ \bibinfo
  {pages} {075143} (\bibinfo {year} {2022})}\BibitemShut {NoStop}%
\bibitem [{\citenamefont {Dressel}\ and\ \citenamefont
  {Gr{\"u}ner}(2002)}]{DresselGruner2002}%
  \BibitemOpen
  \bibfield  {author} {\bibinfo {author} {\bibfnamefont {M.}~\bibnamefont
  {Dressel}}\ and\ \bibinfo {author} {\bibfnamefont {G.}~\bibnamefont
  {Gr{\"u}ner}},\ }\href {https://doi.org/10.1017/CBO9780511606168} {\emph
  {\bibinfo {title} {Electrodynamics of Solids: Optical Properties of Electrons
  in Matter}}}\ (\bibinfo  {publisher} {Cambridge University Press},\ \bibinfo
  {address} {Cambridge},\ \bibinfo {year} {2002})\BibitemShut {NoStop}%
\bibitem [{\citenamefont {Grosso}\ and\ \citenamefont {{Pastori
  Parravicini}}(2000)}]{Grosso}%
  \BibitemOpen
  \bibfield  {author} {\bibinfo {author} {\bibfnamefont {G.}~\bibnamefont
  {Grosso}}\ and\ \bibinfo {author} {\bibfnamefont {G.}~\bibnamefont {{Pastori
  Parravicini}}},\ }\href {https://books.google.de/books?id=L5RrQbbvWn8C}
  {\emph {\bibinfo {title} {Solid State Physics}}}\ (\bibinfo  {publisher}
  {Elsevier Science},\ \bibinfo {year} {2000})\BibitemShut {NoStop}%
\end{thebibliography}%

\end{document}